\documentclass[twocolumn,superscriptaddress, reqno, nobibnotes, aps, prb,showpacs,footinbib]{revtex4-2}
\usepackage{amsmath}
\usepackage{mathtools}
\usepackage{graphicx}
\usepackage{amsfonts}
\usepackage{bbold}
\usepackage{amssymb}
\usepackage{bm}
\usepackage{dsfont}
\usepackage{array}
\usepackage{tikz}
\usepackage{comment}
\usepackage{subcaption}
\usepackage{enumitem}
\usetikzlibrary{math}
\usetikzlibrary{matrix}
\usepackage{pgfplots}
\usepackage{makecell}
\usepackage{csquotes}
\usepackage{multirow}
\usepackage{boldline}
\usepackage{ragged2e}
\pgfplotsset{compat=1.16}
\usepackage{epsfig,float,afterpage}
\usepackage[colorlinks=true,linkcolor=blue,citecolor=blue,urlcolor=black]{hyperref}
\usepackage{pdfcomment}
\usepackage{soul,xcolor}
\allowdisplaybreaks

\newcommand{\beq}{\begin{equation}}
	\newcommand{\eeq}{\end{equation}}
\newcommand{\bes}{\begin{subequations}}
	\newcommand{\ees}{\end{subequations}}
\newcommand{\bea}{\begin{eqnarray}}
	\newcommand{\eea}{\end{eqnarray}}
\newcommand{\ba}{\begin{array}}
	\newcommand{\ea}{\end{array}}
\newcommand{\beqn}{\begin{eqnarray*}}
	\newcommand{\eeqn}{\end{eqnarray*}}

\newcommand{\n}{\eta}

\newcommand{\tr}{\text{tr}}

\begin{document}
    \title{Ergodic and Discrete Time Crystal Phases in Periodically Kicked Many-Body Quantum Systems: An Analytical Study}
	\author{Vijay Kumar}
	\affiliation{Raman Research Institute, Bangalore 560080, India}
	
	\author{Dibyendu Roy}
	\affiliation{Raman Research Institute, Bangalore 560080, India}

    \begin{abstract}
    We analytically study the time evolution of the expectation values of observables in periodically kicked many-body quantum systems. Starting from an initial state, we compute both the transient and the long-time properties of the observables. Our derivation explains the criteria and the mechanism that lead to the infinite-temperature statistical average of observables at long times, irrespective of the initial state. When the criteria are violated, the observables oscillate with time. These oscillations are subharmonic and robust to small perturbations, suggesting the emergence of a discrete time crystal phase. We demonstrate these features explicitly in periodically kicked nonintegrable spin chains. For a spin chain with two kicks per cycle, we show that the kicked chain can exhibit an ergodic or a discrete-time crystal phase for the same kicking strengths, depending on the initial state preparation. We complement our time-evolution study of observables with the spectral form factor of these kicked models.
    \end{abstract}
    \maketitle

    Thermalization in isolated many-body systems is a central problem in statistical physics since its inception. A many-body system thermalizes if (\romannumeral 1) observables reach a constant value at long times, (\romannumeral 2) the constant value is independent of the initial state, and (\romannumeral 3) it matches with statistical physics prediction \cite{rigol2008thermalization}. In classical systems, dynamical chaos leads to phase-space trajectories that uniformly cover the constant-energy hypersurface. This property leads to the long-time average of observables matching the phase-space average—the ergodic hypothesis that underlies the foundation of classical statistical physics \cite{sinai1963foundations,sinai1970dynamical,bunimovich1979ergodic,simanyi2004proof}. 

    In quantum systems, the problem is more complicated because there is no phase-space description due to the Heisenberg uncertainty principle. Nevertheless, following the identification of quantum chaos through random matrix–like level statistics \cite{BohigasPRL1984,McDonaldPRL1979,Casati1980,Berry1977,Berry1981}, the eigenstate thermalization hypothesis (ETH) \cite{Deutsch1991thermalization,Srednicki1994ETH} was proposed, inspired by the apparent randomness of eigenstates. The ETH provides the current framework for understanding thermalization in isolated quantum systems and has been extensively tested numerically and experimentally \cite{rigol2008thermalization,srednicki1999approach,SantosRigol2010}. Additionally, when interacting many-body quantum systems are driven periodically in time, the total energy is no longer conserved. In this case, time-averaged local observables are expected to match the infinite temperature-statistical average (ITSA), which is described by the Floquet ETH \cite{Lazarides2014FETH,DAlessio2014FETH}. To the best of our knowledge, no study has yet shown thermalization of strongly interacting many-body quantum systems by analytically calculating the expectation values of local observables.
    
    In this Letter, we take on this challenge. We consider disordered strongly interacting many-body quantum systems in an arbitrary initial state. We express the expectation value of an observable as a sum over pairs of paths on our computational basis. Following disorder averaging enabled by a random-phase approximation (RPA), we find that, at long times, the total contribution from pairs of identical paths is equal to the ITSA. This happens because the collective dynamics along such pairs of paths is governed by a doubly stochastic matrix $\mathcal{M}$. At long times, only the largest eigenvalue, 1, and the corresponding eigenvector of $\mathcal{M}$ contribute, leading to the ITSA. 
       Additionally, the long-time contribution of other pairs of paths is $\mathcal{O}(1/\mathcal{N})$ or smaller, where $\mathcal{N}$ is the Hilbert space dimension. Therefore, their contribution is only important for finite-size systems. Nevertheless, we computed these corrections and show a good match with direct numerical simulation results. Furthermore, we find that these $\mathcal{O}(1/\mathcal{N})$ corrections depend on the initial state. However, this behavior is expected even in ETH, where the fluctuation term contains initial-state information. Our study also reveals that when the matrix $\mathcal{M}$ has unimodular eigenvalues distinct from 1, the expectation value of an observable does not reach ITSA; instead, it oscillates. These oscillations are subharmonic and robust to small perturbations. Therefore, this is the discrete time crystal phase \cite{wilczek2012quantum_Time_Crystal,khemani2016_Time_Crystal,Else_Time_Crystal_2016,choi2017_Time_Crystal,zhang2017_Time_Crystal,Zaletel_Time_crystal_2023}. We establish general constraints on $\mathcal{M}$ to find disordered interacting spin models that show such behavior. We computed expectation values of the local magnetization and total energy to demonstrate both the ergodic and discrete-time crystal phases. In a model with two kicks per cycle, we show that the kicked system can exhibit the ergodic or discrete-time crystal phase for the same kicking strengths, depending on the initial state preparation. We also computed the spectral form factor for these phases. 
    
    We study a class of periodically kicked interacting many-body quantum systems whose Hamiltonian takes the following form


    \begin{align}
		\hat{H}(t)&=\hat{H}_0+\hat{H}_1\sum_{n\in \mathbb{Z}}\delta\left(\frac{t}{\tau_p}-n\right),
		\label{gen_Ham_paper}
	\end{align}
    where $\hat{H}_0$ is the base Hamiltonian, $\hat{H}_1$ is the driving Hamiltonian, and $\mathbb{Z}$ is the set of integers. This Hamiltonian is periodic in time with period $\tau_p$. We choose $\tau_p=1$. The base Hamiltonian $\hat{H}_0$ contains a coupling of particles/spins with a random classical field. For a fermionic system, this could be random onsite potentials, whereas this could be a random magnetic field in the $z$ spatial direction for a lattice of spins. The base Hamiltonian also contains long-range interactions. In general, if we denote the local degrees of freedom by operators $\hat{O}_i$ where $i$ is the site index and takes values $i=1,...,L$, then the base Hamiltonian can be expressed as follows
	\begin{align}
		\hat{H}_0&=\sum_{i=1}^L \epsilon_i \hat{O}_i+\sum_{i<j}\frac{U_{ij}}{(d_{ij})^\alpha}\hat{O}_i\hat{O}_j,
		\label{S_H0_generic_paper}
	\end{align} 
	where $\epsilon_i$'s represent onsite potentials, $U_{ij}$'s are parameters associated with long-range interactions, and $d_{ij}$ is the distance between a pair of sites labeled $i$ and $j$. For a fermionic chain $\hat{O}_i\equiv \hat{n}_i$ and for a spin chain $\hat{O}_i\equiv \hat{\sigma}^z_i$ where $\hat{n}_i$ is the operator for the number of fermions at a site $i$ and $\hat{\sigma}^z_i$ is the Pauli $z$-operator at the $i^{\text{th}}$ site. We take $\epsilon_i$'s and $U_{ij}$'s as independent Gaussian random numbers with mean $\langle\epsilon_i\rangle=\epsilon$, $\langle U_{ij}\rangle=U_0$ and standard deviation $\sqrt{\langle\epsilon_i^2\rangle-\langle\epsilon_i\rangle^2}=\Delta\epsilon$, and $\sqrt{\langle U_{ij}^2\rangle-\langle U_{ij}\rangle^2}=\Delta U_0$. We also consider periodic boundary conditions, which implies $d_{ij}=min(|i-j|,L-|i-j|)$. From the form of $\hat{H}_0$ in Eq.~(\ref{S_H0_generic_paper}), it is clear that the eigenstates of $\hat{H}_0$ are also the eigenstates of local operators $\hat{O}_i,\forall i$. We denote them by $|\underline{o}\rangle\equiv |o_1,...,o_L\rangle$ where $o_i$ is an eigenvalue of $\hat{O}_i$. The driving Hamiltonian $\hat{H}_1$ causes transitions between different many-body states $|\underline{o}\rangle$ and does not contain any disorder. We take an arbitrary initial state $|\psi\rangle$ at $t=0$ and an observable $\hat{A}$. We then compute the expectation value of $\hat{A}$, $\langle\psi|\hat{A}(t)|\psi\rangle$ at $t=1,2,...$ under stroboscopic evolution generated by the Floquet operator $\hat{U}=\hat{V}\hat{W}$, where $\hat{V}=e^{-i\hat{H}_1}$ and $\hat{W}=e^{-i\hat{H}_0}$. Since $\langle\psi|\hat{A}(t)|\psi\rangle=\langle\psi|\hat{U}^{-t}\hat{A}\hat{U}^t|\psi\rangle$, we first compute the time-evolved state $\hat{U}^t|\psi\rangle$. Inserting identities $\hat{I}=\sum_{\underline{o}_\tau}|\underline{o}_\tau\rangle\langle \underline{o}_\tau|$ for $\tau=0,...,t$, we obtain
    \begin{align}
        \hat{U}^t|\psi\rangle&=\sum_{\underline{o}_t}\langle\underline{o}_t|\psi\rangle|\underline{o}_t\rangle,
        \label{psi_t}
    \end{align}
    where
    \begin{align}
        \langle\underline{o}_t|\psi\rangle=\sum_{\underline{o}_0,...,\underline{o}_{t-1}}\langle\underline{o}_0|\psi\rangle\prod_{\tau=1}^{t-1}U_{\underline{o}_{\tau+1},\underline{o}_\tau},
        \label{psi_coeff_evol}
    \end{align}
    and
    \begin{align}
        U_{\underline{o}_{\tau+1},\underline{o}_{\tau}}&=\langle\underline{o}_{\tau+1}|\hat{U}|\underline{o}_{\tau}\rangle\notag\\
        &=\langle\underline{o}_{\tau+1}|\hat{V}\hat{W}|\underline{o}_{\tau}\rangle=e^{-i\theta_{\underline{o}_\tau}}V_{\underline{o}_{\tau+1},\underline{o}_\tau},
        \label{U_elem}
    \end{align}
    where $\theta_{\underline{o}_\tau}$ are the eigenphases of $\hat{W}$. Following Eq.~(\ref{S_H0_generic_paper}), $\theta_{\underline{o}}$ can be expressed as follows
    \begin{align}
        \theta_{\underline{o}}&=\sum_{j=1}^L\epsilon_j o_j+\sum_{j<k}\frac{U_{jk}}{(d_{jk})^\alpha}o_j o_k.
    \end{align}
    Equation~(\ref{psi_coeff_evol}) expresses the amplitude $\langle\underline{o}_t|\psi\rangle$ of the evolved state $\hat{U}^t|\psi\rangle$ in terms of the amplitudes $\langle\underline{o}_0|\psi\rangle$ of the initial state $|\psi\rangle$ corresponding to different basis states $|\underline{o}_0\rangle$ and the matrix elements of the Floquet operator in the computational basis. Furthermore, Eq.~(\ref{psi_coeff_evol}) has a path integral type pictorial representation (see SM \cite{SMthermalization} for details). A path $\mathbf{\underline{o}}$ is described by a sequence of basis states $\mathbf{\underline{o}}\equiv(\underline{o}_0,...,\underline{o}_t)$. The amplitude along this path is $R[\mathbf{\underline{o}}]\equiv \langle\underline{o}_0|\psi\rangle\prod_{\tau=0}^{t-1}U_{\underline{o}_{\tau+1},\underline{o}_\tau}$. Therefore, Eq.~(\ref{psi_coeff_evol}) can be interpreted as a sum of amplitudes along the path $\mathbf{\underline{o}}$ ending in a particular state $|\underline{o}_t\rangle$. Thus, following Eq.~(\ref{psi_t}), $\langle\psi|\hat{A}(t)|\psi\rangle$ can be expressed as a sum over all pairs of paths $\mathbf{\underline{o}}$ and $\mathbf{\underline{o}'}$ without any restriction as follows
    \begin{align}
        \langle\psi|\hat{A}(t)|\psi\rangle&=\sum_{\underline{o}'_t,\underline{o}_t}\langle\underline{o}_t|\psi\rangle \langle\underline{o}'_t|\psi\rangle^* A_{\underline{o}'_t,\underline{o}_t}\notag\\
        &=\sum_{\mathbf{\underline{o}}}\sum_{\mathbf{\underline{o}'}}R[\mathbf{\underline{o}}]R^*[\mathbf{\underline{o}'}]A_{\underline{o}'_t,\underline{o}_t}.
        \label{At_as_sum_over_paths}
    \end{align}
    At this point, we average over the disorder present in $\hat{H}_0$ to write
    \begin{align}
        \langle\psi|\hat{A}(t)|\psi\rangle_{\text{dis}}&=\sum_{\mathbf{\underline{o}}}\sum_{\mathbf{\underline{o}'}}\langle R[\mathbf{\underline{o}}]R^*[\mathbf{\underline{o}'}]\rangle_{\text{dis}} A_{\underline{o}'_t,\underline{o}_t}.
        \label{At_diss_average}
    \end{align}
    Following Eqs.~(\ref{psi_coeff_evol}) and (\ref{U_elem}), we can express
    \begin{align}
        \langle R[\mathbf{\underline{o}}]R^*[\mathbf{\underline{o}'}]\rangle_{\text{dis}}=&\langle\underline{o}_0|\psi\rangle \langle\underline{o}'_0|\psi\rangle^*\langle e^{-i\sum_{\tau=0}^{t-1}(\theta_{\underline{o}_\tau}-\theta_{\underline{o}'_\tau})}\rangle_{\text{dis}}\notag\\
        &\times\prod_{\tau=0}^{t-1}V_{\underline{o}_{\tau+1},\underline{o}_\tau}V^*_{\underline{o}'_{\tau+1},\underline{o}'_\tau}.
        \label{Amplitude_product}
    \end{align}
    The direct simulation of $\langle\psi|\hat{A}(t)|\psi\rangle$ reveals that its value in the limit of strong disorder $\Delta\epsilon\gg 1$ and $\Delta U_0\gg 1$ matches that of a random phase model (RPM) (see SM \cite{SMthermalization} for details). In the RPM, the matrix $W$ is replaced by a diagonal random matrix, $\text{diag}(e^{-i\theta_1},...,e^{-i\theta_{\mathcal{N}}})$, where $\theta_n$, $\forall n\in\{1,...,\mathcal{N}\}$, is an independent random number distributed uniformly over $[0,2\pi)$ and $\mathcal{N}$ is the dimension of the Hilbert space. Thus, we make this approximation that the phases $\theta_{\underline{o}_\tau}$, $\forall \underline{o}_\tau$, are independent and uniformly distributed random numbers over $[0,2\pi)$. We call this the random phase approximation (RPA) \cite{KosPRX2018,RoyPRE2020,RoyPRE2022,Kumar2024,Kumar2025leading}. Thus,
    \begin{align}
        \langle e^{-i\sum_{\tau=0}^{t-1}(\theta_{\underline{o}_\tau}-\theta_{\underline{o}'_\tau})}\rangle_{\text{dis}}&=\langle e^{-i\sum_{\tau=0}^{t-1}(\theta_{\underline{o}_\tau}-\theta_{\underline{o}'_\tau})}\rangle_{\text{RPA}}\notag\\
        &=\prod_{\tau=0}^{t-1}\delta_{\underline{o}'_{\tau},\underline{o}_{\pi(\tau)}},
        \label{RPA_Eq}
    \end{align}
    \begin{figure*}[t!]
        \centering
        \includegraphics[width=\linewidth]{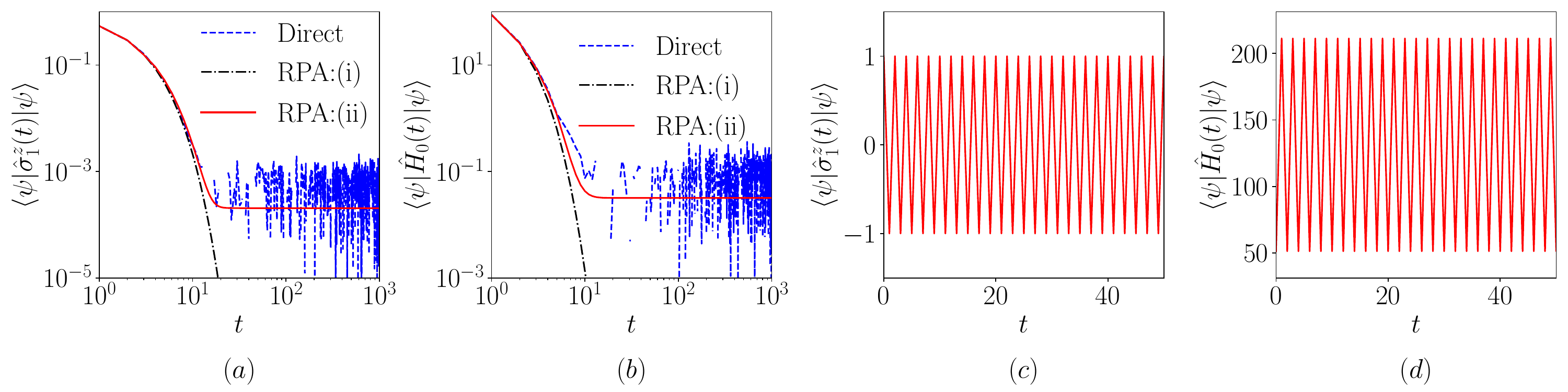}
        \caption{\justifying{\small Time-evolution of (a) local magnetization and (b) total energy from direct numerics and our analytics using the RPA for a periodically kicked long-range transverse-field Ising chain. Here, $L=14, h=0.5, \epsilon=10,\Delta\epsilon=10,U_0=20\sqrt{2},\Delta U_0=10\sqrt{2},\alpha=1.5,|\psi\rangle=|1,1,1,1,1,1,1,-1,-1,-1,-1,-1,-1,-1\rangle$. Averaging over 160 realizations of disorder is performed in each case. Time-evolution of local magnetization (c) and total energy (d) at a special kicking strength $h=\pi/2$ for $L=14$ and initial state $|\psi\rangle=|1,1,1,-1,-1,-1,-1,-1,-1,-1,-1,-1,-1,-1\rangle$}.}
        \label{KIC_observables_thermal_n_DF}
    \end{figure*}
    where $\pi$ is a permutation of $t$ objects. Equation~(\ref{RPA_Eq}) implies that a nonzero contribution in Eq.~\ref{At_diss_average} results only from pairs of paths $\mathbf{\underline{o}},\mathbf{\underline{o}'}$, where a path $\mathbf{\underline{o}'}$ has the same states as the path $\mathbf{\underline{o}}$ for $\tau=0,...,t-1$, allowing only their ordering to be different. Computing the contribution of different pairs of paths, we find that at long times, the leading contribution comes from pairs of paths, where $\mathbf{\underline{o}}\equiv \mathbf{\underline{o}'}$ for $\tau=0,...,t-1$. The contribution of all other pairs of paths is $\mathcal{O}(1/\mathcal{N}^k)$, where $k\geq 1$ at long times. Thus, in the thermodynamic limit, the long time behavior of the expectation value is only determined by pairs of paths satisfying $\mathbf{\underline{o}}\equiv \mathbf{\underline{o}'}$ for $\tau=0,...,t-1$, which corresponds to identity permutation $\pi=I$ in Eq.~(\ref{RPA_Eq}). We denote their contribution by $\langle\psi|\hat{A}(t)|\psi\rangle_{\text{RPA},I}$. Following Eqs.~(\ref{At_as_sum_over_paths}) and (\ref{Amplitude_product}), we obtain
    \begin{align}
        \langle\psi|\hat{A}(t)|\psi\rangle_{\text{RPA},I}=&\sum_{\underline{o}_0,\underline{o}_{t-1},\underline{o}_{t},\underline{o}'_{t}}|\langle\underline{o}_0|\psi\rangle|^2 A_{\underline{o}'_{t},\underline{o}_{t}}\notag\\
        &\times V_{\underline{o}_{t},\underline{o}_{t-1}}V^*_{\underline{o}'_{t},\underline{o}_{t-1}}(\mathcal{M}^{t-1})_{\underline{o}_{t-1},\underline{o}_0},
        \label{At_I}
    \end{align}
    where $\mathcal{M}$ is a doubly stochastic matrix whose elements are related to the elements of the matrix $V$ as $\mathcal{M}_{\underline{o},\underline{o}'}=|V_{\underline{o},\underline{o}'}|^2$. The eigenvalues of $\mathcal{M}$ are $\lambda_0,\lambda_1,...,\lambda_{\mathcal{N}-1}$ that satisfy $\lambda_0=1$ and $1>|\lambda_1|\geq ...\geq |\lambda_{\mathcal{N}-1}|$. Thus, performing eigendecomposition of $\mathcal{M}$, we obtain
    \begin{align}
        \langle\psi|\hat{A}(t)|\psi\rangle_{\text{RPA},I}=&\sum_{\underline{o}_0,\underline{o}_{t-1},\underline{o}_{t},\underline{o}'_{t}}\sum_{i=0}^{\mathcal{N}-1}|\langle\underline{o}_0|\psi\rangle|^2 A_{\underline{o}'_{t},\underline{o}_{t}}\notag\\
        &\times V_{\underline{o}_{t},\underline{o}_{t-1}}V^*_{\underline{o}'_{t},\underline{o}_{t-1}}\lambda_i^{t-1}\mathcal{M}^{(i)}_{\underline{o}_{t-1},\underline{o}_0},
        \label{At_I_eig_decomp}
    \end{align}
    where $\mathcal{M}^{(i)}=|\lambda_i\rangle\langle\lambda_i|$. All terms in Eq.~\ref{At_I_eig_decomp} for $i\neq 0$ are exponentially decaying with time since $|\lambda_i|<1$. Thus, the long-time behavior of $\langle\psi|\hat{A}(t)|\psi\rangle_{\text{RPA},I}$ is determined by the terms corresponding to $i=0$. Since $\langle\lambda_0|\equiv (1/\sqrt{\mathcal{N}})(1,...,1)$, we insert $\mathcal{M}^{(0)}_{\underline{o}_{t-1},\underline{o}_0}=1/\mathcal{N}$. We further use $\sum_{\underline{o}_{t-1}}V_{\underline{o}_{t},\underline{o}_{t-1}}V^*_{\underline{o}'_{t},\underline{o}_{t-1}}=\delta_{\underline{o}_{t},\underline{o}'_{t}}$ and $\sum_{\underline{o}_0}|\langle\underline{o}_0|\psi\rangle|^2=1$ to find 
    \begin{align}
        \langle\psi|\hat{A}(t)|\psi\rangle_{\text{RPA},I}\Bigg|_{t\rightarrow\infty}=\frac{1}{\mathcal{N}}\tr\hat{A}.
        \label{Floquet_thermalization}
    \end{align}
    The right hand side in Eq.~(\ref{Floquet_thermalization}) is exactly the ITSA of an observable $\hat{A}$. Thus, our system heats up to infinite temperature in accordance with the Floquet ETH.

    Equation~(\ref{At_I_eig_decomp}) suggests that the properties of the unimodular eigenvalues of the matrix $\mathcal{M}$ determine the thermalization of our systems. In general, there can be three cases:\\
    \textbf{Case 1}: $\lambda_0=1$ is the only unimodular eigenvalue and is nondegenerate. In this case, the expectation values of observables saturate to infinite temperature statistical average at long times as shown in Eq.~(\ref{Floquet_thermalization}).\\
    \textbf{Case 2}: $\lambda_0=1$ is the only unimodular eigenvalue and is degenerate. This happens when the matrix $\mathcal{M}$ is reducible. Therefore, the matrix $\mathcal{M}$ can be brought to a block diagonal form. In this case, the expectation values of observables saturate to infinite temperature statistical average only if the initial state belongs to one of the blocks. For other choices of initial state, the expectation values of the observables reach a constant value at long times. However, this value depends on the initial state (see SM \cite{SMthermalization} for more details).\\
    \textbf{Case 3}: $\lambda_0=1$ is not the only unimodular eigenvalue. This happens when the matrix $\mathcal{M}$ is irreducible but periodic \cite{Horn_Johnson_1985,seneta2006non}. If the period is $l$, then the unimodular eigenvalues are exactly the $l$-th roots of unity, $e^{i 2\pi k/l},k=0,...,l-1$. In this case, the expectation value of an observable oscillates with time where the period of oscillations is $l\tau_p$. In addition, these oscillations are robust under small perturbations to the driving Hamiltonian $\hat{H}_1$ implying that this is the discrete time crystal phase (see SM \cite{SMthermalization} for details).

    \begin{figure*}[t!]
        \centering
        \includegraphics[width=\linewidth]{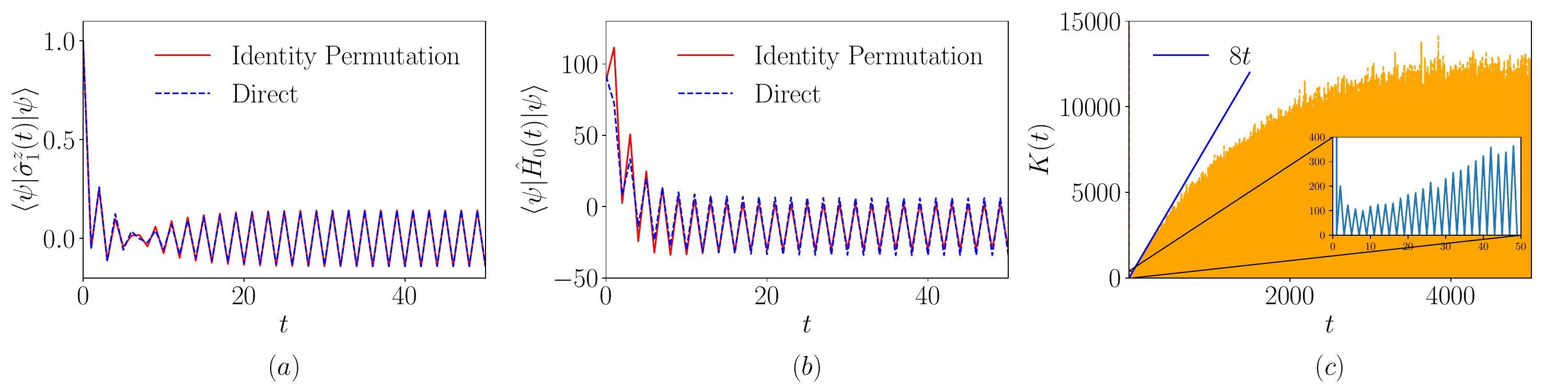}
        \caption{\justifying{\small (a) Time evolution of the local magnetization, (b) time evolution of the total energy, and (c) spectral form factor (SFF) from direct numerics and our analytics using the RPA for a long-range interacting spin chain periodically kicked by the driving Hamiltonians in Eq.~(\ref{H1_2_2kick}). Here, $L=14, J=1,h=\pi/2,\epsilon=10,\Delta\epsilon=10,U_0=10,\Delta U_0=10,\alpha=1.5$ in all plots,  
        $|\psi\rangle=|1,1,1,1,1,1,1,-1,-1,-1,-1,-1,-1,-1\rangle$ in (a,b) and $N=6$ in (c).  Averaging over 320 realizations of disorder is performed for direct numerical simulation in each case.}}
        \label{2kick_DF_obsevrables_n_SFF}
    \end{figure*}
    We now illustrate these cases with explicit examples. We consider a kicked transverse-field Ising chain whose Hamiltonian is described by Eqs.~(\ref{gen_Ham_paper}) and (\ref{S_H0_generic_paper}) with $\hat{O}_i\equiv \hat{\sigma}^z_i$ and $\hat{H}_1=h\sum_{i=1}^L\hat{\sigma}^x_i$ where $\hat{\sigma}_i^x$ is the Pauli $x$-operator at the $i^{\text{th}}$ site. In this case, $\mathcal{M}=m^{\otimes L}$ where
    \begin{align}
        m&=\begin{pmatrix}
            \cos^2(h)& \sin^2(h)\\
            \sin^2(h)& \cos^2(h)
        \end{pmatrix}.
    \end{align}
    The eigenvalues of $m$ are $1$ and $\cos 2h$. Thus, for generic values of $h$, this system corresponds to \textbf{Case 1}. Therefore, at long times, the expectation values of observables such as local magnetization $\hat{\sigma}^z_i$ and total energy $\hat{H}_0$ reach their corresponding ITSA up to the order $\mathcal{O}(1/\mathcal{N})$ corrections. We computed them analytically (see SM \cite{SMthermalization} for details) and compared them with the results of the direct numerical simulations as shown in Figs.~\ref{KIC_observables_thermal_n_DF}a and \ref{KIC_observables_thermal_n_DF}b. The black dashed line in Fig.~\ref{KIC_observables_thermal_n_DF}a is the contribution from identity permutation Eq.~(\ref{At_I}) for local magnetization at site 1. If we take the initial state as a product state $|\psi\rangle=|\psi_1,...,\psi_L\rangle$ where $\psi_i=\pm 1,\forall i$, the contribution to local magnetization resulting from the identity permutation is (see SM \cite{SMthermalization} for details)
    \begin{align}
        \langle\psi|\hat{\sigma}_i^z|\psi\rangle=\langle\psi_i|\hat{\sigma}_i^z|\psi_i\rangle \cos^t(2h),\; \forall i\in\{1,...,L\}.
    \end{align}
    This contribution at a long time limit approaches zero, in agreement with ITSA. However, a direct numerical simulation reveals that local magnetization fluctuates around a non-zero value of  $\mathcal{O}(1/2^L)$. By computing the contribution of other permutations we determine this finite size correction analytically and show an excellent match with the direct numerical simulation result as shown by the red curve in Fig.~\ref{KIC_observables_thermal_n_DF}a. Similar analysis is performed for total energy as well to obtain the red curve in Fig.~\ref{KIC_observables_thermal_n_DF}b.

   The kicked transverse-field Ising chain thermalizes for generic values of $h$. However, when $h=\pm \pi/2,\pm 3\pi/2,...$, the eigenvalues of $m$ are 1 and -1. Therefore, the expectation values of the observables oscillate instead of reaching a constant value at long times (see SM \cite{SMthermalization} for details)
    \begin{align}
        \langle\psi(t)|\hat{\sigma}_j^z|\psi(t)\rangle &=(-1)^t\psi_j,\\
        \langle\psi(t)|\hat{H}_0|\psi(t)\rangle_{\text{dis}}&=(-1)^t\sum_{i=1}^L\epsilon\psi_i+\sum_{i<j}\frac{U_0}{(d_{ij})^\alpha}\psi_i\psi_j.
    \end{align}
    This corresponds to \textbf{Case 3}. As discussed in \textbf{Case 3}, these oscillations are robust to small perturbations in $h$ (see SM \cite{SMthermalization} for details). Therefore, this is a discrete time crystal phase. Furthermore, the amplitude of oscillations of $\langle\psi(t)|\hat{H}_0|\psi(t)\rangle_{\text{dis}}$ depends on the initial state. 
    
    The discrete time crystal phase has been a topic of great interest over the past decade. Therefore, we look for driving Hamiltonians $\hat{H}_1$ that show \textbf{Case 3} type behavior. For simplicity, we look for $\hat{H}_1$ which leads to doubly periodic $\mathcal{M}$, i.e., unimodular eigenvalues are $1$ and $-1$. This is guaranteed to happen if the matrix $\mathcal{M}$ anticommutes with a nonzero matrix $\tau^z$, since that implies that the spectrum of $\mathcal{M}$ is symmetric about the origin of a complex plane. Thus, for each $\lambda_i$, there exists a $\lambda_j=-\lambda_i$, where $j\neq i$. Therefore, both eigenvalues 1 and -1 exist. In particular, block-off diagonal matrices $\mathcal{M}$ anticommute with $\tau^z$ of the form shown below
    \begin{align}
        \mathcal{M}=\begin{pmatrix}
            0& \mathcal{A}\\ 
            \mathcal{B}& 0
        \end{pmatrix},\quad \tau^z=\begin{pmatrix}
            I_{\mathcal{N}/2}& 0\\ 
            0 &-I_{\mathcal{N}/2}
        \end{pmatrix},
    \end{align}
    where both $\mathcal{A}$ and $\mathcal{B}$ are doubly stochastic matrices of size $\mathcal{N}/2\times \mathcal{N}/2$ and $I_{\mathcal{N}/2}$ is the identity matrix of size $\mathcal{N}/2\times \mathcal{N}/2$. Imposing this block-off-diagonal structure on $\mathcal{M}$, we find that the driving Hamiltonian $\hat{H}_1$ must have a spectrum such that if $E_i$ is an eigenvalue, then there exists another eigenvalue $E_j=E_i+(2k+1)\pi$ where $k\in\mathbb{Z}$. In addition, the eigenvectors $|E_i\rangle$ and $|E_j\rangle$ are related to each other in the computational basis by the matrix $\tau^z$ (see SM \cite{SMthermalization} for details). These properties are satisfied by the driving Hamiltonian of the kicked transverse-field Ising chain for $h=\pm\pi/2,\pm 3\pi/2,...$.
    \begin{figure*}[t!]
        \centering
        \includegraphics[width=\linewidth]{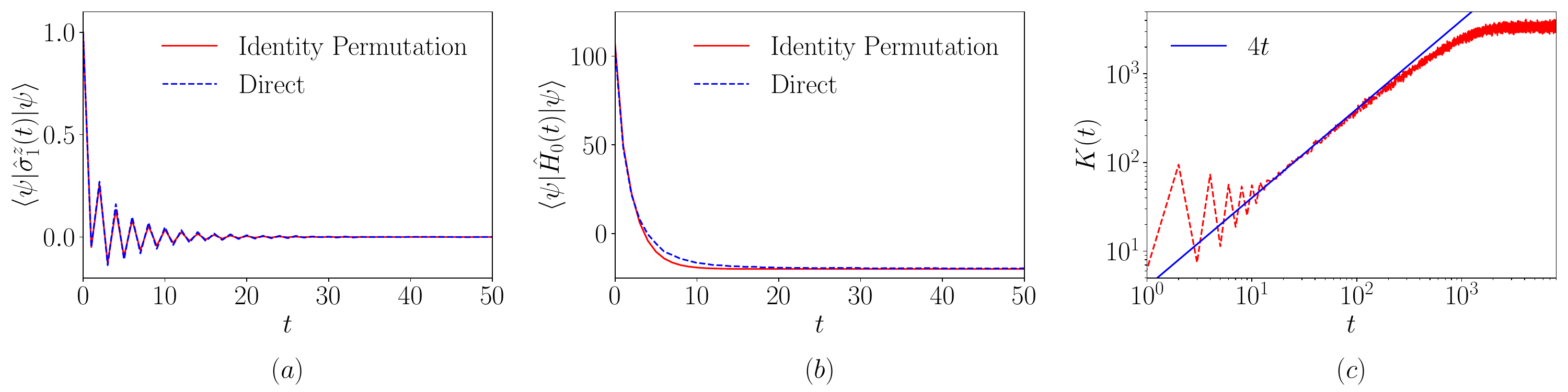}
        \caption{\justifying{\small (a) Time evolution of the local magnetization, (b) time evolution of the total energy, and (c) spectral form factor (SFF) from direct numerics and our analytics using the RPA for a long-range interacting spin chain periodically kicked by the driving Hamiltonians in Eq.~(\ref{H1_2_2kick}). Here, $L=14,J=1,h=\pi/2,\epsilon=10,\Delta\epsilon=10,U_0=10,\Delta U_0=10,\alpha=1.5$ in all plots, $|\psi\rangle=|1,1,1,1,1,1,1,-1,-1,-1,-1,-1,-1,-1\rangle$ in (a,b) and $N=7$ in (c). Averaging over 320 realizations of disorder is performed for direct numerical simulation in each case.}}
        \label{2kick_half_fill_observables_n_SFF}
    \end{figure*}
    In a more generic setup, a block-off-diagonal matrix $\mathcal{M}$ can be obtained by two kicks per cycle:
    \begin{align}
        \hat{H}(t)&=\hat{H}_0+\sum_{n\in\mathbb{Z}}\hat{H}_1\delta\left(\frac{t}{\tau_p}-n-r\right)+\hat{H}_2 \delta\left(\frac{t}{\tau_p}-n \right),
    \end{align}
    where $0<r<1$ and $\tau_p$ is the driving period. To ensure the presence of time reversal symmetry, we choose $r=1/2$. In addition, we choose $\tau_p=1$. The emergence of a doubly stochastic matrix $\mathcal{M}$ for such a system is discussed in SM \cite{SMthermalization}. To demonstrate oscillations of the expectation values of the observables, we take $\hat{H}_0$ as in Eq.~(\ref{S_H0_generic_paper}) with $\hat{O}_i\equiv \hat{\sigma}^z_i$. We take driving $\hat{H}_1$ and $\hat{H}_2$ as follows
    \begin{align}
		\hat{H}_1=&\sum_{i=1}^L J \hat{\sigma}_i^+\hat{\sigma}_{i+1}^-+h.c.,\quad \hat{H}_2=&\sum_{i=1}^L h \hat{\sigma}_{i}^x,
        \label{H1_2_2kick}
	\end{align}
    where $\hat{\sigma}^\pm_i=(\hat{\sigma}^x_i\pm \hat{\sigma}^y_i)/2$, and $\hat{\sigma}_i^y$ is the Pauli $y$-operator at site $i$. In this case, the Floquet operator is $\hat{U}=\hat{X}\hat{W}\hat{V}\hat{W}$ where $\hat{X}=e^{-i \hat{H}_2}$, $\hat{W}=e^{-i \hat{H}_0/2}$, and $\hat{V}=e^{-i \hat{H}_1}$. The operators $\hat{H}_0$ and $\hat{H}_1$ commute with $\hat{N}=(1/2)\sum_{i=1}^L (\hat{\sigma}_i^z+1)$, where $\hat{N}$ measures the total number of spins in $|\uparrow\rangle$ state along the $z$ direction. However, $\hat{N}$ does not commute with $\hat{H}_2$. Therefore, this Hamiltonian does not have any $U(1)$ symmetry. Nevertheless, the Hilbert space $\mathcal{H}$ can still be decomposed into degenerate eigenspaces of $\hat{N}$, $\mathcal{H}=\bigoplus_{N=0}^{L}\mathcal{H}_N$, where $\mathcal{H}_N$ is the degenerate eigenspace associated with an eigenvalue $N$ of $\hat{N}$. The operator $\hat{X}$ couples different $\mathcal{H}_N$.
    Similarly to the one kick per cycle case, the expectation values of observables can also be computed following the RPA in this case (see SM \cite{SMthermalization}). We find that observables saturate to infinite temperature statistical average for arbitrary $J$ and $h\neq \pm \pi/2,\pm 3\pi/2,...$. When $h=\pm\pi/2,\pm3\pi/2,...$, $\hat{X}=(-i)^L\hat{F}$ where $\hat{F}$ is a flip operator that flips all spins in a state. In this case, if $|\psi\rangle\in \mathcal{H}_N$ then $\hat{U}^t|\psi\rangle\in \mathcal{H}_N$ if $t$ is even and $\hat{U}^t|\psi\rangle\in \mathcal{H}_{L-N}$ if $t$ is odd. This feature leads to the emergence of a doubly stochastic block off-diagonal matrix $\mathcal{M}$. As discussed, this matrix structure leads to eigenvalues 1 and -1. Consequently, at long times, the expectation value of an observable $\hat{A}$ oscillates between $(\tr_{\mathcal{H}_N}\hat{A}/\text{dim}(\mathcal{H}_N))$ and $(\tr_{\mathcal{H}_{L-N}}\hat{A}/\text{dim}(\mathcal{H}_{L-N}))$ for even and odd $t$, respectively, as shown in Fig.~\ref{2kick_DF_obsevrables_n_SFF}(a) and Fig.~\ref{2kick_DF_obsevrables_n_SFF}(b). Furthermore, these oscillations are robust to small perturbations in $h$ suggesting that this is a discrete time crystal phase. In addition, we analytically compute the spectral form factor (SFF) for this case (see SM \cite{SMthermalization} for details). We find that after Thouless time, the SFF oscillates between $8t$ and 0 for even and odd $t$, respectively. Direct numerical simulation also reveals the same behavior as shown in Fig.~\ref{2kick_DF_obsevrables_n_SFF}(c). In contrast, when $L$ is even and $\psi\in\mathcal{H}_{L/2}$, the observables saturate to infinite temperature statistical average $\tr_{\mathcal{H}_{L/2}}\hat{A}/\text{dim}(\mathcal{H}_{L/2})$ as shown in Figs.~\ref{2kick_half_fill_observables_n_SFF}(a)-(b). Surprisingly, the SFF has a $4t$ ramp after Thouless time, as shown in Fig.~\ref{2kick_half_fill_observables_n_SFF}(c). The linear ramp in SFF differs from the expected random matrix prediction by a factor of 2 \cite{Dyson_1970,Haake2001,Mehta2004}. Studying the distribution of the eigenphase spacings of the Floquet operator, we find that it matches the Brody distribution for $\beta=0.33$ suggesting that this system has a mixed behavior for $N=L/2$.

     We have analytically calculated the expectation values of observables in periodically kicked many-body quantum systems. We explain how a periodically kicked nonintegrable quantum system, starting from an arbitrary initial state, approaches the expected infinite-temperature state at long times. Our formalism also reveals scenarios in which the expectation values of the observables oscillate rather than reaching a constant value at long times. We also find that these oscillations are subharmonic and robust to small perturbations, indicating the emergence of a discrete-time crystal phase in our analytical study of disordered strongly interacting Floquet models. We also derived the conditions a driving Hamiltonian must satisfy to exhibit a discrete-time crystal phase with a $2\tau_p$ period. Under these conditions, we found a periodically kicked interacting spin chain with two kicks per cycle, which exhibits an ergodic or a discrete-time crystal phase depending on the initial state.


    \bibliography{bibliographyRMT}
    
    \onecolumngrid
    \vspace{10cm}
    
    \begin{center}
    	{\Large \textbf{Supplementary Material for ``Ergodic and Discrete Time Crystal Phases in Periodically Kicked Many-Body Quantum Systems: An Analytical Study''}}\\[1em]
    	
    	Vijay Kumar$^{1}$, Dibyendu Roy$^{1}$\\[0.5em]
    	
    	{\small
    		$^{1}$Raman Research Institute, Bangalore 560080, India
    	}
    \end{center}

   \setcounter{equation}{0}
   \renewcommand{\theequation}{S\arabic{equation}}
   
   \setcounter{figure}{0}
   \renewcommand{\thefigure}{S\arabic{figure}}
   
   \setcounter{table}{0}
   \renewcommand{\thetable}{S\arabic{table}}

    \vspace{1cm}
    	\title{Supplementary Material for ``Ergodic and Discrete Time Crystal Phases in Periodically Kicked Many-Body Quantum Systems: An Analytical Study''}
	\author{Vijay Kumar}
	\affiliation{Raman Research Institute, Bangalore 560080, India}
	
	\author{Dibyendu Roy}
	\affiliation{Raman Research Institute, Bangalore 560080, India}
	\maketitle
	
	\section{Time-evolution of observables in periodically kicked many-body quantum systems}\label{Intro}
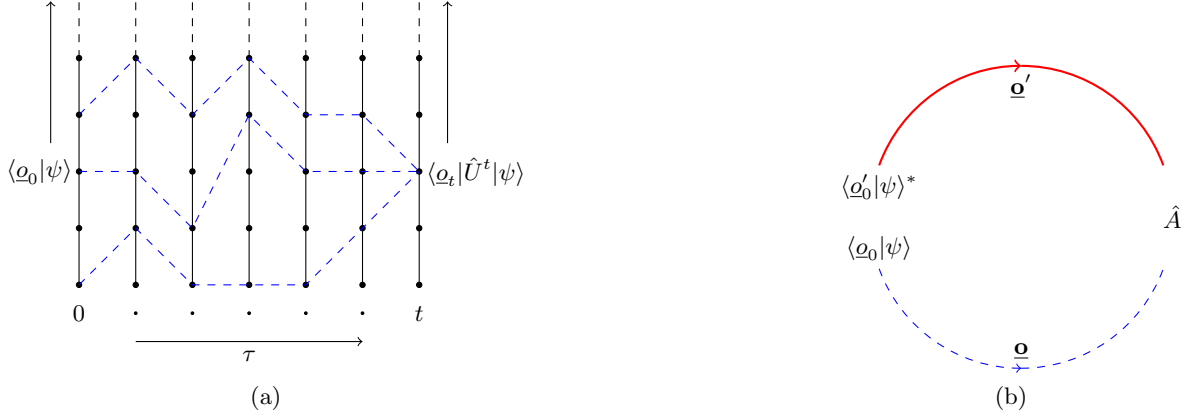
\begin{figure*}[t!]
		\centering
		\begin{subfigure}[t]{0.45\linewidth}
			\centering
			\begin{tikzpicture}
				\def\vstep{0.75}
				\def\nstates{5}
				\def\ntslices{7}
				\def\hstep{0.75}
				\def\dotthickness{1pt}
				\foreach \x in {1,...,\ntslices}{\foreach \n in {1,...,\nstates}{\draw[fill=black] ({\hstep*\x},{\vstep*\n}) circle(\dotthickness);}\draw[black] ({\x*\hstep},\vstep)--({\x*\hstep},\nstates*\vstep);\draw[dashed] ({\x*\hstep},{\nstates*\vstep})--({\x*\hstep},{(\nstates+1)*\vstep});}
				
				\draw ({0.3*\hstep},{3*\vstep}) node[]{$\langle\underline{o}_0|\psi\rangle$};
				\draw[->] ({0.5*\hstep},{(3+0.5)*\vstep})--({0.5*\hstep},{(\nstates+1)*\vstep});
				
				\draw ({(\ntslices+1)*\hstep},{3*\vstep}) node[]{$\langle\underline{o}_t|\hat{U}^t|\psi\rangle$};
				\draw[->] ({(\ntslices+0.5)*\hstep},{(3+0.5)*\vstep})--({(\ntslices+0.5)*\hstep},{(\nstates+1)*\vstep});
				
				\draw (\hstep,{0.5*\vstep}) node[]{$0$};
				\draw ({\hstep*\ntslices},{0.5*\vstep}) node[]{$t$};
				\foreach \x in {2,...,6}{\draw[fill=black] ({\x*\hstep},{0.5*\vstep}) circle(0.5pt);}
				\draw[->] ({2*\hstep},0)--({\hstep*(\ntslices-1)},0);
				\draw ({4*\hstep},0) node[below]{$\tau$};
				
				\draw[blue,dashed] (\hstep,\vstep)--({2*\hstep},{2*\vstep})--({3*\hstep},\vstep)--({4*\hstep},\vstep)--({5*\hstep},\vstep)--({6*\hstep},{2*\vstep})--({7*\hstep},{3*\vstep});
				
				\draw[blue,dashed] (\hstep,{3*\vstep})--({2*\hstep},{3*\vstep})--({3*\hstep},{2*\vstep})--({4*\hstep},{4*\vstep})--({5*\hstep},{3*\vstep})--({6*\hstep},{3*\vstep})--({7*\hstep},{3*\vstep});
				
				\draw[blue,dashed] (\hstep,{4*\vstep})--({2*\hstep},{5*\vstep})--({3*\hstep},{4*\vstep})--({4*\hstep},{5*\vstep})--({5*\hstep},{4*\vstep})--({6*\hstep},{4*\vstep})--({7*\hstep},{3*\vstep});
				
			\end{tikzpicture}
			\caption{}
			\label{psi_evol_diag}
		\end{subfigure}
		\hfill
		\begin{subfigure}[t]{0.45\linewidth}
			\centering
			\begin{tikzpicture}
				\def\rad{2}
				\draw[blue,dashed,->] (200:\rad) arc(200:270:\rad);\draw[blue,dashed] (270:\rad) arc(270:340:\rad);
				\draw[red,thick,->] (160:\rad) arc(160:90:\rad);\draw[red,thick] (90:\rad) arc(90:20:\rad);
				\draw (200:\rad) node[above]{$\langle\underline{o}_0|\psi\rangle$};
				\draw (160:\rad) node[below]{$\langle\underline{o}'_0|\psi\rangle^*$};
				\draw (0:\rad) node[]{$\hat{A}$};
				\draw (90:\rad) node[below]{$\mathbf{\underline{o}'}$};
				\draw (270:\rad) node[above]{$\mathbf{\underline{o}}$};
			\end{tikzpicture}
			\caption{}
			\label{At_summand_as_paths}
		\end{subfigure}
		\caption{\justifying{\small Path integral type decomposition of (a) time evolution of a state $|\psi\rangle$ and (b) expectation value of an observable $\hat{A}$ in the time evolved state. In (a), each vertical line is a time slice with solid black dots on them representing basis states $|\underline{o}\rangle$. An arbitrary initial state $|\psi\rangle$ is described by its amplitudes in the computational basis $\langle\underline{o}_0|\psi\rangle$. These amplitudes sit on the left most time slice labeled $\tau=0$. The amplitudes of its time evolved counterparts are $\langle\underline{o}_t|\hat{U}^t|\psi\rangle=\sum_{\underline{o}_0,...,\underline{o}_{t-1}}\left(\prod_{\tau=0}^{t-1}U_{\underline{o}_{\tau+1},\underline{o}_\tau}\right)\langle\underline{o}_0|\psi\rangle$. These amplitudes sit on the right most time slice labeled $t$. From the above expression, amplitude $\langle\underline{o}_t|\psi\rangle$ can be expressed as a sum of contributions resulting from all the blue dashed paths starting from different black dots on the left most slice and ending on a black dot corresponding to a basis state $\underline{o}_t$ on the right most slice. A path is denoted by $\boldsymbol{\underline{o}}\equiv(\underline{o}_0,...,\underline{o}_t)$ and the amplitude along this is $\left(\prod_{\tau=0}^{t-1}U_{\underline{o}_{\tau+1},\underline{o}_\tau}\right)\langle\underline{o}_0|\psi\rangle$. Since expectation value of an observable contains both ket $\hat{U}^{t}|\psi\rangle$ and bra $\langle\psi|\hat{U}^{-t}$, we have blue dashed paths $\boldsymbol{\underline{o}}\equiv (\underline{o}_0,...,\underline{o}_t)$ corresponding to the ket and solid red paths $\boldsymbol{\underline{o}'}\equiv (\underline{o}'_0,...,\underline{o}'_t)$ corresponding to the bra in (b). Expectation value $\langle\psi|\hat{U}^{-t}\hat{A}\hat{U}^t|\psi\rangle$ can be expressed as sum over pairs of paths $\boldsymbol{\underline{o}}$ and $\boldsymbol{\underline{o}'}$.}}
	\end{figure*}

\begin{figure*}[t!]
		\centering
		\begin{subfigure}[t]{0.45\linewidth}
			\centering
			\includegraphics[width=\linewidth]{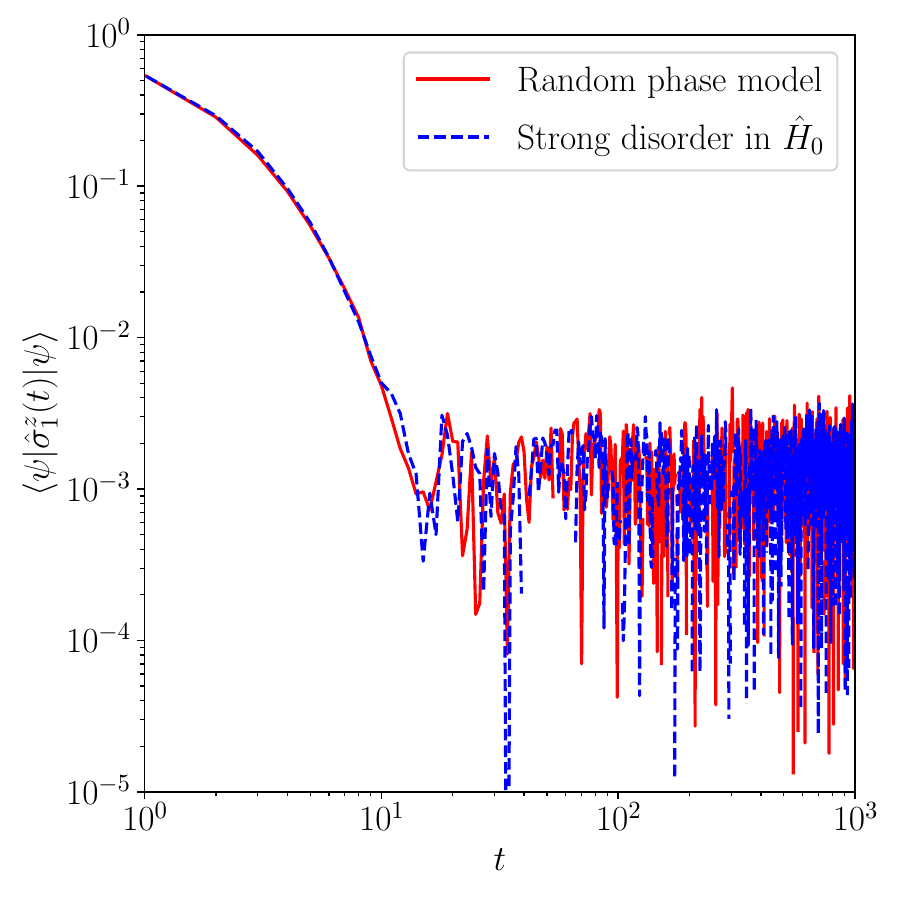}
			\caption{}
		\end{subfigure}
		\hfill
		\begin{subfigure}[t]{0.45\linewidth}
			\centering
			\includegraphics[width=\linewidth]{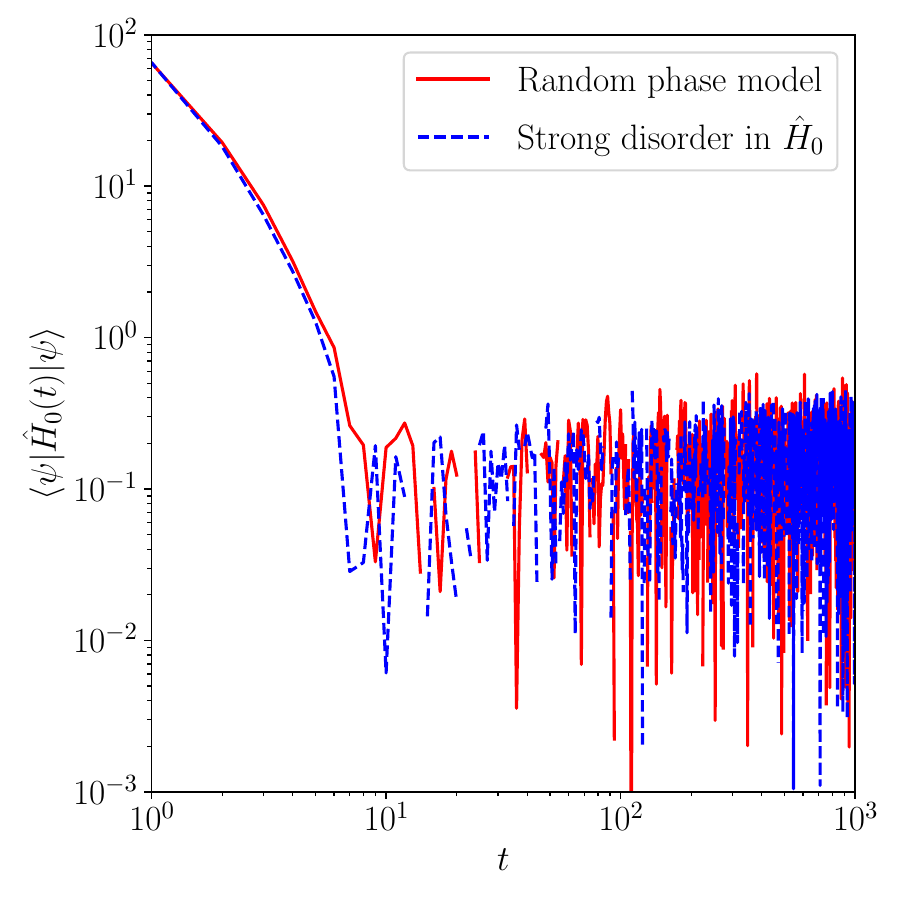}
			\caption{}
		\end{subfigure}
		\caption{\justifying{\small Time evolution of (a) local magnetization and (b) total energy from direct numerical simulation for a periodically kicked long-range transverse-field Ising chain with strong disorder. The results are compared with direct simulation of corresponding random phase model. Here $L=12, h=0.5, \epsilon=10,\Delta\epsilon=10,U_0=20\sqrt{2},\Delta U_0=10\sqrt{2},\alpha=1.5,|\psi\rangle=|1,1,1,1,1,1,-1,-1,-1,-1,-1,-1\rangle$. Averaging over 320 realizations of disorder is performed in each case.}}
		\label{RPM_vs_strong_dis_KIC}
	\end{figure*}

	We take a periodically kicked system, whose Hamiltonian can be written as
	\begin{align}
		\hat{H}(t)&=\hat{H}_0+\hat{H}_1\sum_{n\in \mathbb{Z}}\delta\left(\frac{t}{\tau_p}-n\right),
		\label{gen_Ham}
	\end{align}
	where $\hat{H}_0$ is the base Hamiltonian, $\hat{H}_1$ is the driving Hamiltonian, and $\mathbb{Z}$ is a set of integers. This Hamiltonian is periodic in time with period $\tau_p$. We choose $\tau_p=1$. 
	The stroboscopic time evolution generated by this Hamiltonian is governed by the Floquet operator
	\begin{align}
		\hat{U}&=\lim_{\varepsilon\rightarrow 0}\mathcal{T} e^{-i\int_\varepsilon^{1+\varepsilon}dt \hat{H}(t)}\notag\\
		&=\hat{V}\hat{W},
		\label{Floquet_operator}
	\end{align}
	where $\mathcal{T}$ represents time-ordering, and 
	\begin{align}
		\hat{V}&=e^{-i\hat{H}_1},\\
		\hat{W}&=e^{-i\hat{H}_0}.
	\end{align} 
	Let us assume that the system is initially (at $t=0$) in a state $|\psi\rangle$. Consider an observable $\hat{A}$. We want to study the expectation value of this observable $\langle\psi|\hat{A}(t)|\psi\rangle$, at an arbitrary later time. To analytically calculate this quantity, we choose the eigenstates of $\hat{H}_0$ as the computational basis. The base Hamiltonian $\hat{H}_0$ contains a coupling of particles/spins with a random classical field. For a fermionic system, this could be random onsite potentials, whereas this could be random magnetic fields in the $z$ spatial direction for a lattice of spins. The base Hamiltonian also contains long-range interactions. Let us say that the local degrees of freedom are described by operators $\hat{O}_i$ where $i$ is the site index and takes values $i=1,...,L$. For a fermionic system, $\hat{O}_i\equiv \hat{n}_i$, and $\hat{O}_i\equiv \hat{S}_i^z$ for a spin system. Then
	\begin{align}
		\hat{H}_0&=\sum_{i=1}^L \epsilon_i \hat{O}_i+\sum_{i<j}\frac{U_{ij}}{(d_{ij})^\alpha}\hat{O}_i\hat{O}_j,
		\label{S_H0_generic}
	\end{align} 
	where $\epsilon_i$'s and $U_{ij}$'s are independent Gaussian random numbers with mean $\langle\epsilon_i\rangle=\epsilon$, $\langle U_{ij}\rangle=U_0$ and standard deviation $\sqrt{\langle\epsilon_i^2\rangle-\langle\epsilon_i\rangle^2}=\Delta\epsilon$, and $\sqrt{\langle U_{ij}^2\rangle-\langle U_{ij}\rangle^2}=\Delta U_0$. The distance between the $i^{\text{th}}$ and $j^{\text{th}}$ site is $d_{ij}$ and the exponent $\alpha$ takes values from the interval $(1,2]$. We take periodic boundary conditions; therefore, $d_{ij}=min(|i-j|,L-|i-j|)$. From the form of $\hat{H}_0$ in Eq.~(\ref{S_H0_generic}), it is clear that the eigenstates of $\hat{H}_0$ are also the eigenstates of local operators $\hat{O}_i,\forall i$. We denote them by $|\underline{o}\rangle\equiv |o_1,...,o_L\rangle$ where $o_i$ is an eigenvalue of $\hat{O}_i$. Therefore,
	\begin{align}
		\hat{O}_i|\underline{o}\rangle&\equiv o_i|\underline{o}\rangle,\\
		\hat{W}|\underline{o}\rangle&\equiv e^{-i\theta_{\underline{o}}}|\underline{o}\rangle,
		\label{W_eig}
	\end{align}
	where 
	\begin{align}
		\theta_{\underline{o}}&=\sum_{i=1}^L\epsilon_i o_i+\sum_{i<j}\frac{U_{ij}}{(d_{ij})^\alpha}o_i o_j.
	\end{align}
	We have
	\begin{align}
		\langle\psi|\hat{A}(t)|\psi\rangle&=\langle\psi|\hat{U}^{-t}\hat{A} \hat{U}^t|\psi\rangle.
	\end{align}
	To compute this expression analytically, we employ a path integral type decomposition. This can be understood from Figs.~\ref{psi_evol_diag} and \ref{At_summand_as_paths}.
	We insert identities $\sum_{\underline{o}_\tau}|\underline{o}_\tau\rangle\langle\underline{o}_\tau|$ and $\sum_{\underline{o}'_\tau}|\underline{o}'_\tau\rangle\langle\underline{o}'_\tau|$ for $\tau=0,t$ to obtain
	\begin{align}
		\langle\psi|\hat{A}(t)|\psi\rangle&=\sum_{\underline{o}_0,\underline{o}_{t}}\sum_{\underline{o}'_0,\underline{o}'_{t}}\langle\psi|\underline{o}'_0\rangle\langle\underline{o}'_0|\hat{U}^{-t}|\underline{o}'_{t}\rangle\langle\underline{o}'_{t}|\hat{A}|\underline{o}_{t}\rangle\langle\underline{o}_{t}|\hat{U}^t|\underline{o}_0\rangle\langle\underline{o}_0|\psi\rangle.
	\end{align}
	We further insert identities $\sum_{\underline{o}_\tau}|\underline{o}_\tau\rangle\langle\underline{o}_\tau|$ between different $\hat{U}$ operators and $\sum_{\underline{o}'_\tau}|\underline{o}'_\tau\rangle\langle\underline{o}'_\tau|$ between different $\hat{U}^{-1}$ operators for $\tau=1,...,t-1$ to obtain
	\begin{align}
		\langle\psi|\hat{A}(t)|\psi\rangle=&\sum_{\{\underline{o}_\tau\}_{\tau=0}^{t}}\sum_{\{\underline{o}'_\tau\}_{\tau=0}^{t}}\langle\psi|\underline{o}'_0\rangle\langle\underline{o}'_0|\hat{U}^\dagger|\underline{o}'_1\rangle...\langle\underline{o}'_{t-1}|\hat{U}^\dagger|\underline{o}'_{t}\rangle\langle\underline{o}'_{t}|\hat{A}|\underline{o}_{t}\rangle\langle\underline{o}_{t}|\hat{U}|\underline{o}_{t-1}\rangle...\langle\underline{o}_{1}|\hat{U}|\underline{o}_{0}\rangle\langle\underline{o}_0|\psi\rangle,
		\label{At_1}
	\end{align}
	where $\sum_{\{\underline{o}_\tau\}_{\tau=0}^{t}}=\sum_{\underline{o}_0,...,\underline{o}_{t}}$ and $\sum_{\{\underline{o}'_\tau\}_{\tau=0}^{t}}=\sum_{\underline{o}'_0,...,\underline{o}'_{t}}$. Substituting Eq.~(\ref{Floquet_operator}) into Eq.~(\ref{At_1}) and using Eq.~(\ref{W_eig}), we obtain
	\begin{align}
		\langle\psi|\hat{A}(t)|\psi\rangle=&\sum_{\{\underline{o}_\tau\}_{\tau=0}^{t}}\sum_{\{\underline{o}'_\tau\}_{\tau=0}^{t}}\langle\psi|\underline{o}'_0\rangle\langle\underline{o}_0|\psi\rangle A_{\underline{o}'_{t},\underline{o}_{t}}e^{-i\sum_{\tau=0}^{t-1}(\theta_{\underline{o}_\tau}-\theta_{\underline{o}'_{\tau}})}\prod_{\tau=0}^{t-1}V_{\underline{o}_{\tau+1},\underline{o}_\tau}V^*_{\underline{o}'_{\tau+1},\underline{o}'_\tau},
	\end{align}
	where $A_{\underline{o}'_{t+1},\underline{o}_{t+1}}= \langle \underline{o}'_{t+1}|\hat{A}|\underline{o}_{t+1}\rangle$ and $V_{\underline{o}_{\tau+1},\underline{o}_\tau}=\langle {\underline{o}_{\tau+1}| \hat{V}|\underline{o}_\tau}\rangle$. A direct numerical simulation of $\langle\psi| \hat{A}(t)|\psi\rangle$ reveals that in the limit of strong disorder $\Delta\epsilon,\Delta U_0\gg 1$, it matches a random phase model (RPM) in  Fig.~\ref{RPM_vs_strong_dis_KIC}. In RPM, $\hat{W}$ is replaced by a diagonal random matrix in the computational basis, $\text{diag}(e^{-i\theta_1},...,e^{-i\theta_{\mathcal{N}}})$, where $\theta_1,...,\theta_{\mathcal{N}}$ are independent random numbers uniformly distributed over $[0,2\pi)$. Therefore, for analytical computations, we make this assumption that the phases $\theta_{\underline{o}_\tau}, \forall \underline{o}_\tau$ are independent random numbers uniformly distributed over $[0,2\pi)$. We call this the random phase approximation (RPA). Therefore,
	\begin{samepage}
		\begin{align}
			\langle\psi|\hat{A}(t)|\psi\rangle_{\text{dis}}=&\sum_{\{\underline{o}_\tau\}_{\tau=0}^{t}}\sum_{\{\underline{o}'_\tau\}_{\tau=0}^{t}}\langle\psi|\underline{o}'_0\rangle\langle\underline{o}_0|\psi\rangle A_{\underline{o}'_{t},\underline{o}_{t}}\langle e^{-i\sum_{\tau=0}^{t-1}(\theta_{\underline{o}_\tau}-\theta_{\underline{o}'_{\tau}})}\rangle_{\text{dis}}\prod_{\tau=0}^{t-1}V_{\underline{o}_{\tau+1},\underline{o}_\tau}V^*_{\underline{o}'_{\tau+1},\underline{o}'_\tau},
		\end{align}
	\end{samepage}
	where
	\begin{align}
		\langle e^{-i\sum_{\tau=0}^{t-1}(\theta_{\underline{o}_\tau}-\theta_{\underline{o}'_{\tau}})}\rangle_{\text{dis}}\simeq &\langle e^{-i\sum_{\tau=0}^{t-1}(\theta_{\underline{o}_\tau}-\theta_{\underline{o}'_{\tau}})}\rangle_{\text{RPA}}=\prod_{\tau=0}^{t-1}\delta_{\underline{o}'_\tau,\underline{o}_{\pi(\tau)}},
	\end{align}
	where $\pi$ is a permutation of $t$ objects. Therefore,
	\begin{align}
		\langle\psi|\hat{A}(t)|\psi\rangle_{\text{dis}}=&\langle\psi|\hat{A}(t)|\psi\rangle_{\text{RPA}}\notag\\
		=&\sum_{\{\underline{o}_\tau\}_{\tau=0}^{t}}\sum_\pi\sum_{\underline{o}'_{t}}\langle\psi|\underline{o}_{\pi(0)}\rangle\langle\underline{o}_0|\psi\rangle A_{\underline{o}'_{t},\underline{o}_{t}}V_{\underline{o}_{t},\underline{o}_{t-1}}V^*_{\underline{o}'_{t},\underline{o}_{\pi(t-1)}}\prod_{\tau=0}^{t-2}V_{\underline{o}_{\tau+1},\underline{o}_\tau}V^*_{\underline{o}_{\pi(\tau+1)},\underline{o}_{\pi(\tau)}}.
		\label{At_2}
	\end{align}
	Consider only the identity permutation $I$. We denote its contribution as $\langle\psi|\hat{A}(t)|\psi\rangle_{\text{RPA},I}$. From Eq.~(\ref{At_2}) we write
	\begin{align}
		\langle\psi|\hat{A}(t)|\psi\rangle_{\text{RPA},I}=&\sum_{\{\underline{o}_\tau\}_{\tau=0}^{t}}\sum_{\underline{o}'_{t}}|\langle\underline{o}_0|\psi\rangle|^2 A_{\underline{o}'_{t},\underline{o}_{t}}V_{\underline{o}_{t},\underline{o}_{t-1}}V^*_{\underline{o}'_{t},\underline{o}_{t-1}}\prod_{\tau=0}^{t-2}\mathcal{M}_{\underline{o}_{\tau+1},\underline{o}_\tau},
		\label{At_RPA_identity}
	\end{align}
	where $\mathcal{M}_{\underline{o}_{\tau+1},\underline{o}_\tau}=|V_{\underline{o}_{\tau+1},\underline{o}_\tau}|^2$. Therefore,
	\begin{align}
		\langle\psi|\hat{A}(t)|\psi\rangle_{\text{RPA},I}=&\sum_{\underline{o}_0,\underline{o}_{t-1},\underline{o}_{t},\underline{o}'_{t}}|\langle\underline{o}_0|\psi\rangle|^2 A_{\underline{o}'_{t},\underline{o}_{t}}V_{\underline{o}_{t},\underline{o}_{t-1}}V^*_{\underline{o}'_{t},\underline{o}_{t-1}}(\mathcal{M}^{t-1})_{\underline{o}_{t-1},\underline{o}_0}.
	\end{align}
	The matrix $\mathcal{M}$ is doubly stochastic. Therefore, its eigenvalues are $\lambda_0,...,\lambda_{\mathcal{N}-1}$ which satisfy $\lambda_0(=1)> |\lambda_1|\geq ...\geq|\lambda_{\mathcal{N}-1}|$. Performing eigendecomposition of the matrix $\mathcal{M}$ we write
	\begin{align}
		\langle\psi|\hat{A}(t)|\psi\rangle_{\text{RPA},I}=&\sum_{\underline{o}_0,\underline{o}_{t-1},\underline{o}_{t},\underline{o}'_{t}}\sum_{i=0}^{\mathcal{N}-1}|\langle\underline{o}_0|\psi\rangle|^2 A_{\underline{o}'_{t},\underline{o}_{t}}V_{\underline{o}_{t},\underline{o}_{t-1}}V^*_{\underline{o}'_{t},\underline{o}_{t-1}}\lambda_i^{t-1}\mathcal{M}^{(i)}_{\underline{o}_{t-1},\underline{o}_0}
	\end{align}
	where $\mathcal{M}^{(i)}=|\lambda_i\rangle\langle\lambda_i|$. Due to the doubly stochastic nature of the matrix $\mathcal{M}$, $\langle\lambda_0|\equiv (1/\sqrt{\mathcal{N}})(1,...,1)$. Therefore, $\mathcal{M}^{(0)}_{\underline{o}_{t-1},\underline{o}_0}=1/\mathcal{N}$. Using this, we find
	\begin{align}
		\langle\psi|\hat{A}(t)|\psi\rangle_{\text{RPA},I}&=\sum_{\underline{o}_0,\underline{o}_{t-1},\underline{o}_{t},\underline{o}'_{t}}|\langle\underline{o}_0|\psi\rangle|^2 A_{\underline{o}'_{t},\underline{o}_{t}}V_{\underline{o}_{t},\underline{o}_{t-1}}V^*_{\underline{o}'_{t},\underline{o}_{t-1}}\left(\frac{1}{\mathcal{N}}+\sum_{i=1}^{\mathcal{N}-1}\lambda_i^{t-1}\mathcal{M}^{(i)}_{\underline{o}_{t-1},\underline{o}_0}\right).
		\label{At_RPA_identity_eig_dec_expan}
	\end{align}
	We notice that $\sum_{\underline{o}_0}|\langle\underline{o}_0|\psi\rangle|^2=1$ and $\sum_{\underline{o}_{t-1}}V_{\underline{o}_{t},\underline{o}_{t-1}}V^*_{\underline{o}'_{t},\underline{o}_{t-1}}=\delta_{\underline{o}_{t},\underline{o}'_{t}}$. Therefore, the first term on the right hand side in Eq.~(\ref{At_RPA_identity_eig_dec_expan}) simplifies and we obtain
	\begin{align}
		\langle\psi|\hat{A}(t)|\psi\rangle_{\text{RPA},I}&=\frac{1}{\mathcal{N}}\tr A+\sum_{\underline{o}_0,\underline{o}_{t-1},\underline{o}_{t},\underline{o}'_{t}}\sum_{i=1}^{\mathcal{N}-1}|\langle\underline{o}_0|\psi\rangle|^2 A_{\underline{o}'_{t},\underline{o}_{t}} V_{\underline{o}_{t},\underline{o}_{t-1}}V^*_{\underline{o}'_{t},\underline{o}_{t-1}}\lambda_i^{t-1}\mathcal{M}^{(i)}_{\underline{o}_{t-1},\underline{o}_0}.
		\label{At_RPA_identity_generic}
	\end{align}
	In Eq.~(\ref{At_RPA_identity_generic}), the first term on right hand side is the infinite temperature statistical average (ITSA) expected from a periodically driven nonintegrable system. Since $|\lambda_i|<1$, $\forall i\neq 0$, the second term decays exponentially with time. Therefore, within identity permutation, we find that the expectation value of an observable $\hat{A}$ reaches the ITSA at long times, irrespective of the initial state.
	
	For further analysis, we only consider observables $\hat{A}$ which are diagonal in the computational basis. In addition, initial states are chosen from the basis states. With these restrictions, Eq.~(\ref{At_2}) takes a simpler form
	\begin{align}
		\langle\psi|\hat{A}(t)|\psi\rangle_{\text{RPA}}=\sum_{\{\underline{o}_\tau\}_{\tau=1}^{t}}\sum_\pi A_{\underline{o}_{t},\underline{o}_{t}}V_{\underline{o}_{t},\underline{o}_{t-1}}V^*_{\underline{o}_{t},\underline{o}_{\pi(t-1)}}\left(\prod_{\tau=1}^{t-2}V_{\underline{o}_{\tau+1},\underline{o}_\tau}V^*_{\underline{o}_{\pi(\tau+1)},\underline{o}_{\pi(\tau)}}\right)V_{\underline{o}_{1},\psi}V^*_{\underline{o}_{1},\psi}.
		\label{A_diag_psi_basis}
	\end{align}
	In this case, the contribution of identity permutation can be expressed as
	\begin{align}
		\langle\psi|\hat{A}(t)|\psi\rangle_{\text{RPA},I}&=\sum_{\underline{o}_{t}}A_{\underline{o}_{t},\underline{o}_{t}}\left(\mathcal{M}^t\right)_{\underline{o}_{t},\psi}\notag\\
		&=\frac{1}{\mathcal{N}}\tr A+\sum_{\underline{o}_{t}}\sum_{i=1}^{\mathcal{N}-1} A_{\underline{o}_{t},\underline{o}_{t}}\lambda_i^{t}\mathcal{M}^{(i)}_{\underline{o}_{t},\psi}.
		\label{At_I_diag_A_psi_basis}
	\end{align}
	Now we consider some many-body Hamiltonians, and compute expectations of some observables.
	\subsection{Kicked transverse-field Ising chain}
\begin{figure}[H]
		\centering
		\begin{subfigure}[t]{0.45\linewidth}
			\centering
			\includegraphics[width=\linewidth]{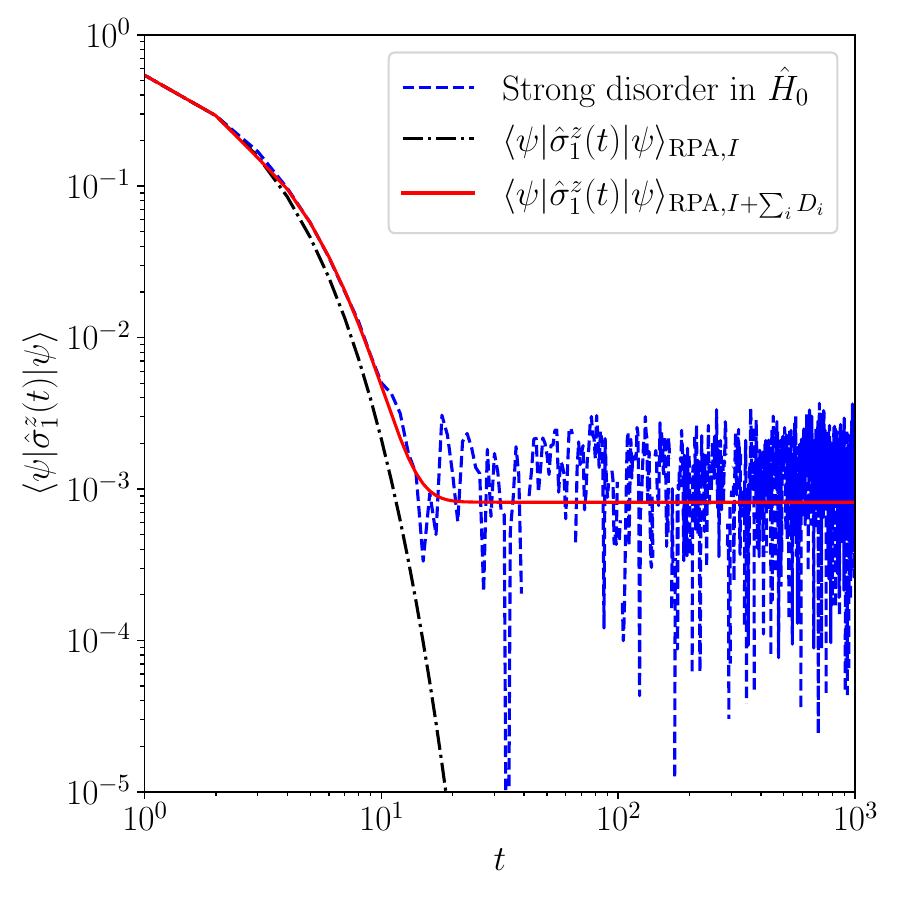}
			\caption{}
		\end{subfigure}
		\hfill
		\begin{subfigure}[t]{0.45\linewidth}
			\centering
			\includegraphics[width=\linewidth]{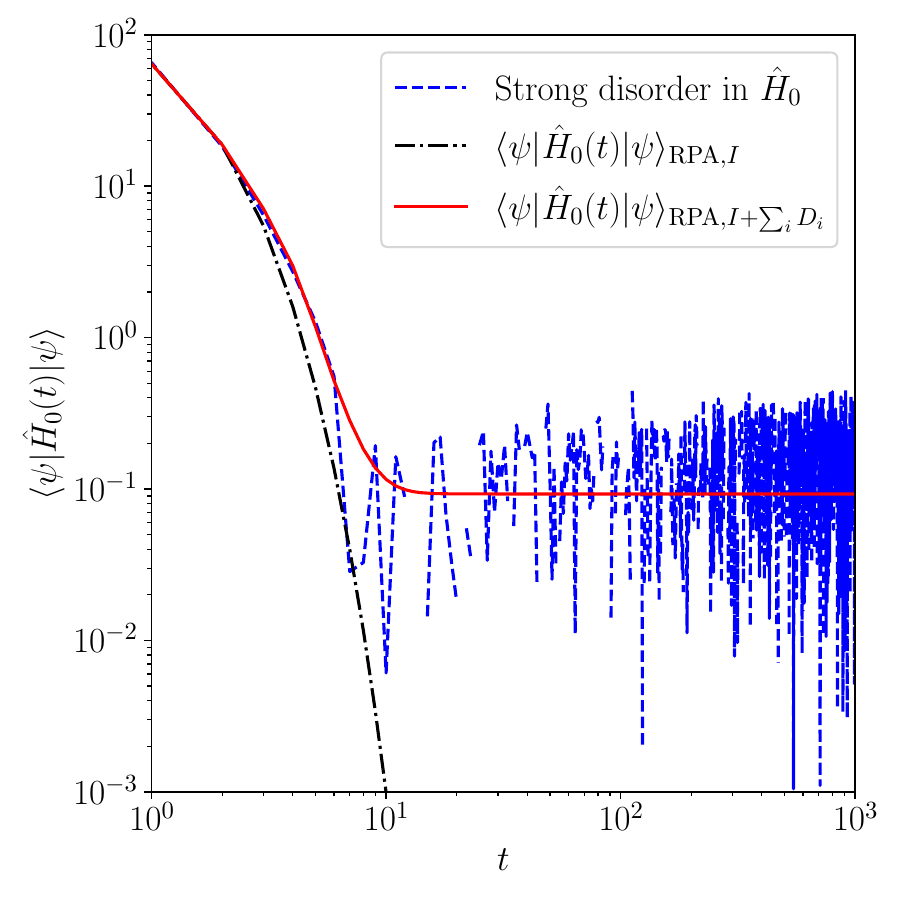}
			\caption{}
		\end{subfigure}
		\caption{Time evolution of (a) local magnetization and (b) total energy from direct numerics and analytics using the RPA for a periodically kicked long-range interacting transverse-field Ising chain. Here $L=12,h=0.5,\epsilon=10,\Delta\epsilon=10,U_0=10,\Delta U_0=10,\alpha=1.5,|\psi\rangle=|1,1,1,1,1,1,-1,-1,-1,-1,-1,-1\rangle$ and periodic boundary condition. Averaging over 320 realizations of disorder is performed.}
		\label{KIC_local_magnetization}
	\end{figure}

\begin{figure*}[t!]
		\centering
		\begin{subfigure}[t]{0.45\linewidth}
			\centering
			\includegraphics[width=\linewidth]{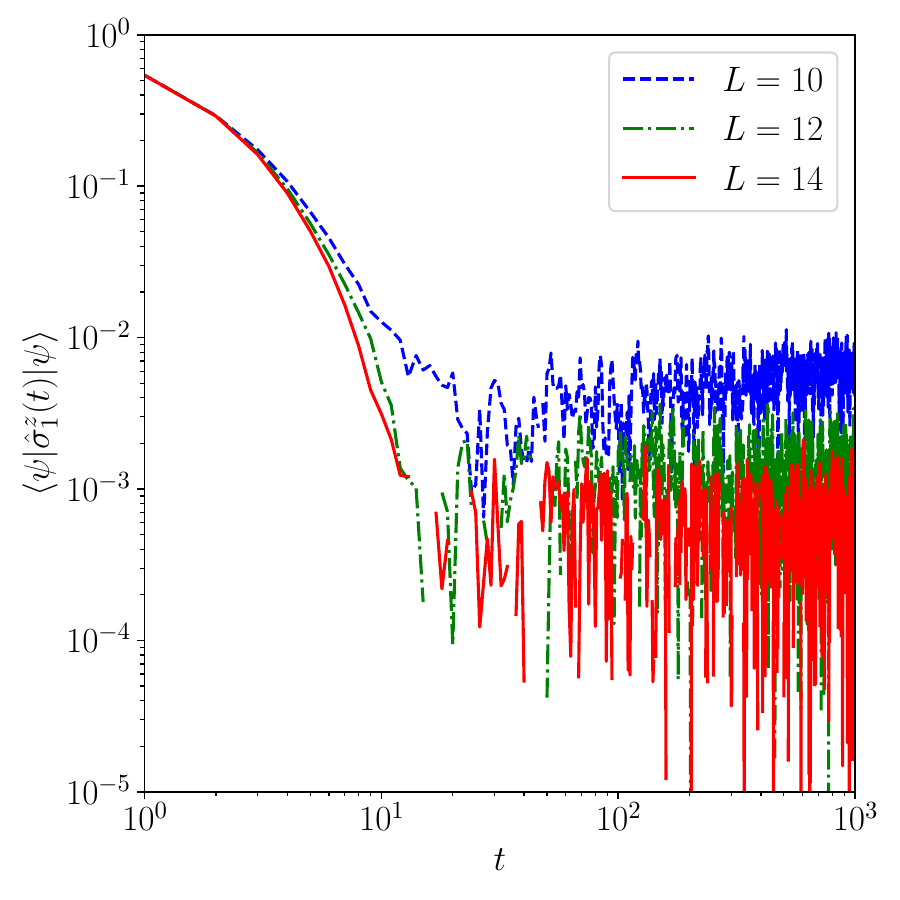}
			\caption{}
		\end{subfigure}
		\hfill
		\begin{subfigure}[t]{0.45\linewidth}
			\centering
			\includegraphics[width=\linewidth]{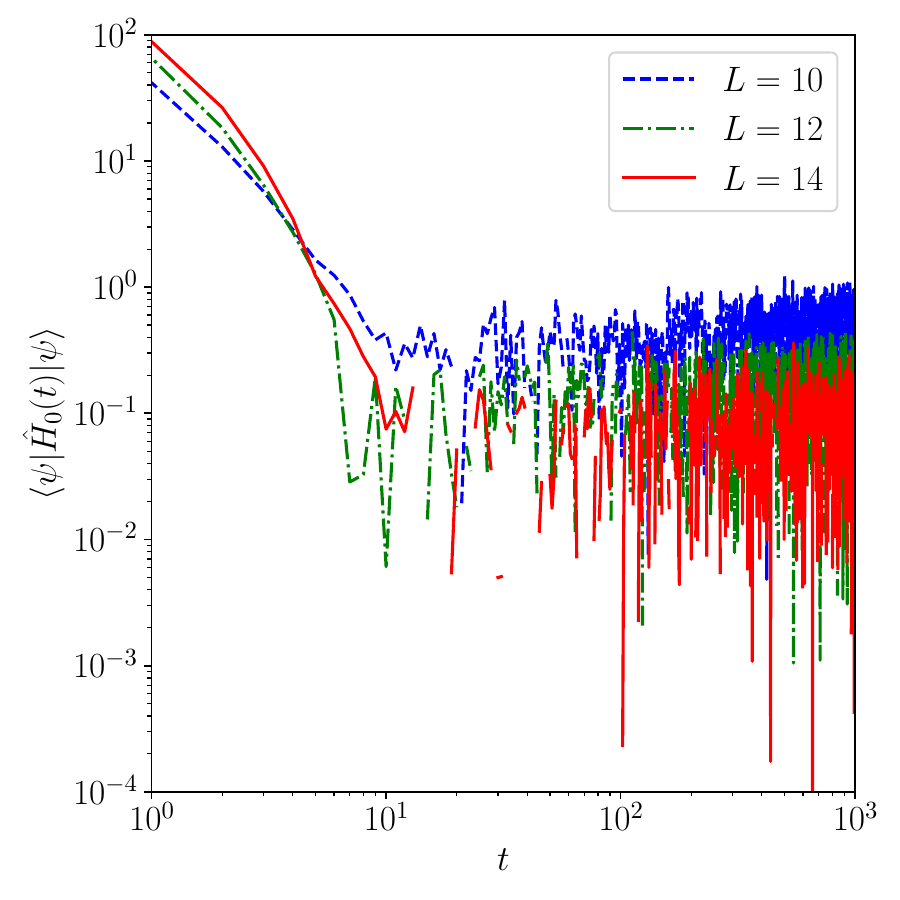}
			\caption{}
		\end{subfigure}
		\caption{\justifying{\small Time evolution of (a) local magnetization and (b) total energy from direct numerical simulation for a periodically kicked long-range interacting transverse-field Ising chain with strong disorder. Here $h=0.5, \epsilon=10,\Delta\epsilon=10,U_0=20\sqrt{2},\Delta U_0=10\sqrt{2},\alpha=1.5$. In each case the initial state $|\psi\rangle\equiv |\psi_1,...,\psi_L\rangle$ where $\psi_i=1$, $\forall i\in\{1,...,L/2\}$ and $\psi_i=-1$, $\forall i\in\{L/2+1,...,L\}$. Averaging over 160-320 realizations of disorder is performed in each case.}}
		\label{KIC_sz_n_E_vs_L}
	\end{figure*}

	Here, we take a chain of spin-1/2's with the following base and driving Hamiltonians: 
	\begin{align}
		\hat{H}_0&=\sum_{i=1}^L\epsilon_i \hat{\sigma}_i^z+\sum_{i<j}\frac{U_{ij}}{(d_{ij})^\alpha}\hat{\sigma}_i^z\hat{\sigma}_j^z,
		\label{KIC_H0}\\
		\hat{H}_1&=h\sum_{i=1}^L\hat{\sigma}_i^x,
		\label{KIC_H1}
	\end{align}
	where $L$ is the size of the system or the total number of spins and $\hat{\sigma}_i^x$ and $\hat{\sigma}_i^z$ are the Pauli-$x$ and $z$ operators, respectively. We consider $\hat{\sigma}_j^z$ as a local observable at site $j$. We consider an initial state $|\underline{s}\rangle$ which is an eigenstate of local observables $\hat{\sigma}_i^z$, $\forall i\in\{1,...,L\}$
	\begin{align}
		\hat{\sigma}_i^z|\underline{s}\rangle=s_i|\underline{s}\rangle,\; s_i=\pm 1.
	\end{align}
	Therefore, $|\underline{s}\rangle$ can be expressed as $|\underline{s}\rangle=|s_1,...,s_L\rangle$. The disorder-averaged expectation value of $\hat{\sigma}_i^z$ under stroboscopic evolution at times $t=0,1,2,...$ can be computed using Eq.~(\ref{A_diag_psi_basis}).  In this case,
	\begin{align}
		\hat{A}&\equiv \hat{\sigma}_j^z,\\
		V&=u^{\otimes L},\; u=\begin{pmatrix}
			\cos (h)& -i \sin(h)\\
			-i \sin(h)& \cos (h)
		\end{pmatrix},
		\label{KIC_V}\\
		|\underline{o}\rangle &\equiv |\underline{s}\rangle
	\end{align}
	First, we compute the contribution resulting from the identity permutation. Following Eq.~(\ref{At_I_diag_A_psi_basis}), we obtain
	\begin{align}
		\langle\psi|\hat{\sigma}_j^z(t)|\psi\rangle_{\text{RPA},I}&=\sum_{\underline{s}_{t+1}}(\sigma_j^z)_{\underline{s}_{t+1},\underline{s}_{t+1}}\left(\mathcal{M}^t\right)_{\underline{s}_{t+1},\psi},
		\label{sigma_j^z_t_I}
	\end{align}
	where
	\begin{align}
		\mathcal{M}=m^{\otimes L},\; m=\begin{pmatrix}
			\cos^2(h)& \sin^2(h)\\
			\sin^2(h)& \cos^2(h)
		\end{pmatrix}.
		\label{M_KIC}
	\end{align}
	Since $\underline{s}_{t+1}$ is just a dummy index in Eq.~(\ref{sigma_j^z_t_I}), we replace it by $a$. This represents a many-body state $|a\rangle\equiv |a_1,...,a_L\rangle$. Therefore, substituting Eq.~(\ref{M_KIC}) in Eq.~(\ref{sigma_j^z_t_I}), we obtain
	\begin{align}
		\langle\psi|\hat{\sigma}_j^z(t)|\psi\rangle_{\text{RPA},I}&=\sum_{a}(\sigma_j^z)_{a,a}\left((m^{\otimes L})^t\right)_{a,\psi}\notag\\
		&=\sum_{a_1,...,a_L}\left\{(\sigma_j^z)_{a_j,a_j}(m^t)_{a_j,\psi_j}\right\}\prod_{k\neq j}(m^t)_{a_k,\psi_k}.
		\label{sigma_j^z_t_I_prod_expan}
	\end{align}
	Since $|\psi\rangle$ is one of the basis states, it can be expressed as $|\psi_1,...,\psi_L\rangle$. We have used this to obtain Eq.~(\ref{sigma_j^z_t_I_prod_expan}). We simplify the right hand side in Eq.~(\ref{sigma_j^z_t_I_prod_expan}) using Eq.~(\ref{M_KIC}) and obtain
	\begin{align}
		\langle\psi|\hat{\sigma}_j^z(t)|\psi\rangle_{\text{RPA},I}=\begin{cases}
			\cos^t(2h),\;\text{if $\psi_j=1$},\\
			-\cos^t(2h),\;\text{if $\psi_j=-1$}
		\end{cases}.
		\label{sigma_j^z_t_I_contri}
	\end{align}
	According to Eq.~(\ref{sigma_j^z_t_I_contri}), the expectation value of a local observable $\hat{\sigma}_j^z$ approaches the ITSA ($(\tr \sigma_j^z)/2^L=0$) at long times. This is the expected behavior for nonintegrable periodically kicked many-body quantum systems, in accordance with the Floquet ETH. However, when we compare Eq.~(\ref{sigma_j^z_t_I_contri}) with the direct simulation result, we find that they match only at early times, as shown by black dashed-dotted and blue dashed curves in Fig. \ref{KIC_local_magnetization}. This happens because these plots are for finite-size systems, which are accessible numerically. For finite system sizes, permutations other than identity also have a nonzero contribution at long times. The contributions of these permutations are inversely proportional to the Hilbert-space dimension $\mathcal{O}(1/2^L)$. This feature can be seen in the direct simulation as well as shown in Fig.~\ref{KIC_sz_n_E_vs_L}. Therefore, it decays exponentially with system size, and we expect only the contribution resulting from the identity permutation to survive in the thermodynamic limit. We call here the contribution from the identity permutation the leading-order contribution. Nevertheless, we also compute next to the leading-order contribution resulting from other non-trivial permutations. Inclusion of these higher-order contributions leads to a better match with the direct simulation result, as shown by the red full curve in Fig.~\ref{KIC_local_magnetization}.
	Now we present our calculation of next to leading order contribution which together with leading order result Eq.~(\ref{sigma_j^z_t_I_contri}) determines the red curve in Fig.~\ref{KIC_local_magnetization}.
	\subsubsection{Next to leading-order contribution}
	The next to leading-order contribution involves evaluating the contribution of various nontrivial permutations in Eq.~(\ref{A_diag_psi_basis}). However, doing so reveals an important difficulty. In priciple, we have a many-body trajectory given by a set of basis states $\{\psi,\underline{s}_1,...,\underline{s}_{t}\}$ and a permutation of states $\underline{s}_1,...,\underline{s}_{t-1}$ denoted by $\pi$. Therefore, following Eq.~(\ref{A_diag_psi_basis}), the contribution of a permutation $\pi$ of a trajectory can be expressed as
	\begin{align}
		&(\sigma_j^z)_{\underline{s}_{t},\underline{s}_{t}}V_{\underline{s}_{t},\underline{s}_{t-1}}V^*_{\underline{s}_{t},\underline{s}_{\pi(t-1)}} \left(\prod_{\tau=1}^{t-2}V_{\underline{s}_{\tau+1},\underline{s}_\tau}V^*_{\underline{s}_{\pi(\tau+1)},\underline{s}_{\pi(\tau)}}\right)V_{\underline{s}_{1},\psi}V^*_{\underline{s}_{\pi(1)},\psi}.
	\end{align}
	There are $2^{Lt}$ number of trajectories for different choices of basis states $\underline{s}_1,...,\underline{s}_{t}$. The contribution of a permutation $\pi$ of all the trajectories can be included by summing over basis states $\underline{s}_1,...,\underline{s}_{t}$ as follows
	\begin{align}
		\langle\psi|\hat{\sigma}_j^z(t)|\psi\rangle_{\text{RPA},\pi}&=\sum_{\{\underline{s}_\tau\}_{\tau=1}^{t}}(\sigma_j^z)_{\underline{s}_{t},\underline{s}_{t}}V_{\underline{s}_{t},\underline{s}_{t-1}}V^*_{\underline{s}_{t},\underline{s}_{\pi(t-1)}}\left(\prod_{\tau=1}^{t-2}V_{\underline{s}_{\tau+1},\underline{s}_\tau}V^*_{\underline{s}_{\pi(\tau+1)},\underline{s}_{\pi(\tau)}}\right)V_{\underline{s}_{1},\psi}V^*_{\underline{s}_{\pi(1)},\psi}.
	\end{align}
	Although this expression seems to be correctly adding contributions resulting from a permutation $\pi$ of all the trajectories, a feature that will be clarified later causes contribution of some of the trajectories to be considered multiple times. This feature can be understood as follows. Let us say that $\pi$ can be expressed as a product of two permutations $T_{\tau_1,\tau_2}$ and $\pi'$, $\pi=T_{\tau_1,\tau_2}\pi'$. The permutation $T_{\tau_1,\tau_2}$ transposes basis states with subscript $\tau_1$ and $\tau_2$ which are $\underline{s}_{\tau_1}$ and $\underline{s}_{\tau_2}$ whereas $\pi'$ can be any arbitrary permutation that leaves $\underline{s}_{\tau_1}$ and $\underline{s}_{\tau_2}$ unchanged. Now consider a trajectory for which $\underline{s}_{\tau_1}$ and $\underline{s}_{\tau_2}$ are identical. For such a trajectory, both permutations $\pi$ and $\pi'$ give the same permutation and consequently the same contribution. Therefore, if both $\pi$ and $\pi'$ are considered in calculating next to leading-order correction, contribution of trajectories with repeated basis states will be counted two times. To fix this error we explicitly calculate contribution of such trajectories which we denote by $\langle\psi|\hat{\sigma}_j^z(t)|\psi\rangle_{\text{RPA},\pi'}^{\{\tau_1,\tau_2\}}$. We than subtract it from the contribution of permutations $\pi$ and $\pi'$ as follows
	\begin{align}
		\langle\psi|\hat{\sigma}_j^z(t)|\psi\rangle_{\text{RPA},\pi}+\langle\psi|\hat{\sigma}_j^z(t)|\psi\rangle_{\text{RPA},\pi'}-\langle\psi|\hat{\sigma}_j^z(t)|\psi\rangle_{\text{RPA},\pi'}^{\{\tau_1,\tau_2\}},
	\end{align}
	where
	\begin{samepage}
		\begin{align}
			\langle\psi|\hat{\sigma}_j^z(t)|\psi\rangle_{\text{RPA},\pi'}^{\{\tau_1,\tau_2\}}&=\sum_{\{\underline{s}_\tau\}_{\tau=1}^t}\delta_{\underline{s}_{\tau_1},\underline{s}_{\tau_2}}(\sigma_j^z)_{\underline{s}_{t},\underline{s}_{t}}V_{\underline{s}_{t},\underline{s}_{t-1}}V^*_{\underline{s}_{t},\underline{s}_{\pi(t-1)}}\left(\prod_{\tau=1}^{t-2}V_{\underline{s}_{\tau+1},\underline{s}_\tau}V^*_{\underline{s}_{\pi(\tau+1)},\underline{s}_{\pi(\tau)}}\right)V_{\underline{s}_{1},\psi}V^*_{\underline{s}_{\pi(1)},\psi}.
		\end{align}
	\end{samepage}
	Similarly, other correction terms need to be included for a higher number of repetitions of basis states. However, for next to leading-order correction we only need to consider up to minimal repetition. Next, we compute the contributions from various permutations. However, to avoid lengthy expressions in what follows, we first introduce a diagrammatic representation.\\
	
	\textbf{Diagrammatic representation}:
	\begin{enumerate}
		\item We denote states $\psi,\underline{s}_1,...,\underline{s}_{t-1},\underline{s}_{t}$ by a dashed blue arc as shown in Fig.~\ref{dashed_blue_curve}. The states $\psi$ and $\underline{s}_{t}$ are denoted by the left and the right ends of the arc, respectively. Other states $\underline{s}_1,...,\underline{s}_{t-1}$ are equidistant points on the arc between the end points. 
		\begin{figure}[H]
			\centering
			\begin{tikzpicture}
				\def\rad{1.5}
				\draw[dashed,blue,->] (120:\rad) arc(120:270:\rad);\draw[dashed,blue] (270:\rad) arc(270:420:\rad);
				\draw (120:\rad) node[above]{$\psi$};
				\draw (60:\rad) node[above]{$\underline{s}_{t}$};
			\end{tikzpicture}
			\caption{}
			\label{dashed_blue_curve}
		\end{figure}
		\item We denote the states after permutation $\psi, \underline{s}_{\pi(1)},...,\underline{s}_{\pi(t-1)},\underline{s}_{t}$ by a solid red curve as shown in Fig.~\ref{dashed_blue_and_solid_red_curve}. 
		\begin{figure}[H]
			\centering
			\begin{tikzpicture}
				\def\rad{1.5}
				\draw[dashed,blue] (120:\rad) arc(120:420:\rad);
				\draw (120:\rad) node[above]{$\psi$};
				\draw (60:\rad) node[above]{$\underline{s}_{t}$};
				\draw[red,thick,->] (120:\rad) arc(120:150:\rad)--(0:\rad)--(210:\rad) arc(210:270:\rad);\draw[red,thick] (270:\rad) arc(270:330:\rad)--(180:\rad)--(30:\rad) arc(30:60:\rad);
			\end{tikzpicture}
			\caption{Red curve represents transposition of two states.}
			\label{dashed_blue_and_solid_red_curve}
		\end{figure}
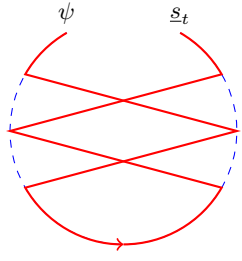
	\end{enumerate}
	Next, we introduce rules to evaluate these diagrams.\\
	\noindent (\romannumeral 1) For a red arc of length $n$ time steps, insert a factor as \\
	\noindent \raisebox{-0.5\height}{\tikz{\draw[red,thick,->] (30:1) arc(30:90:1);\draw[red,thick] (90:1) arc(90:150:1);
			\draw (30:1) node[right]{$a$};
			\draw (150:1) node[left]{$b$};
			\draw (90:1) node[above]{\scriptsize$n$ time steps};}} $\equiv$  \raisebox{-0.5\height}{\tikz{\draw[red,thick] (30:1) arc(30:90:1);\draw[red,thick,<-] (90:1) arc(90:150:1);
			\draw (30:1) node[right]{$a$};
			\draw (150:1) node[left]{$b$};
			\draw (90:1) node[above]{\scriptsize$n$ time steps};}} 
	$\equiv \left(\mathcal{M}^{n}\right)_{a,b}$.\\
	\noindent (\romannumeral 2) \raisebox{-0.5\height}{\tikz{\draw[blue,dashed] (0:1)--(0,0);\draw[blue,dashed,<-] (0,0)--(180:1);
			\draw (0:1) node[right]{$b$};
			\draw (180:1) node[left]{$a$};}} $\equiv$ \raisebox{-0.5\height}{\tikz{\draw[blue,dashed] (-1,0)--(0,0);\draw[blue,dashed,<-] (0,0)--(1,0);
			\draw (0:1) node[right]{$b$};
			\draw (180:1) node[left]{$a$};}}
	$\equiv V_{a,b}$.\\
	\noindent (\romannumeral 3) \raisebox{-0.5\height}{\tikz{\draw[red,thick] (0:1)--(0,0);\draw[red,thick,<-] (0,0)--(180:1);
			\draw (0:1) node[right]{$b$};
			\draw (180:1) node[left]{$a$};}} $\equiv$ \raisebox{-0.5\height}{\tikz{\draw[red,thick] (-1,0)--(0,0);\draw[red,thick,<-] (0,0)--(1,0);
			\draw (0:1) node[right]{$b$};
			\draw (180:1) node[left]{$a$};}}
	$\equiv V^*_{a,b}$.\\
	\noindent (\romannumeral 4) Sum over all the matrix indices except $\psi$.\\
	Next, we present all the diagrams that contribute at next to leading-order. We evaluate these diagrams using the above mentioned rules and Eqs.~(\ref{KIC_V}) and (\ref{M_KIC}). Instead of giving explicitly the permutation represented by these diagrams we just label them as $D_1, D_2,...$ for simplicity. 
	\begin{figure}[H]
		\centering
		\begin{tikzpicture}
			\def\rad{2}
			\draw[blue,dashed] (120:\rad) arc(120:420:\rad);
			\draw[red,thick,->] (120:\rad)--(30:\rad) arc(30:-90:\rad);\draw[red,thick] (-90:\rad) arc(-90:-210:\rad)--(60:\rad);
			\draw (120:{\rad}) node[above]{$\psi$};
			\draw (150:\rad) node[left]{$b$};
			\draw (30:\rad) node[right]{$c$};
			\draw (420:{\rad}) node[above]{$d$};
		\end{tikzpicture}
		\caption{Diagram $D_1$}
		\label{D_1}
	\end{figure}
	Contribution of the diagram shown in Fig.~\ref{D_1} is
	\begin{align}
		\langle\psi|\hat{\sigma}^z_{j}(t)|\psi\rangle_{\text{RPA},D_1}&=\left((\sigma_j^z)_{d_j,d_j} m^{t-2}_{b_j,c_j}u_{\psi_j,b_j}u_{c_j,d_j}u^*_{\psi_j,c_j}u^*_{b_j,d_j}\right)\left\{\prod_{k\neq j}\left(m^{t-2}_{b_k,c_k}u_{\psi_k,b_k}u_{c_k,d_k}u^*_{\psi_k,c_k}u^*_{b_k,d_k}\right)\right\}\theta(t-3)\notag\\
		&=\frac{\langle\psi_j|\hat{\sigma}_j^z|\psi_j\rangle}{2^L}(1+\cos 4h \cos^{t-2}2h)(1+\cos^{t-2}2h)^{L-1}\theta(t-3),
	\end{align}
	where
	\begin{align}
		\theta(x)=\begin{cases}
			1, \quad x\geq 0\\
			0, \quad x<0
		\end{cases}.
	\end{align}
	\begin{figure}[H]
		\centering
		\begin{tikzpicture}
			\def\rad{2}
			\draw[blue,dashed] (120:\rad) arc(120:420:\rad);
			\draw[red,thick,->] (120:\rad)--(-30:\rad) arc(-30:-90:\rad);\draw[red,thick] (-90:\rad) arc(-90:-210:\rad)--(0:\rad) arc(0:60:\rad);
			\draw (120:{\rad}) node[above]{$\psi$};
			\draw (150:\rad) node[left]{$b$};
			\draw (-30:\rad) node[right]{$c$};
			\draw (0:{\rad}) node[right]{$d$};
			\draw (420:{\rad}) node[above]{$e$};
		\end{tikzpicture}
		\caption{Diagram $D_2$. State $c$ appears at time step $\tau_1$ where $4\leq\tau_1\leq t-2$.}
		\label{D_2}
	\end{figure}
	Contribution of the diagram $D_2$ as shown in Fig.~\ref{D_2} is
	\begin{align}
		\langle\psi|\hat{\sigma}^z_{j}(t)|\psi\rangle_{\text{RPA},D_2}&=\sum_{\tau_1=4}^{t-2}\left((\sigma_j^z)_{e_j,e_j}m^{\tau_1-1}_{b_j,c_j}m^{t-\tau_1-1}_{d_j,e_j}u_{a_j,b_j}u_{c_j,d_j}u^*_{a_j,c_j}u^*_{b_j,d_j}\right)\notag\\
		&\times\left\{\prod_{k\neq j}\left(m^{\tau_1-1}_{b_k,c_k}m^{t-\tau_1-1}_{d_k,e_k}u_{a_k,b_k}u_{c_k,d_k}u^*_{a_k,c_k}u^*_{b_k,d_k}\right)\right\}\theta(t-4)\notag\\
		&=\frac{\langle\psi_j|\hat{\sigma}_j^z|\psi_j\rangle}{2^L}\sum_{\tau_1=4}^{t-2}\cos^{t-\tau_1-1}2h\left(1+\cos 4h\cos^{\tau_1-1} 2h\right)\left(1+\cos^{\tau_1-1}2h\right)^{L-1}\theta(t-4)
	\end{align}
	\begin{figure}[H]
		\centering
		\begin{tikzpicture}
			\def\rad{2}
			\draw[blue,dashed] (120:\rad) arc(120:420:\rad);
			\draw[red,thick,->] (120:\rad) arc(120:180:\rad)--(30:\rad) arc(30:-90:\rad);\draw[red,thick](-90:\rad) arc(-90:-150:\rad)--(60:\rad);
			\draw (120:{\rad}) node[above]{$\psi$};
			\draw (180:\rad) node[left]{$b$};
			\draw (-150:\rad) node[left]{$c$};
			\draw (30:{\rad}) node[right]{$d$};
			\draw (60:{\rad}) node[above]{$e$};
		\end{tikzpicture}
		\caption{Diagram $D_3$}
		\label{D_3}
	\end{figure}
	Diagrams $D_2$ and $D_3$ are just reflections of each other about a vertical line. Therefore, their contributions are identical,
	\begin{align}
		\langle\psi|\hat{\sigma}^z_{j}(t)|\psi\rangle_{\text{RPA},D_3}=\langle\psi|\hat{\sigma}^z_{j}(t)|\psi\rangle_{\text{RPA},D_2}.
	\end{align}
	\begin{figure}[H]
		\centering
		\begin{tikzpicture}
			\def\rad{2}
			\draw[blue,dashed] (120:\rad) arc(120:420:\rad);
			\draw[red,thick,->] (120:\rad) arc(120:150:\rad)--(0:\rad) arc(0:-90:\rad);\draw[red,thick](-90:\rad) arc(-90:-180:\rad)--(30:\rad) arc(30:60:\rad);
			\draw (120:{\rad}) node[above]{$\psi$};
			\draw (150:\rad) node[left]{$b$};
			\draw (180:\rad) node[left]{$c$};
			\draw (0:{\rad}) node[right]{$d$};
			\draw (30:{\rad}) node[right]{$e$};
			\draw (420:{\rad}) node[above]{$f$};
		\end{tikzpicture}
		\caption{Diagram $D_4$. State $b$ appears at time step $\tau_1$ and state $d$ appears at time step $\tau_2$ where $1\leq\tau_1\leq t-6$ and $\tau_1+4\leq\tau_2\leq t-2$.}
		\label{D_4}
	\end{figure}
	Contribution of the diagram $D_4$ as shown in Fig.~\ref{D_4} is
	\begin{align}
		\langle\psi|\hat{\sigma}^z_{j}(t)|\psi\rangle_{\text{RPA},D_4}&=\sum_{\tau_1=1}^{t-6}\sum_{\tau_2=\tau_1+4}^{t-2}\left((\sigma_j^z)_{f_j,f_j}m^{\tau_1}_{\psi_j,b_j}m^{\tau_2-\tau_1-1}_{c_j,d_j}m^{t-\tau_2-1}_{e_j,f_j}u_{b_j,c_j}u_{d_j,e_j}u^*_{b_j,d_j}u^*_{c_j,e_j}\right)\notag\\
		&\quad\times \left\{\prod_{k\neq j}\left(m^{\tau_1}_{\psi_k,b_k}m^{\tau_2-\tau_1-1}_{c_k,d_k}m^{t-\tau_2-1}_{e_k,f_k}u_{b_k,c_k}u_{d_k,e_k}u^*_{b_k,d_k}u^*_{c_k,e_k}\right)\right\}\theta(t-5)\notag\\
		&=\frac{\langle\psi_j|\hat{\sigma}_j^z|\psi_j\rangle}{2^L}\sum_{\tau_1=1}^{t-6}\sum_{\tau_2=\tau_1+4}^{t-2}\left((\cos2h)^{t-\tau_2+\tau_1-1}(1+\cos 4h(\cos2h)^{\tau_2-\tau_1-1})\right)\left(1+(\cos2h)^{\tau_2-\tau_1-1}\right)^{L-1}\notag\\
		&\quad\times\theta(t-5).
	\end{align}
	\begin{figure}[H]
		\centering
		\begin{tikzpicture}
			\def\rad{2}
			\draw[blue,dashed] (120:\rad) arc(120:420:\rad);
			\draw[red,thick,->] (120:\rad)--(-30:\rad) arc(-30:-90:\rad);\draw[red,thick] (-90:\rad) arc(-90:-210:\rad)--(0:\rad) arc(0:60:\rad);
			\draw (120:{\rad}) node[above]{$\psi$};
			\draw[fill=green] (150:\rad) circle(2pt) node[left]{$b$};
			\draw[fill=green] (-30:\rad) circle(2pt) node[right]{$b$};
			\draw (0:{\rad}) node[right]{$d$};
			\draw (420:{\rad}) node[above]{$e$};
		\end{tikzpicture}
		\caption{Diagram $D_5$. State $b$ appears at time step $1$ and $\tau_1$ where $4\leq\tau_1\leq t-2$.}
		\label{D_5}
	\end{figure}
	Contribution of the diagram $D_5$ as shown in Fig.~\ref{D_5} is
	\begin{align}
		\langle\psi|\hat{\sigma}^z_{j}(t)|\psi\rangle_{\text{RPA},D_5}&=\sum_{\tau_1=4}^{t-2}\langle\psi|\hat{\sigma}^z_{j}(t)|\psi\rangle^{\{1,\tau_1\}}_{\text{RPA},D_5}\notag\\
		&=\sum_{\tau_1=4}^{t-2}\left((\sigma_j^z)_{e_j,e_j}m^{\tau_1-1}_{b_j,b_j}m^{t-\tau_1-1}_{d_j,e_j}u_{\psi_j,b_j}u_{b_j,d_j}u^*_{\psi_j,b_j}u^*_{b_j,d_j}\right)\prod_{k\neq j}\left(m^{\tau_1-1}_{b_k,b_k}m^{t-\tau_1-1}_{d_k,e_k}u_{\psi_k,b_k}u_{b_k,d_k}u^*_{\psi_k,b_k}u^*_{b_k,d_k}\right)\notag\\
		&\quad\times\theta(t-4)\notag\\
		&=\frac{\langle\psi_j|\hat{\sigma}_j^z|\psi_j\rangle}{2^L}\sum_{\tau_1=4}^{t-2}(\cos 2h)^{t-\tau_1+1}\left(1+(\cos 2h)^{\tau_1-1}\right)^{L}\theta(t-4).
	\end{align}
	\begin{figure}[H]
		\centering
		\begin{tikzpicture}
			\def\rad{2}
			\draw[blue,dashed] (120:\rad) arc(120:420:\rad);
			\draw[red,thick,->] (120:\rad) arc(120:180:\rad)--(30:\rad) arc(30:-90:\rad);\draw[red,thick](-90:\rad) arc(-90:-150:\rad)--(60:\rad);
			\draw (120:{\rad}) node[above]{$\psi$};
			\draw (180:\rad) node[left]{$b$};
			\draw[fill=green] (-150:\rad) circle(2pt) node[left]{$c$};
			\draw[fill=green] (30:{\rad}) circle(2pt) node[right]{$c$};
			\draw (60:{\rad}) node[above]{$e$};
		\end{tikzpicture}
		\caption{Diagram $D_6$.}
		\label{D_6}
	\end{figure}
	Diagram $D_6$ as shown in Fig.~\ref{D_6} is just a reflection of Diagram $D_5$ as shown in Fig.~\ref{D_5} about a vertical line passing from the middle of the diagram. Therefore, their contributions are identical. Thus,
	\begin{align}
		\langle\psi|\hat{\sigma}^z_{j}(t)|\psi\rangle_{\text{RPA},D_6}&=\langle\psi|\hat{\sigma}^z_{j}(t)|\psi\rangle_{\text{RPA},D_5}.
	\end{align}
	\begin{figure}[H]
		\centering
		\begin{tikzpicture}
			\def\rad{2}
			\draw[blue,dashed] (120:\rad) arc(120:420:\rad);
			\draw[red,thick,->] (120:\rad) arc(120:150:\rad)--(0:\rad) arc(0:-90:\rad);\draw[red,thick](-90:\rad) arc(-90:-180:\rad)--(30:\rad) arc(30:60:\rad);
			\draw (120:{\rad}) node[above]{$\psi$};
			\draw (150:\rad) node[left]{$b$};
			\draw[fill=green] (180:\rad) circle(2pt) node[left]{$c$};
			\draw[fill=green] (0:{\rad}) circle(2pt) node[right]{$c$};
			\draw (30:{\rad}) node[right]{$e$};
			\draw (420:{\rad}) node[above]{$f$};
		\end{tikzpicture}
		\caption{Diagrams $D_7$. State $b$ appears at time step $\tau_1$ and state $c$ appears at time steps $\tau_1+1$ and $\tau_2$ where $1\leq\tau_1\leq t-6$ and $\tau_1+4\leq\tau_2\leq t-2$.}
		\label{D_7}
	\end{figure}
	Contribution of Diagram $D_7$ as shown in Fig.~\ref{D_7} is
	\begin{align}
		\langle\psi|\hat{\sigma}^z_{j}(t)|\psi\rangle_{\text{RPA},D_7}&=\sum_{\tau_1=1}^{t-6}\sum_{\tau_2=\tau_1+4}^{t-2}\langle\psi|\hat{\sigma}^z_{j}(t)|\psi\rangle^{\{\tau_1,\tau_2\}}_{\text{RPA},D_7}\notag\\
		&=\sum_{\tau_1=1}^{t-6}\sum_{\tau_2=\tau_1+4}^{t-2}\left((\sigma_j^z)_{f_j,f_j}m^{\tau_1}_{\psi_j,b_j}m^{\tau_2-\tau_1-1}_{c_j,c_j}m^{t-\tau_2-1}_{e_j,f_j}u_{b_j,c_j}u_{c_j,e_j}u^*_{b_j,c_j}u^*_{c_j,e_j}\right)\notag\\
		&\quad\times\prod_{k\neq j}\left(m^{\tau_1}_{\psi_k,b_k}m^{\tau_2-\tau_1-1}_{c_k,c_k}m^{t-\tau_2-1}_{e_k,f_k}u_{b_k,c_k}u_{c_k,e_k}u^*_{b_k,c_k}u^*_{c_k,e_k}\right)\theta(t-5)\notag\\
		&=\frac{\langle\psi_j|\hat{\sigma}_j^z|\psi_j\rangle}{2^L}\sum_{\tau_1=1}^{t-6}\sum_{\tau_2=\tau_1+4}^{t-2}(\cos 2h)^{t-\tau_2+\tau_1+1}\left(1+(\cos 2h)^{\tau_2-\tau_1-1}\right)^{L}\theta(t-5).
	\end{align}
	\begin{figure}[H]
		\centering
		\begin{tikzpicture}
			\def\rad{2}
			\draw[blue,dashed] (120:\rad) arc(120:420:\rad);
			\draw[red,thick,->] (120:\rad)--(30:\rad) arc(30:-90:\rad);\draw[red,thick] (-90:\rad) arc(-90:-210:\rad)--(60:\rad);
			\draw (120:{\rad}) node[above]{$\psi$};
			\draw[fill=green] (150:\rad) circle(2pt) node[left]{$b$};
			\draw[fill=green] (30:\rad) circle(2pt) node[right]{$b$};
			\draw (420:{\rad}) node[above]{$d$};
		\end{tikzpicture}
		\caption{Diagram $D_8$. State $b$ appears at time steps $1$ and $t-1$.}
		\label{D_8}
	\end{figure}
	Contribution of Diagram $D_8$ as shown in Fig.~\ref{D_8} is
	\begin{align}
		\langle\psi|\hat{\sigma}^z_{j}(t)|\psi\rangle_{\text{RPA},D_8}&=\langle\psi|\hat{\sigma}^z_{j}(t)|\psi\rangle^{\{1,t-1\}}_{\text{RPA},D_8}\notag\\
		&=\left((\sigma_j^z)_{d_j,d_j} m^{t-2}_{b_j,b_j}u_{\psi_j,b_j}u_{b_j,d_j}u^*_{\psi_j,b_j}u^*_{b_j,d_j}\right)\prod_{k\neq j}\left(m^{t-2}_{b_k,b_k}u_{\psi_k,b_k}u_{b_k,d_k}u^*_{\psi_k,b_k}u^*_{b_k,d_k}\right)\theta(t-3)\notag\\
		&=\frac{\langle\psi_j|\hat{\sigma}_j^z|\psi_j\rangle}{2^L}\cos^2 2h\left(1+(\cos 2h)^{t-2}\right)^{L}\theta(t-3).
	\end{align}
	\begin{figure}[H]
		\centering
		\begin{tikzpicture}
			\def\rad{2}
			\draw[blue,dashed] (120:\rad) arc(120:420:\rad);
			\draw[red,thick,->] (120:\rad)--(30:\rad)--(180:\rad) arc(180:270:\rad);\draw[red,thick] (270:\rad) arc(270:360:\rad)--(150:\rad)--(60:\rad);
			\draw (150:\rad) node[left]{$b$};
			\draw (180:\rad) node[left]{$c$};
			\draw (0:{\rad}) node[right]{$d$};
			\draw (30:\rad) node[right]{$e$};
			\draw (60:\rad) node[above]{$f$};
			\draw (120:{\rad}) node[above]{$\psi$};
		\end{tikzpicture}
		\caption{Diagram $D_9$}
		\label{D_9}
	\end{figure}
	
	\begin{align}
		\langle\psi|\hat{\sigma}^z_{j}(t)|\psi\rangle_{\text{RPA},D_9}&=\left((\sigma_j^z)_{f_j,f_j} (m^{t-4})_{c_j,d_j}u_{\psi_j,b_j}u_{b_j,c_j}u_{d_j,e_j}u_{e_j,f_j}u^*_{\psi_j,e_j}u^*_{e_j,c_j}u^*_{d_j,b_j}u^*_{b_j,f_j}\right)\notag\\
		&\quad\times\left\{\prod_{k\neq j}\left( (m^{t-4})_{c_k,d_k}u_{\psi_k,b_k}u_{b_k,c_k}u_{d_k,e_k}u_{e_k,f_k}u^*_{\psi_k,e_k}u^*_{e_k,c_k}u^*_{d_k,b_k}u^*_{b_k,f_k}\right)\right\}\theta(t-5)\\
		&=\frac{\langle\psi_j|\hat{\sigma}_j^z|\psi_j\rangle}{2^L} \left(\frac{1}{4} (\cos (8 h)+3) \cos ^{t-4}(2 h)+\cos ^2(2 h)\right) \left(\cos
		^{t-2}(2 h)+1\right)^{L-1}\theta(t-5)
	\end{align}
	\begin{figure}[H]
		\centering
		\begin{subfigure}[t]{0.3\linewidth}
			\centering
			\begin{tikzpicture}
				\def\rad{2}
				\draw[blue,dashed] (120:\rad) arc(120:420:\rad);
				\draw[red,thick,->] (120:\rad) arc(120:150:\rad)--(30:\rad)--(210:\rad) arc(210:270:\rad);\draw[red,thick] (270:\rad) arc(270:360:\rad)--(180:\rad)--(60:\rad);
				\node[left] at (150:\rad) {$b$};
				\node[left] at (180:\rad) {$c$};
				\node[left] at (-150:\rad) {$d$};
				\node[right] at (0:\rad) {$e$};
				\node[right] at (30:\rad) {$f$};
				\node[above] at (60:\rad) {$g$};
				\draw (120:{\rad}) node[above]{$\psi$};
			\end{tikzpicture}
			\caption{Diagram $D_{10}$. State $b$ appears at time step $\tau_1$ where $1\leq\tau_1\leq t-5$.}
			\label{D_10}
		\end{subfigure}
		\hfill
		\begin{subfigure}[t]{0.3\linewidth}
			\centering
			\begin{tikzpicture}
				\def\rad{2}
				\draw[blue,dashed] (120:\rad) arc(120:360:\rad)--(30:\rad) arc(30:60:\rad);
				\draw[red,thick,->] (120:\rad) arc(120:330:\rad);\draw[red,thick] (330:\rad)--(30:\rad)--(0:\rad)--(60:\rad);
				\node[right] at (-30:\rad) {$b$};
				\node[right] at (0:\rad) {$c$};
				\node[right] at (30:\rad) {$d$};
				\node[above] at (60:\rad) {$e$};
				\draw (120:{\rad}) node[above]{$\psi$};
			\end{tikzpicture}
			\caption{Diagram $D_{11}$. State $b$ appears at time step $t-3$.}
			\label{D_11}
		\end{subfigure}
		\hfill
		\begin{subfigure}[t]{0.3\linewidth}
			\centering
			\begin{tikzpicture}
				\def\rad{2}
				\draw[blue,dashed] (120:\rad) arc(120:330:\rad)--(0:\rad)--(30:\rad) arc(30:60:\rad);
				\draw[red,thick,->] (120:\rad) arc(120:300:\rad);\draw[red,thick] (300:\rad)--(30:\rad)--(0:\rad)--(330:\rad)--(60:\rad);
				\node[below] at (-60:\rad) {$b$};
				\node[right] at (-30:\rad) {$c$};
				\node[right] at (0:\rad) {$d$};
				\node[right] at (30:\rad) {$e$};
				\node[above] at (60:\rad) {$f$};
				\draw (120:{\rad}) node[above]{$\psi$};
			\end{tikzpicture}
			\caption{Diagram $D_{12}$. State $b$ appears at time step $t-4$.}
			\label{D_12}
		\end{subfigure}
	\end{figure}
	
	\begin{align}
		\langle\psi|\hat{\sigma}^z_{j}(t)|\psi\rangle_{\text{RPA},D_{10}}&=\sum_{\tau_1=1}^{t-5}(\sigma_j^z)_{g_j,g_j}(m^{\tau_1})_{\psi_j,b_j}(m^{t-\tau_1-4})_{d_j,e_j}u_{b_j,c_j}u_{c_j,d_j}u_{e_j,f_j}u_{f_j,g_j}u^*_{b_j,f_j}u^*_{f_j,d_j}u^*_{e_j,c_j}u^*_{c_j,g_j}\notag\\
		&\qquad\times\left\{\prod_{k\neq j}\left((m^{\tau_1})_{\psi_k,b_k}(m^{t-\tau_1-4})_{d_k,e_k}u_{b_k,c_k}u_{c_k,d_k}u_{e_k,f_k}u_{f_k,g_k}u^*_{b_k,f_k}u^*_{f_k,d_k}u^*_{e_k,c_k}u^*_{c_k,g_k}\right)\right\}\theta(t-6)\notag\\
		&=\frac{\langle\psi_j|\hat{\sigma}_j^z|\psi_j\rangle}{2^{L}} \sum _{\tau_1=1}^{t-5} \frac{1}{4} \cos ^{\tau_1}(2 h)
		\left((\cos (8 h)+3) \cos ^{t-\tau_1-4}(2 h)+4 \cos ^2(2 h)\right) \left(\cos^{t-\tau_1-2}(2 h)+1\right)^{L-1}\notag\\
		&\qquad\times\theta(t-6)
	\end{align}
	
	\begin{align}
		\langle\psi|\hat{\sigma}^z_{j}(t)|\psi\rangle_{\text{RPA},D_{11}}&=(\sigma_j^z)_{e_j,e_j}(m^{t-3})_{\psi_j,b_j}m_{c_j,d_j}u_{b_j,c_j}u_{d_j,e_j}u^*_{b_j,d_j}u^*_{c_j,e_j}\notag\\
		&\quad\times\left\{\prod_{k\neq j}\left((m^{t-3})_{\psi_k,b_k}m_{c_k,d_k}u_{b_k,c_k}u_{d_k,e_k}u^*_{b_k,d_k}u^*_{c_k,e_k}\right)\right\}\theta(t-4)\notag\\
		&=\frac{\langle\psi_j|\hat{\sigma}_j^z|\psi_j\rangle}{2^{L+1}}\left(2+\cos(2h)+\cos(6h)\right)\cos^{t-3}(2h)(1+\cos(2h))^{L-1}\theta(t-4)
	\end{align}
	
	\begin{align}
		\langle\psi|\hat{\sigma}^z_{j}(t)|\psi\rangle_{\text{RPA},D_{12}}&=(\sigma_j^z)_{f_j,f_j}(m^{t-4})_{a_j,b_j}(m^2)_{c_j,e_j}u_{b_j,c_j}u_{e_j,f_j}u^*_{b_j,e_j}u^*_{c_j,f_j}\notag\\
		&\quad\times\left\{\prod_{k\neq j}\left((m^{t-4})_{a_k,b_k}(m^2)_{c_k,e_k}u_{b_k,c_k}u_{e_k,f_k}u^*_{b_k,e_k}u^*_{c_k,f_k}\right)\right\}\theta(t-5)\notag\\
		&=\frac{\langle\psi_j|\hat{\sigma}_j^z|\psi_j\rangle}{2^{L}}\left(1+\cos(4h)\cos^2(2h)\right)\cos^{t-4}(2h)(1+\cos^2(2h))^{L-1}\theta(t-5)
	\end{align}
	\begin{figure}[H]
		\centering
		\begin{subfigure}[t]{0.3\linewidth}
			\centering
			\begin{tikzpicture}
				\def\rad{2}
				\draw[blue,dashed] (120:\rad) arc(120:420:\rad);
				\draw[red,thick,->] (120:\rad)--(0:\rad)--(180:\rad) arc(180:270:\rad);\draw[red,thick] (270:\rad) arc(270:330:\rad)--(150:\rad)--(30:\rad) arc(30:60:\rad);
				\node[left] at (150:\rad) {$b$};
				\node[left] at (180:\rad) {$c$};
				\node[below right] at (-30:\rad) {$d$};
				\node[right] at (0:\rad) {$e$};
				\node[right] at (30:\rad) {$f$};
				\node[above] at (60:\rad) {$g$};
				\draw (120:{\rad}) node[above]{$\psi$};
			\end{tikzpicture}
			\caption{Diagram $D_{13}$. State $f$ appears at time step $\tau_1$.}
			\label{D_13}
		\end{subfigure}
		\hfill
		\begin{subfigure}[t]{0.3\linewidth}
			\centering
			\begin{tikzpicture}
				\def\rad{2}
				\draw[blue,dashed] (120:\rad)--(150:\rad)--(180:\rad)--(210:\rad) arc(210:420:\rad);
				\draw[red,thick,->] (120:\rad)--(180:\rad)--(150:\rad)--(210:\rad) arc(210:270:\rad);\draw[red,thick] (270:\rad) arc(270:420:\rad);
				\node[left] at (150:\rad) {$b$};
				\node[left] at (180:\rad) {$c$};
				\node[left] at (-150:\rad) {$d$};
				\node[above] at (60:\rad) {$e$};
				\draw (120:{\rad}) node[above]{$\psi$};
			\end{tikzpicture}
			\caption{Diagram $D_{14}$. State $d$ appears at time step 4.}
			\label{D_14}
		\end{subfigure}
		\hfill
		\begin{subfigure}[t]{0.3\linewidth}
			\centering
			\begin{tikzpicture}
				\def\rad{2}
				\draw[blue,dashed] (120:\rad)--(150:\rad)--(180:\rad)--(210:\rad)--(240:\rad) arc(240:420:\rad);
				\draw[red,thick,->] (120:\rad)--(210:\rad)--(180:\rad)--(150:\rad)--(240:\rad) arc(240:270:\rad);\draw[red,thick] (270:\rad) arc(270:420:\rad);
				\node[left] at (150:\rad) {$b$};
				\node[left] at (180:\rad) {$c$};
				\node[left] at (-150:\rad) {$d$};
				\node[left] at (-120:\rad) {$e$};
				\node[above] at (60:\rad) {$f$};
				\draw (120:{\rad}) node[above]{$\psi$};
			\end{tikzpicture}
			\caption{Diagram $D_{15}$. State $e$ appears at time step 5.}
			\label{D_15}
		\end{subfigure}
	\end{figure}
	Diagrams $D_{13}, D_{14}$, and $D_{15}$ are mirror images of diagrams $D_{10}, D_{11}$, and $D_{12}$, respectively, up to direction of arrow along red arcs. Since the $m$ and $v$ matrices are symmetric, change in arrows along red arcs does not lead to any change in the contribution of diagrams. Therefore, we conclude
	\begin{align}
		\langle\psi|\hat{\sigma}^z_{j}(t)|\psi\rangle_{\text{RPA},D_{13}}&=\langle\psi|\hat{\sigma}^z_{j}(t)|\psi\rangle_{\text{RPA},D_{10}}\\
		\langle\psi|\hat{\sigma}^z_{j}(t)|\psi\rangle_{\text{RPA},D_{14}}&=\langle\psi|\hat{\sigma}^z_{j}(t)|\psi\rangle_{\text{RPA},D_{11}}\\
		\langle\psi|\hat{\sigma}^z_{j}(t)|\psi\rangle_{\text{RPA},D_{15}}&=\langle\psi|\hat{\sigma}^z_{j}(t)|\psi\rangle_{\text{RPA},D_{12}}
	\end{align}
	\begin{figure}[H]
		\centering
		\begin{subfigure}[t]{0.3\linewidth}
			\centering
			\begin{tikzpicture}
				\def\rad{2}
				\draw[blue,dashed] (120:\rad) arc(120:420:\rad);
				\draw[red,thick,->] (120:\rad) arc(120:150:\rad)--(0:\rad)--(210:\rad) arc(210:270:\rad);\draw[red,thick] (270:\rad) arc(270:330:\rad)--(180:\rad)--(30:\rad) arc(30:60:\rad);
				\node[left] at (150:\rad) {$b$};
				\node[left] at (180:\rad) {$c$};
				\node[left] at (210:\rad) {$d$};
				\node[right] at (-30:\rad) {$e$};
				\node[right] at (0:\rad) {$f$};
				\node[right] at (30:\rad) {$g$};
				\node[above] at (60:\rad) {$h$};
				\draw (120:{\rad}) node[above]{$\psi$};
			\end{tikzpicture}
			\caption{Diagram $D_{16}$. State $c$ appears at time step $\tau_1$ and state $f$ appears at time step $\tau_2$ where $2\leq\tau_1\leq t-5$ and $\tau_1+3\leq\tau_2\leq t-2$.}
		\end{subfigure}
		\hfill
		\begin{subfigure}[t]{0.3\linewidth}
			\centering
			\begin{tikzpicture}
				\def\rad{2}
				\draw[blue,dashed] (120:\rad) arc(120:225:\rad)--(255:\rad)--(285:\rad)--(315:\rad) arc(315:420:\rad);
				\draw[red,thick,->] (120:\rad) arc(120:180:\rad); \draw[red,thick] (180:\rad) arc(180:225:\rad)--(285:\rad)--(255:\rad)--(315:\rad) arc(315:420:\rad);
				\node[left] at (225:\rad) {$b$};
				\node[below] at (255:\rad) {$c$};
				\node[below] at (285:\rad) {$d$};
				\node[right] at (315:\rad) {$e$};
				\node[above] at (60:\rad) {$f$};
				\draw (120:{\rad}) node[above]{$\psi$};
			\end{tikzpicture}
			\caption{Diagram $D_{17}$. State $c$ appears at time step $\tau_1$ and state $d$ appears at time step $\tau_1+1$ where $2\leq\tau_1\leq t-3$.}
		\end{subfigure}
		\hfill
		\begin{subfigure}[t]{0.3\linewidth}
			\centering
			\begin{tikzpicture}
				\def\rad{2}
				\draw[blue,dashed] (120:\rad) arc(120:210:\rad)--(240:\rad)--(270:\rad)--(300:\rad)--(330:\rad) arc(330:420:\rad);
				\draw[red,thick,->] (120:\rad) arc(120:180:\rad);\draw[red,thick] (180:\rad) arc(180:210:\rad)--(300:\rad)--(270:\rad)--(240:\rad)--(330:\rad) arc(330:420:\rad);
				\node[left] at (210:\rad) {$b$};
				\node[below] at (240:\rad) {$c$};
				\node[below] at (270:\rad) {$d$};
				\node[below] at (300:\rad) {$e$};
				\node[right] at (-30:\rad) {$f$};
				\node[above] at (60:\rad) {$g$};
				\draw (120:{\rad}) node[above]{$\psi$};
			\end{tikzpicture}
			\caption{Diagram $D_{18}$. State $c$ appears at time step $\tau_1$ and state $e$ appears at time step $\tau_1+2$ where $2\leq\tau_1\leq t-4$.}
		\end{subfigure}
	\end{figure}
	
	\begin{samepage}
		\begin{align}
			\langle\psi|\hat{\sigma}^z_{j}(t)|\psi\rangle_{\text{RPA},D_{16}}&=\sum_{\tau_1=2}^{t-3}\sum_{\tau_2=\tau_1+3}^{t-2}(\sigma_j^z)_{h_j,h_j}(m^{\tau_1-1})_{a_j,b_j}(m^{\tau_2-\tau_1-2})_{d_j,e_j}(m^{t-\tau_2-1})_{g_j,h_j}u_{b_j,c_j}u_{c_j,d_j}u_{e_j,f_j}u_{f_j,g_j}\notag\\
			&\quad\times u^*_{b_j,f_j}u^*_{f_j,d_j}u^*_{e_j,c_j}u^*_{c_j,g_j}\notag\\
			&\quad\times\prod_{k\neq j}\left((m^{\tau_1-1})_{a_k,b_k}(m^{\tau_2-\tau_1-2})_{d_k,e_k}(m^{t-\tau_2-1})_{g_k,h_k}u_{b_k,c_k}u_{c_k,d_k}u_{e_k,f_k}u_{f_k,g_k}u^*_{b_k,f_k}u^*_{f_k,d_k}u^*_{e_k,c_k}u^*_{c_k,g_k}\right)\notag\\
			&\quad\times\theta(t-7)\notag\\
			=&\frac{\langle\psi_j|\hat{\sigma}_j^z|\psi_j\rangle}{2^{L}}\sum_{\tau_1=2}^{t-5}\sum_{\tau_2=\tau_1+3}^{t-2}\left(\cos^{t-\tau_2+\tau_1}(2h)+(1-(1/2)\sin^2(4h))\cos^{t-4}(2h)\right)\left(1+\cos^{\tau_2-\tau_1}(2h)\right)^{L-1}\notag\\
			&\times \theta(t-7)
		\end{align}
	\end{samepage}
	
	\begin{align}
		\langle\psi|\hat{\sigma}^z_{j}(t)|\psi\rangle_{\text{RPA},D_{17}}=&\sum_{\tau_1=2}^{t-3}(\sigma_j^z)_{f_j,f_j}(m^{\tau_1-1})_{a_j,b_j}(m^{t-\tau_1-2})_{e_j,f_j}m_{c_j,d_j}u_{b_j,c_j}u_{d_j,e_j}u^*_{b_j,d_j}u^*_{c_j,e_j}\notag\\
		&\times\left\{\prod_{k\neq j}\left((m^{\tau_1-1})_{a_k,b_k}(m^{t-\tau_1-2})_{e_k,f_k}m_{c_k,d_k}u_{b_k,c_k}u_{d_k,e_k}u^*_{b_k,d_k}u^*_{c_k,e_k}\right)\right\}\theta(t-5)\notag\\
		=&\frac{\langle\psi_j|\hat{\sigma}_j^z|\psi_j\rangle}{2^L}(t-4)(1+2\cos^3(2h)-\cos(2h))\cos^{t-2}(2h)\left(1+\cos(2h)\right)^{L-1}\theta(t-5)
	\end{align}
	
	\begin{align}
		\langle\psi|\hat{\sigma}^z_{j}(t)|\psi\rangle_{\text{RPA},D_{18}}=&\sum_{\tau_1=2}^{t-4}(\sigma_j^z)_{g_j,g_j}(m^{\tau_1-1})_{a_j,b_j}(m^{t-\tau_1-3})_{f_j,g_j}m^2_{c_j,e_j}u_{b_j,c_j}u_{e_j,f_j}u^*_{b_j,e_j}u^*_{c_j,f_j}\notag\\
		&\times \left\{\prod_{k\neq j}\left((m^{\tau_1-1})_{a_k,b_k}(m^{t-\tau_1-3})_{f_k,g_k}m^2_{c_k,e_k}u_{b_k,c_k}u_{e_k,f_k}u^*_{b_k,e_k}u^*_{c_k,f_k}\right)\right\}\theta(t-6)\notag\\
		=&\frac{\langle\psi_j|\hat{\sigma}_j^z|\psi_j\rangle}{2^L}(t-5)(1+\cos(4h)\cos^2(2h))\cos^{t-4}(2h)(1+\cos^2(2h))^{L-1}\theta(t-6)
	\end{align}
	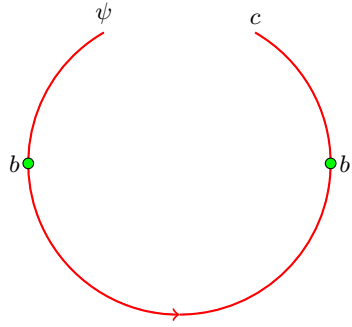
\begin{figure}[H]
		\centering
		\begin{tikzpicture}
			\def\rad{2}
			\draw[blue,dashed] (120:\rad) arc(120:420:\rad);
			\draw[red,thick,->] (120:\rad) arc(120:270:\rad);\draw[red,thick] (270:\rad) arc(270:420:\rad);
			\draw[fill=green] (180:\rad) circle(2pt);
			\draw[fill=green] (0:\rad) circle(2pt);
			\node[left] at (180:\rad) {$b$};
			\node[right] at (0:\rad) {$b$};
			\node[above] at (60:\rad) {$c$};
			\draw (120:{\rad}) node[above]{$\psi$};
		\end{tikzpicture}
		\caption{Diagram $D_{19}$. State $b$ appears at time steps $\tau_1$ and $\tau_2$ where $1\leq\tau_1\leq t-2$ and $\tau_1+1\leq\tau_2\leq t-1$.}
	\end{figure}
	\begin{samepage}
		\begin{align}
			\langle\psi|\hat{\sigma}^z_{j}(t)|\psi\rangle_{\text{RPA},D_{19}}&=\sum_{\tau_1=1}^{t-2}\sum_{\tau_2=\tau_1+1}^{t-1}\langle\psi|\hat{\sigma}^z_{j}(t)|\psi\rangle^{\{\tau_1,\tau_2\}}_{\text{RPA},D_{18}}\notag\\
			&=\sum_{\tau_1=1}^{t-2}\sum_{\tau_2=\tau_1+1}^{t-1}(\sigma_j^z)_{c_j,c_j} (m^{\tau_1})_{a_j,b_j}(m^{\tau_2-\tau_1})_{b_j,b_j}(m^{t-\tau_2})_{b_j,c_j}\notag\\
			&\quad\times\left\{\prod_{k\neq j}\left((m^{\tau_1})_{a_k,b_k}(m^{\tau_2-\tau_1})_{b_k,b_k}(m^{t-\tau_2})_{b_k,c_k}\right)\right\}\theta(t-3)\notag\\
			&=\frac{\langle\psi_j|\hat{\sigma}_j^z|\psi_j\rangle}{2^L}\sum_{\tau_1=1}^{t-2}\sum_{\tau_2=\tau_1+1}^{t-1}\cos^{t-\tau_2+\tau_1}(2h)(1+\cos^{\tau_2-\tau_1}(2h))^{L}\theta(t-3).
		\end{align}
	\end{samepage}
	Next to leading-order correction is
	\begin{align}
		\langle\psi|\hat{\sigma}^z_j(t)|\psi\rangle_{\mathrm{RPA},\sum_i D_i}=\sum_{k=1}^4\langle\psi|\hat{\sigma}^z_{j}(t)|\psi\rangle_{\text{RPA},D_{k}}-\sum_{k=5}^8\langle\psi|\hat{\sigma}^z_{j}(t)|\psi\rangle_{\text{RPA},D_{k}}+\sum_{k=9}^{18}\langle\psi|\hat{\sigma}^z_{j}(t)|\psi\rangle_{\text{RPA},D_{k}}-\langle\psi|\hat{\sigma}^z_{j}(t)|\psi\rangle_{\text{RPA},D_{19}}.
	\end{align}
	In Fig.~\ref{KIC_local_magnetization}a, the red curve is obtained by calculating this expression in addition to the leading order result Eq.~(\ref{sigma_j^z_t_I_contri}) which shows a really good match with the direct simulation result even at long times. Similar calculations can also be performed for the total energy which determines the red curve in Fig.~\ref{KIC_local_magnetization}b.
	
	\section{Discrete time crystal}
	In Eq.~(\ref{At_RPA_identity_eig_dec_expan}), the first term on right hand side is the infinite temperature statistical mechanics value. Within identity permutation the observable $\hat{A}$ saturates to this value at long times. This term results from largest eigenvalue $\lambda_0=1$ and corresponding eigenvector $\langle\lambda_0|\equiv(1,...,1)/\sqrt{\mathcal{N}}$ of the matrix $\mathcal{M}$. This eigenvalue is typically nondegenerate when $\mathcal{M}$ is irreducible. However, for certain hermitian matrices $H_1$, $\mathcal{M}$ is reducible and decomposes into direct sum of smaller doubly stochastic matrices $\mathfrak{m}_1,...,\mathfrak{m}_q$
	\begin{align}
		\mathcal{M}=&\mathfrak{m}_1\oplus...\oplus\mathfrak{m}_q.
	\end{align} 
	In this case, we label the eigenvalues of $\mathfrak{m}_{q'}$ as $\lambda_{q',i}$ where $q'=1,...,q$. Since each $\mathfrak{m}_{q'}$ is a doubly stochastic matrix, $\lambda_{q',0}=1,\forall q'$. Therefore, eigenvalue $1$ is $q$-fold degenerate. In this case, system thermalizes to infinite temperature state only if the initial state belongs to only one of these sectors. If the initial state has support over multiple sectors then at long times the expectation value of an observable $\hat{A}$ saturates to
	\begin{align}
		\langle\psi|\hat{A}|\psi\rangle_{\text{RPA},I}&=\sum_{q'=1}^qp_{q'}(|\psi\rangle)\frac{\tr_{q'}\hat{A}}{\mathcal{N}_{q'}},
	\end{align}
	where $p_{q'}(|\psi\rangle)$ is the probability of finding the initial state in the $q'$-th block and $\mathcal{N}_{q'}$ is the dimension of that block. There exists a special case where even if the matrix $\mathcal{M}$ is irreducible, multiple distinct eigenvalues can have unit magnitude. They are called maximum-modulus eigenvalues. According to Perron-Frobenius theorem for nonnegative irreducible matrices, if there are $l$ maximum-modulus eigenvalues then they are $r e^{2i\pi k/l}$ where $k=0,...,l-1$ and $r$ is the magnitude of these eigenvalues. For our doubly-stochastic matrix $\mathcal{M}$, $r=1$. Therefore, the maximum-modulus eigenvalues are exactly the $l^{\text{th}}$ roots of unity. In this case, expectation values of observables do not reach a constant value at long times, instead they oscillate with a period of $l\tau_p$. A simple case is when $l=2$ and the eigenvalues $1$ and $-1$ appear. Now, we find hermitian matrices $H_1$ which can give rise to the $-1$ eigenvalue of the matrix $\mathcal{M}$. It is well known that a matrix has a spectrum symmetric around the origin in complex plane if it anticommutes with a matrix $\tau^z$, $\{\tau^z,\mathcal{M}\}=0$. If $|\lambda\rangle$ is an eigenvector of $\mathcal{M}$ with eigenvalue $\lambda$ then $\tau^z|\lambda\rangle$ is also an eigenvector of $\mathcal{M}$ with eigenvalue $-\lambda$. In general, it is hard to find the set of all doubly stochastic matrices $\mathcal{M}$ and corresponding $\tau^z$ that anticommute with each other. Nevertheless, in general, a block-off diagonal matrix
	\begin{align}
		\mathcal{M}=\begin{pmatrix}
			0& \mathcal{A}\\
			\mathcal{B}& 0
		\end{pmatrix},
		\label{M_dyn_freez}
	\end{align}
	anticommutes with 
	\begin{align}
		\tau^z=\begin{pmatrix}
			I_{\frac{\mathcal{N}}{2}}& 0\\
			0& -I_{\frac{\mathcal{N}}{2}}
		\end{pmatrix},
		\label{sigma^z}
	\end{align}
	where $\mathcal{A}$ and $\mathcal{B}$ are doubly stochastic matrices of $\mathcal{N}/2\times \mathcal{N}/2$ and $I_{\frac{\mathcal{N}}{2}}$ is the identity matrix of size $\mathcal{N}/2\times \mathcal{N}/2$. Such a matrix $\mathcal{M}$ has both $1$ and $-1$ as eigenvalues. Since elements of $\mathcal{M}$ are related to elements of $V$ by $\mathcal{M}_{ab}=|V_{ab}|^2$, for $\mathcal{M}$ to be block off diagonal $V$ should also be block off diagonal. This implies that the matrix $V$ also anticommutes with $\tau^z$. Therefore, the eigenvalues of $V$ are also symmetrically located around the origin of a complex plane. Since $\hat{V}=e^{-i\hat{H}_1}$, if $E_j$ is an eigenvalue of $\hat{H}_1$ with eigenvector $|E_j\rangle$ then $e^{-i E_j}$ is an eigenvalue of $\hat{V}$ with the same eigenvector $|E_j\rangle$. As argued earlier, based on anticommutation of $V$ with $\tau^z$, there must exist some $E_{j'}$ such that $e^{-i E_{j'}}=-e^{-i E_j}$. Therefore, $E_{j'}-E_j=(2n+1)\pi$ where $n$ is some integer. Additionally, if
	\begin{align}
		|E_j\rangle\equiv\begin{pmatrix}
			e_j\\
			e'_j
		\end{pmatrix}
	\end{align}
	then
	\begin{align}
		|E_{j'}\rangle\equiv \tau^z\begin{pmatrix}
			e_j\\
			e'_j
		\end{pmatrix}=\begin{pmatrix}
			e_j\\
			-e'_j
		\end{pmatrix}
	\end{align}
	These properties are satisfied by the spectrum of any half-integer spin in a magnetic field of strength $\pi$. For example, the spectrum of a spin-$s$ hamiltonian $\pi S^x$ is just $\pi k$ where $k=-s,...,s$ and $s\in\{1/2,3/2,5/2,...\}$. Thus, $E_k-E_{-k}=2k\pi$. Since, $k$ is half-integer, $2k$ is an odd number. Since a many-body Hamiltonian has eigenvalues which grow linearly with system size, this result also implies that for $H_1$ to be a many-body Hamiltonian, it should be expressed as a sum of local Hamiltonians, e.g.
	\begin{align}
		\hat{H}_1=&\pi\sum_{j=1}^L \hat{S}_{j}^{x}.
	\end{align}
	Therefore, with a single kick in one period of driving, most likely only $\hat{H}_1$ that can be expressed as a sum of local hamiltonians of the kind discussed above can lead to the discrete time crystal phase. Now we consider some examples. 
	\subsection{Kicked transverse-field Ising chain}
	The Hamiltonian for this model was given in Eqs.~(\ref{KIC_H0}) and (\ref{KIC_H1}). In this case, when $h=\pi/2$, following Eq.~(\ref{M_KIC}), we obtain
	\begin{align}
		m&=\begin{pmatrix}
			0& 1\\
			1& 0
		\end{pmatrix}.
	\end{align}
	Therefore, eigenvalues of $m$ are $\pm1$ leading to discrete time crystal phase. We obtain deeper insights into the reasons behind this behavior by analyzing the matrix $V$. Following Eq.~(\ref{KIC_V}), we find that when $h=\pi/2$
	\begin{align}
		u&=-i\begin{pmatrix}
			0& 1\\
			1& 0
		\end{pmatrix}.
	\end{align}
	Therefore, $\hat{V}$ has the effect of flipping all the spins and changing the phase of a many-body state by $(-i)^L$. Thus, we can express it as follows
	\begin{align}
		\hat{V}&=(-i)^L\hat{F},\; \hat{F}=\hat{\sigma}_1^x\otimes...\otimes\hat{\sigma}_L^x.
		\label{Dyn_freez_V_oper}
	\end{align}
	Equation (\ref{Dyn_freez_V_oper}) implies that $\hat{V}^2=(-i)^{2L}\hat{I}$ where $\hat{I}$ is the identity operator. Therefore, if $|\psi\rangle$ is a basis state, then $\hat{F}|\psi\rangle$ is also a basis state. We denote this new basis state as $|F\psi\rangle$. Following these properties of the operators $\hat{V}$ and $\hat{W}$, we obtain the time evolved state 
	\begin{align}
		\hat{U}^t|\psi\rangle&=\begin{cases}
			(-i)^{tL}e^{-i(t/2)(\theta_{\psi}+\theta_{F\psi})}|\psi\rangle,\;\text{if $t$ is even}\\
			(-i)^{tL}e^{-i(\lceil t/2\rceil \theta_{\psi}+\lfloor t/2\rfloor \theta_{F\psi})}|F\psi\rangle,\; \text{if $t$ is odd}
		\end{cases}.
	\end{align}
	Therefore, for any observable $\hat{A}$
	\begin{align}
		\langle\psi|\hat{A}(t)|\psi\rangle&=\begin{cases}
			\langle\psi|\hat{A}|\psi\rangle,\;\text{if $t$ is even},\\
			\langle F\psi|\hat{A}|F\psi\rangle,\;\text{if $t$ is odd}
		\end{cases}.
	\end{align}
	\begin{figure*}[t!]
		\centering
		\begin{subfigure}[t]{0.45\linewidth}
			\centering
			\includegraphics[width=\linewidth]{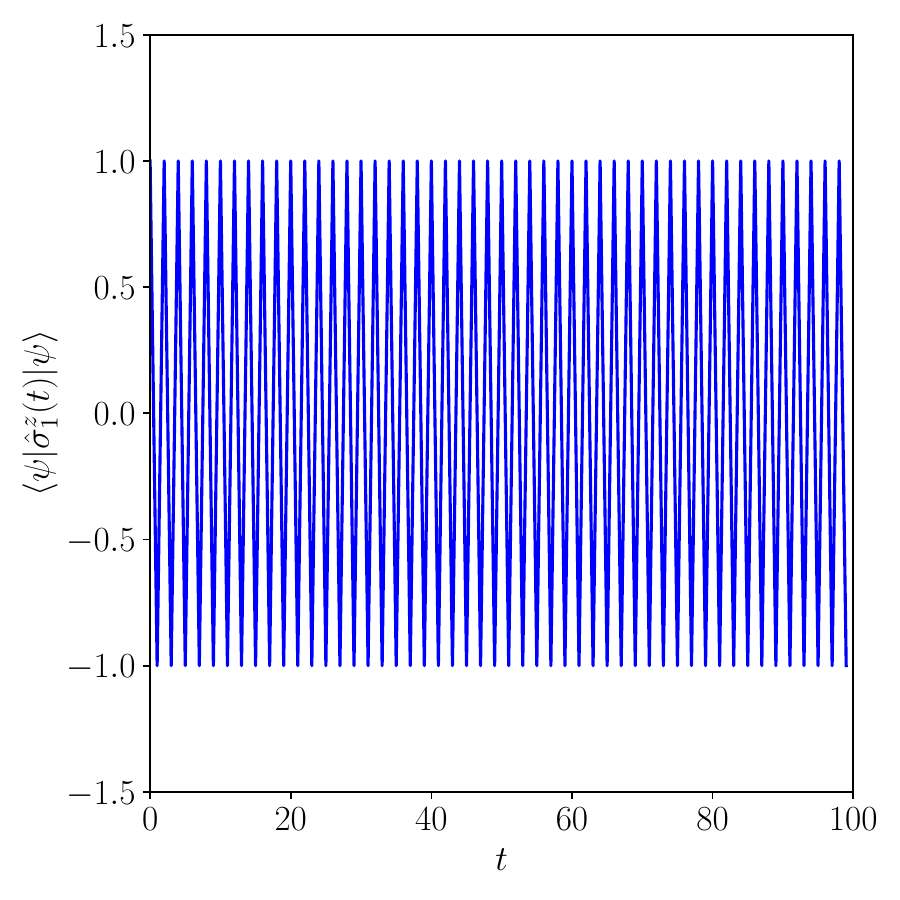}
			\caption{}
		\end{subfigure}
		\hfill
		\begin{subfigure}[t]{0.45\linewidth}
			\centering
			\includegraphics[width=\linewidth]{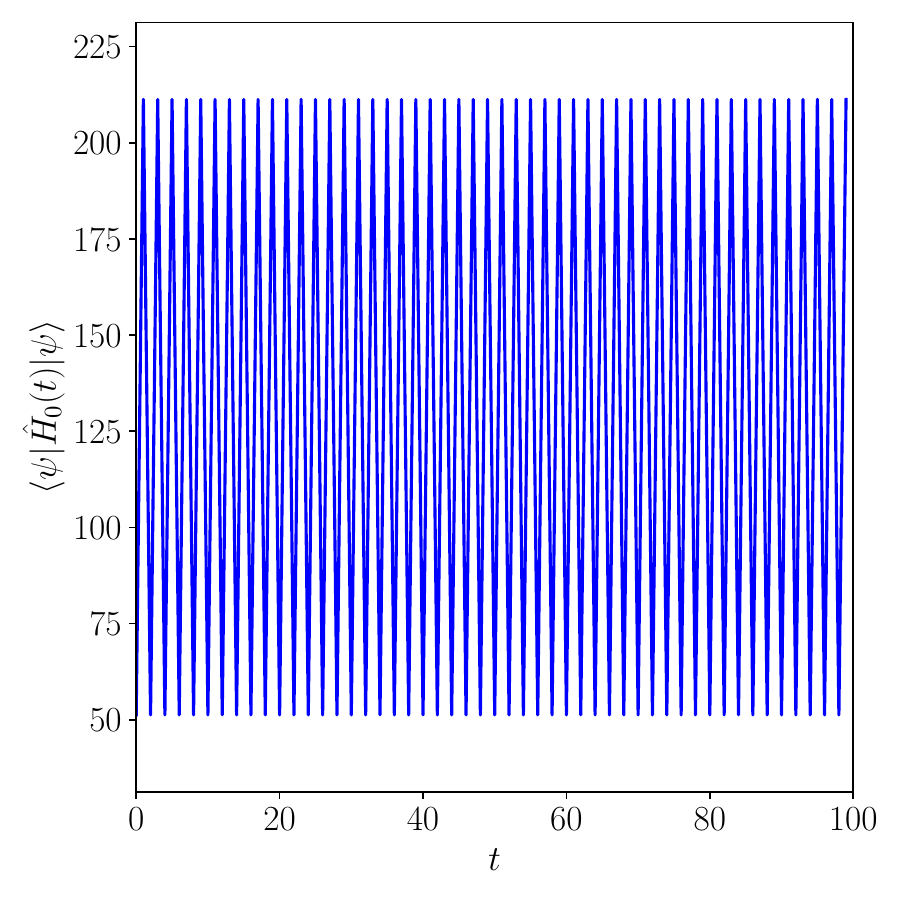}
			\caption{}
		\end{subfigure}
		\caption{\justifying{\small Time evolution of (a) local magnetization and (b) total energy calculated in the discrete time crystal phase at $h=\pi/2$ for the Hamiltonian described by Eqs.~(\ref{KIC_H0}) and (\ref{KIC_H1}). We take initial state $|\psi\rangle=|1,1,1,-1,-1,-1,-1,-1,-1,-1,-1,-1,-1,-1\rangle$.}}
		\label{KIC_szt_H0t_Dyn_freez}
	\end{figure*}
	Therefore, the expectation value of each observable oscillates with time, depends on the initial state, and does not match the statistical mechanics prediction, implying the absence of thermalization. For example, let us consider a local observable $\hat{\sigma}_j^z$ and an initial state $|\psi\rangle=|\psi_1,...,\psi_L\rangle$. The time evolved state up to a global phase is
	\begin{align}
		|\psi(t)\rangle&=\begin{cases}
			|\psi_1,...,\psi_L\rangle,\;\text{if $t$ is even},\\
			|-\psi_1,...,-\psi_L\rangle,\;\text{if $t$ is odd}
		\end{cases},
	\end{align}
	where $\psi_j=\pm 1$, $\forall j$. Therefore, 
	\begin{align}
		\langle\psi(t)|\hat{\sigma}_j^z|\psi(t)\rangle=(-1)^t\psi_j.
		\label{KIC_szt_Dyn_freez}
	\end{align}
	We next consider the operator of total energy $\hat{H}_0$
	\begin{align}
		\langle\psi(t)|\hat{H}_0|\psi(t)\rangle&=(-1)^t\sum_{i=1}^L\epsilon_i\psi_i+\sum_{i<j}\frac{U_{ij}}{(d_{ij})^\alpha}\psi_i\psi_j.
	\end{align}
	Performing average over disorder, we obtain
	\begin{align}
		\langle\psi(t)|\hat{H}_0|\psi(t)\rangle_{\text{dis}}&=(-1)^t\sum_{i=1}^L\epsilon\psi_i+\sum_{i<j}\frac{U_0}{(d_{ij})^\alpha}\psi_i\psi_j.
		\label{KIC_H0t_Dyn_freez}
	\end{align}
	We plot these results in Eqs.~(\ref{KIC_szt_Dyn_freez}) and (\ref{KIC_H0t_Dyn_freez}) in Fig.~\ref{KIC_szt_H0t_Dyn_freez}.
	\subsection{Kicked transverse-field Ising chain with a nonlinear drive}
	Here, we consider a periodically kicked chain of $L$ spin-$s$ degrees of freedom, whose base Hamiltonian $\hat{H}_0$ and driving Hamiltonian $\hat{H}_1$ are
	\begin{align}
		\hat{H}_0&=\sum_{i=1}^L\epsilon_i \hat{S}_i^z+\sum_{i<j}\frac{U_{ij}}{(d_{ij})^\alpha}\hat{S}_i^z\hat{S}_j^z,\\
		\hat{H}_1&=h\sum_{i=1}^L(\hat{S}_i^x)^r,
	\end{align}
	where $r$ takes some positive integer value. In this case, the driving is local but nonlinear. The nonlinear nature of the drive leads to some nontrivial values of $h$ for which a discrete time crystal phase occurs, as discussed below. The matrix $V$ is
	\begin{align}
		V&=v^{\otimes L},\; v=e^{-ih (S^x)^r}.
	\end{align}
	Consider $s=2$. For this case, in the computational basis
	\begin{align}
		S^x=\begin{pmatrix}
			0& 1& 0& 0& 0\\
			1& 0& \sqrt{3/2}& 0& 0\\
			0& \sqrt{3/2}& 0& \sqrt{3/2}& 0\\
			0& 0& \sqrt{3/2}& 0& 1\\
			0& 0& 0& 1& 0
		\end{pmatrix}.
	\end{align}
	The eigenvalues $s_x$ of $S^x$ with the corresponding eigenvectors $w_{s_x}$ in the computational basis are
	\begin{table}[H]
		\centering
		\begin{tabular}{|c|c|}
			\hline
			$s_x$ & $w_{s_x}$ \\[10pt]
			\hline
			$-2$ & $w_{-2}= \displaystyle\frac{1}{4}(1, -2, \sqrt{6}, -2, 1)^T$\\[10pt]
			\hline
			$-1$ & $w_{-1}= \displaystyle\frac{1}{2}(-1, 1, 0, -1, 1)^T$\\[10pt]
			\hline
			$0$ & $w_0= \displaystyle\frac{1}{2\sqrt{2}}(\sqrt{3}, 0, -\sqrt{2}, 0, \sqrt{3})^T$\\[10pt]
			\hline
			$1$ & $w_1= \displaystyle\frac{1}{2}(-1, -1, 0, 1, 1)^T$\\[10pt]
			\hline
			$2$ & $w_2= \displaystyle\frac{1}{4}(1, 2, \sqrt{6}, 2, 1)^T$\\[10pt]
			\hline
		\end{tabular}.
		\caption{Eigenvalues $s_x$ and eigenvectors $w_{s_x}$ of $\hat{S}^x$ for a spin-$s$ in the eigenbasis of $\hat{S}^z$.}
		\label{tab:placeholder}
	\end{table}
	For even value of $r$, $(-2)^r=2^r$. Thus, $w_{-2}$ and $w_{2}$ belong to a degenerate eigenspace of $(S^x)^r$. Similarly, $w_{-1}$ and $w_1$ also belong to a degenerate eigenspace of $(S^x)^r$. Therefore, we can take their linear combinations as follows
	\begin{align}
		w_{2^r,1}&=\frac{1}{\sqrt{2}}(w_2+w_{-2})=\frac{1}{2\sqrt{2}}(1,0,\sqrt{6},0,1)^T,\\
		w_{2^r,2}&=\frac{1}{\sqrt{2}}(w_2-w_{-2})=\frac{1}{\sqrt{2}}(0,1,0,1,0)^T,\\
		w_{1,1}&=\frac{1}{\sqrt{2}}(w_1+w_{-1})=\frac{1}{\sqrt{2}}(-1,0,0,0,1)^T,\\
		w_{1,2}&=\frac{1}{\sqrt{2}}(w_1-w_{-1})=\frac{1}{\sqrt{2}}(0,-1,0,1,0)^T.
	\end{align}
	We notice that the eigenvectors $w_{2^r,2}$ and $w_{1,2}$ have nonzero elements at positions two and four, while the elements of the other eigenvectors $w_{2^r,1}$, $w_{1,1}$, and $w_0$ are zero at the corresponding positions. Thus, the matrix $w$ is block-diagonal in the computational basis. One block denoted by $v_1$ acts on the subspace spanned by $w_{2^r,2}$ and $w_{1,2}$ while the other block denoted by $v_2$ acts on the subspace spanned by $w_{2^r,1}$, $w_{1,1}$, and $w_0$. We only compute the block $v_1$ as follows
	\begin{align}
		v_1&=e^{-ih2^r}w_{2^r,2}w_{2^r,2}^T+e^{-ih}w_{1,2}w_{1,2}^T\notag\\
		&=\frac{1}{2}\begin{pmatrix}
			0& 0& 0& 0& 0\\
			0& e^{-ih2^r}+e^{-ih}& 0& e^{-ih2^r}-e^{-ih}& 0\\
			0& 0& 0& 0& 0\\
			0& e^{-ih2^r}-e^{-ih}& 0& e^{-ih2^r}+e^{-ih}& 0\\
			0& 0& 0& 0& 0
		\end{pmatrix}.
	\end{align}
	To have a discrete time crystal phase, the diagonal elements of $v_1$ must be zero. Therefore, we solve the following equation
	\begin{align}
		e^{-ih2^r}+e^{-ih}&=0\notag\\
		\implies e^{-i h 2^r}=e^{-ih -i(2k+1)\pi}\notag\\
		\implies h&=\frac{2k+1}{2^r-1}\pi,\; k=0,\pm1,\pm2,...
	\end{align}
	The difference between two consecutive values of $h$ corresponding to the discrete time crystal phase is $2\pi/(2^r-1)$. Therefore, for large values of $r$, this phase will be robust for any value of $h$.
	
	Discrete time crystal phase with more generic many-body systems can be obtained with two kicks in one period. We explain this in the following subsection.
	\subsection{Two kicks per cycle}
	Here, we consider a periodically kicked chain of spin-1/2's, whose Hamiltonian takes the following form:
	\begin{align}
		\hat{H}(t)&=\hat{H}_0+\sum_{n\in\mathbb{Z}}\hat{H}_1\delta\left(\frac{t}{\tau_p}-n-r\right)+\hat{H}_2 \delta\left(\frac{t}{\tau_p}-n \right),
		\label{H_2kick}
	\end{align}
	where $0<r<1$ and $\tau_p$ is the driving period. To ensure the presence of time reversal symmetry, we choose $r=1/2$. In addition, we choose $\tau_p=1$. The driving Hamiltonians $\hat{H}_1$ and $\hat{H}_2$ are active at $t=(n+1/2)$ and $t=n$, respectively, where $n=0,\pm 1,...$. We take the following:
	\begin{align}
		\hat{H}_0=&\sum_{i=1}^L \epsilon_i \hat{\sigma}_i^z+\sum_{i<j}\frac{U_{ij}}{(d_{ij})^\alpha}\hat{\sigma}_i^z\hat{\sigma}_j^z,
		\label{H0_2kick}\\
		\hat{H}_1=&\sum_{i=1}^L J \hat{\sigma}_i^+\hat{\sigma}_{i+1}^-+h.c.,
		\label{H1_2kick}\\
		\hat{H}_2=& h\sum_{i=1}^L \hat{\sigma}_{i}^x,
		\label{H2_2kick}
	\end{align}
	where $\hat{\sigma}_i^y$ is the Pauli-$y$ operator at the $i^{\text{th}}$ site and $\sigma^{\pm}_i=(\sigma^x_i\pm\sigma^y_i)/2$. The Floquet operator for this system can be expressed as follows
	\begin{align}
		\hat{U}&=\lim_{\varepsilon\rightarrow 0}\mathcal{T} e^{-i\int_\varepsilon^{1+\varepsilon}dt \hat{H}(t)}\notag\\
		&=\hat{X}\hat{W}\hat{V}\hat{W},
		\label{U_2kick}
	\end{align}
	where $\hat{X}=e^{-i \hat{H}_2}$, $\hat{W}=e^{-i\hat{H}_0/2}$, and $\hat{V}=e^{-i\hat{H}_1}$. The operators $\hat{H}_0$ and $\hat{H}_1$ commute with $\hat{N}=\sum_{i=1}^L (\hat{\sigma}_i^z+1)/2$, where $\hat{N}$ measures the total number of spins in $|\uparrow\rangle$ state along the $z$ direction. However, the operator $\hat{H}_2$ does not commute with $\hat{N}$. Therefore, this Hamiltonian does not have any $U(1)$ symmetry. Nevertheless, the Hilbert space $\mathcal{H}$ can still be decomposed into degenerate eigenspaces of $\hat{N}$ as follows
	\begin{align}
		\mathcal{H}=\bigoplus_{N=0}^{L}\mathcal{H}_N,
	\end{align}
	where $\mathcal{H}_N$ is the degenerate eigenspace associated with an eigenvalue $N$ of $\hat{N}$. The operator $\hat{X}$ couples different $\mathcal{H}_N$. In particular, when $h=\pi/2$, $\hat{X}$ only couples eigenspaces $\mathcal{H}_N$ and $\mathcal{H}_{L-N}$. This can be understood from the matrix form of $\hat{X}$ in the computational basis
	\begin{align}
		X=&u^{\otimes L},
		\label{X_in_comp_basis}
	\end{align}  
	where
	\begin{align}
		u=&\begin{pmatrix}
			\cos(h)& -i\sin(h)\\
			-i \sin(h)& \cos(h)
		\end{pmatrix}.
		\label{u_in_comp_basis}
	\end{align}
	We notice that when $h=\pi/2$, $u=-i\sigma^x$ where $\sigma^x$ is the $x$ component of Pauli spin matrix. In this case, the action of $\hat{X}$ just flips all spins. Therefore, if a state $|\psi\rangle\in \mathcal{H}_N$ then $\hat{X}|\psi\rangle\in \mathcal{H}_{L-N}$. This leads to a very interesting stroboscopic dynamics. The action of the Floquet operator $\hat{U}$ on the state $|\psi\rangle$ can be understood in two steps as follows:\\
	(\romannumeral 1) Since $\hat{H}_0$ and $\hat{H}_1$ commute with $\hat{N}$, $\hat{W}\hat{V}\hat{W}|\psi\rangle\in\mathcal{H}_N$.\\
	(\romannumeral 2) For $h=\pi/2$, $\hat{X}$ flips all spins. Therefore, $\hat{X}\hat{W}\hat{V}\hat{W}|\psi\rangle\in\mathcal{H}_{L-N}$.\\
	Thus, $\hat{U}|\psi\rangle\in\mathcal{H}_{L-N}$ and $\hat{U}^2|\psi\rangle\in\mathcal{H}_N$. In general,
	\begin{align}
		\hat{U}^t|\psi\rangle\in&\begin{cases}
			\text{$\mathcal{H}_N$, if $t$ is even}\\
			\text{$\mathcal{H}_{L-N}$, if $t$ is odd}
		\end{cases}.
		\label{jump_btw_blocks}
	\end{align}
	Consequently, the state of the system at arbitrary time is trapped in the subspace $\mathcal{H}_N\oplus\mathcal{H}_{L-N}$. Therefore, while computing expectation value of observables $\langle\psi|\hat{A}(t)|\psi\rangle$ for an initial state $|\psi\rangle\in\mathcal{H}_N$ we must only consider $\hat{U}$ on the subspace $\mathcal{H}_N\oplus\mathcal{H}_{L-N}$. Our next goal is to calculate
	\begin{align}
		\langle\psi|\hat{A}(t)|\psi\rangle=&\langle\psi|\hat{U}^{-t}\hat{A}\hat{U}^t|\psi\rangle.
	\end{align}
	As earlier, we choose eigenstates of $\hat{H}_0$ as computational basis and denote them as $|\underline{s}\rangle=|s_1,...,s_L\rangle$ where $s_j$ is an eigenvalue of $\hat{\sigma}_j^z,\forall j\in\{1,...,L\}$.
	\begin{align}
		\hat{\sigma}_j^z|\underline{s}\rangle =s_j |\underline{s}\rangle,\quad s_j=\pm 1.
	\end{align}
	Inserting identities $\sum_{\underline{s}_\tau}|\underline{s}_\tau\rangle\langle\underline{s}_\tau|$ and $\sum_{\underline{s}'_\tau}|\underline{s}'_\tau\rangle\langle\underline{s}'_\tau|$, where $\tau=0,t$, we express
	\begin{align}
		\langle\psi|\hat{A}(t)|\psi\rangle&=\sum_{\underline{s}_0,\underline{s}_{t},\underline{s}'_0,\underline{s}'_{t}}\langle\psi|\underline{s}'_0\rangle\langle\underline{s}'_0|\hat{U}^{-t}|\underline{s}'_{t}\rangle\langle\underline{s}'_{t}|\hat{A}|\underline{s}_{t}\rangle\langle\underline{s}_{t}|\hat{U}^t|\underline{s}_0\rangle\langle\underline{s}_0|\psi\rangle.
		\label{At_2kick}
	\end{align}
	Inserting more identities $\sum_{\underline{s}_\tau}|\underline{s}_\tau\rangle\langle\underline{s}_\tau|$ for $\tau=1,...,t-1$, we express
	\begin{align}
		\langle\underline{s}_{t}|\hat{U}^t|\underline{s}_0\rangle&=\sum_{\underline{s}_1,...,\underline{s}_{t-1}}\prod_{\tau=0}^{t-1}U_{\underline{s}_{\tau+1},\underline{s}_\tau},
		\label{Ut_2kick_elem_1}
	\end{align}
	where
	\begin{align}
		U_{\underline{s}_{\tau+1},\underline{s}_\tau}=\langle \underline{s}_{\tau+1}|\hat{U}|\underline{s}_\tau\rangle.
		\label{U_2kick_elem_1}
	\end{align}
	Substituting Eq.~(\ref{U_2kick}) in (\ref{U_2kick_elem_1}), we obtain
	\begin{align}
		U_{\underline{s}_{\tau+1},\underline{s}_\tau}&=\langle \underline{s}_{\tau+1}|\hat{X}\hat{W}\hat{V}\hat{W}|\underline{s}_\tau\rangle.
		\label{U_2kick_elem_2}
	\end{align}
	The operator $\hat{X}$ flips all spins. However, it also changes the phase of a state by $(-i)^L$. This can be understood from Eqs.~(\ref{X_in_comp_basis}) and (\ref{u_in_comp_basis}) by substituting $h=\pi/2$. We define another operator $\hat{F}$ that flips all spins without any change in the phase of a state. Therefore,
	\begin{align}
		\hat{X}&=(-i)^L\hat{F},
	\end{align}
	where $\hat{F}$ satisfies $\hat{F}^\dagger=\hat{F}$ and $\hat{F}^2=\hat{I}$ as $\hat{F}=\hat{\sigma}_1^x\otimes...\otimes\hat{\sigma}_L^x$.
	Therefore,
	\begin{align}
		\langle\underline{s}_{\tau+1}|\hat{X}=(\hat{X}^\dagger |\underline{s}_{\tau+1}\rangle)^\dagger=(-i)^L(\hat{F}|\underline{s}_{\tau+1}\rangle)^\dagger\notag\\
		=(-i)^L\langle F\underline{s}_{\tau+1}|.
		\label{X_s_tau}
	\end{align}
	Using Eq.~(\ref{X_s_tau}) in (\ref{U_2kick_elem_2}), we obtain
	\begin{align}
		U_{\underline{s}_{\tau+1},\underline{s}_\tau}&=i^{-L}\langle F\underline{s}_{\tau+1}|\hat{W}\hat{V}\hat{W}|\underline{s}_\tau\rangle.
	\end{align}
	Both $|\underline{s}_\tau\rangle$ and $|F\underline{s}_\tau\rangle$ are eigenstates of $\hat{H}_0$ and $\hat{W}$. Therefore,
	\begin{align}
		\hat{W}|\underline{s}_\tau\rangle&=e^{-i\theta_{\underline{s}_\tau}}|\underline{s}_\tau\rangle,
		\label{W_s_bar}\\
		\langle F\underline{s}_{\tau+1}|\hat{W}&=(\hat{W}^\dagger |F\underline{s}_{\tau+1}\rangle)^\dagger=e^{-i\theta_{F\underline{s}_{\tau+1}}}\langle F\underline{s}_{\tau+1}|.
		\label{W_F_s_bar}
	\end{align}
	Using Eqs.~(\ref{W_s_bar}) and (\ref{W_F_s_bar}), we obtain
	\begin{align}
		U_{\underline{s}_{\tau+1},\underline{s}_\tau}&=i^{-L} e^{-i\theta_{F\underline{s}_{\tau+1}}-i\theta_{\underline{s}_\tau}}\langle F\underline{s}_{\tau+1}|\hat{V}|\underline{s}_\tau\rangle\notag\\
		&=i^{-L} e^{-i\theta_{F\underline{s}_{\tau+1}}-i\theta_{\underline{s}_\tau}} V_{F\underline{s}_{\tau+1},\underline{s}_\tau},
		\label{U_2kick_elem_3}
	\end{align}
	Substituting Eq.~(\ref{U_2kick_elem_3}) in (\ref{Ut_2kick_elem_1}), we obtain
	\begin{align}
		\langle\underline{s}_{t}|\hat{U}^t|\underline{s}_0\rangle&=i^{-Lt}\sum_{\underline{s}_1,...,\underline{s}_{t-1}}\prod_{\tau=0}^{t-1} e^{-i(\theta_{F\underline{s}_{\tau+1}}+\theta_{\underline{s}_\tau})}V_{F\underline{s}_{\tau+1},\underline{s}_\tau}\notag\\
		&=i^{-Lt}\sum_{\underline{s}_1,...,\underline{s}_{t-1}}e^{-i\sum_{\tau=0}^{t-1}(\theta_{F\underline{s}_{\tau+1}}+\theta_{\underline{s}_\tau})}\prod_{\tau=0}^{t-1} V_{F\underline{s}_{\tau+1},\underline{s}_\tau}.
		\label{Ut_2kick_elem_2}
	\end{align}
	Similarly,
	\begin{align}
		\langle\underline{s}'_{0}|\hat{U}^{-t}|\underline{s}'_{t}\rangle&=i^{Lt}\sum_{\underline{s}'_1,...,\underline{s}'_{t-1}}e^{i\sum_{\tau=0}^{t-1}(\theta_{F\underline{s}'_{\tau+1}}+\theta_{\underline{s}'_\tau})}\prod_{\tau=0}^{t-1} V^*_{F\underline{s}'_{\tau+1},\underline{s}'_\tau}.
		\label{U-t_2kick_elem_1}
	\end{align}
	Substituting Eqs.~(\ref{Ut_2kick_elem_2}) and (\ref{U-t_2kick_elem_1}) in Eq.~(\ref{At_2kick}), we obtain
	\begin{align}
		\langle\psi|\hat{A}(t)|\psi\rangle&=\sum_{\{\underline{s}_\tau\}}\sum_{\{\underline{s}'_\tau\}}\langle\psi|\underline{s}'_0\rangle \langle\underline{s}_0|\psi\rangle\langle\underline{s}'_{t}|\hat{A}|\underline{s}_{t}\rangle e^{i\sum_{\tau=0}^{t-1}(\theta_{F\underline{s}_{\tau+1}}+\theta_{\underline{s}_\tau}-\theta_{F\underline{s}'_{\tau+1}}-\theta_{\underline{s}'_\tau})}\prod_{\tau=0}^{t-1} V_{F\underline{s}_{\tau+1},\underline{s}_\tau}V^*_{F\underline{s}'_{\tau+1},\underline{s}'_\tau}.
	\end{align}
	Now, we perform disorder average using the RPA as follows
	\begin{align}
		\langle\psi|\hat{A}(t)|\psi\rangle_{\text{diss}}&=\langle\psi|\hat{A}(t)|\psi\rangle_{\text{RPA}}\notag\\
		&=\sum_{\{\underline{s}_\tau\}}\sum_{\{\underline{s}'_\tau\}}\langle\psi|\underline{s}'_0\rangle \langle\underline{s}_0|\psi\rangle\langle\underline{s}'_{t}|\hat{A}|\underline{s}_{t}\rangle\langle e^{i\sum_{\tau=0}^{t-1}(\theta_{F\underline{s}_{\tau+1}}+\theta_{\underline{s}_\tau}-\theta_{F\underline{s}'_{\tau+1}}-\theta_{\underline{s}'_\tau})}\rangle_{\text{RPA}}\prod_{\tau=0}^{t-1} V_{F\underline{s}_{\tau+1},\underline{s}_\tau}V^*_{F\underline{s}'_{\tau+1},\underline{s}'_\tau},
	\end{align}
	where
	\begin{align}
		\langle e^{i\sum_{\tau=0}^{t-1}(\theta_{F\underline{s}_{\tau+1}}+\theta_{\underline{s}_\tau}-\theta_{F\underline{s}'_{\tau+1}}-\theta_{\underline{s}'_\tau})}\rangle_{\text{RPA}}&=1
		\label{RPA_av_2kick}
	\end{align}
	if $\underline{s}'_\tau=\underline{s}_{\pi(\tau)}$ or $F\underline{s}_{\pi(\tau)}$, $\forall \tau\in\{1,...,t-1\}$ and $\underline{s}'_0=\underline{s}_0,F\underline{s}'_{t}=F\underline{s}_{t}$ or $\underline{s}'_0=F\underline{s}_{t},F\underline{s}'_{t}=\underline{s}_{0}$. For other set of states $\{\underline{s}'_0,F\underline{s}'_1,...,\underline{s}'_{t-1},F\underline{s}'_{t}\}$ the left hand side of Eq.~(\ref{RPA_av_2kick}) evaluates to zero. Here the symbol $\pi$ represents permutation. For simplicity, we consider the initial state $|\psi\rangle$ to be one of the basis states. Therefore,
	\begin{align}
		\langle\psi|\underline{s}'_0\rangle &=\delta_{\psi,\underline{s}'_0},\\
		\langle\underline{s}_0|\psi\rangle &=\delta_{\underline{s}_0,\psi}.
	\end{align}
	This eliminates the possibility $\underline{s}'_0=F\underline{s}_{t},F\underline{s}'_{t}=\underline{s}_{0}$ and we get
	\begin{align}
		\langle\psi|\hat{A}(t)|\psi\rangle_{\text{RPA}}&=\sum_{\{\underline{s}_\tau\}_{\tau=1}^{t}}\sum_{\pi}\sum_{\vec{\mu}}\langle\underline{s}_{t}|\hat{A}|\underline{s}_{t}\rangle V_{F\underline{s}_{t},\underline{s}_{t-1}}V^*_{F\underline{s}_{t},\underline{s}^{(\mu_{t-1})}_{\pi(t-1)}}\left(\prod_{\tau=1}^{t-2} V_{F\underline{s}_{\tau+1},\underline{s}_\tau}V^*_{F\underline{s}^{(\mu_{\tau+1})}_{\pi(\tau+1)},\underline{s}^{(\mu_\tau)}_{\pi(\tau)}}\right)V_{F\underline{s}_{1},\psi} V^*_{F\underline{s}^{(\mu_{1})}_{\pi(1)},\psi},
	\end{align}
	where $\vec{\mu}=(\mu_1,...,\mu_{t-1})$ and $\mu_\tau=0,1$, $\forall\tau\in\{1,...,t-1\}$. The superscript $\mu_\tau$ can be understood as follows
	\begin{align}
		\underline{s}^{(0)}_{\pi(\tau)}&=\underline{s}_{\pi(\tau)},\\
		\underline{s}^{(1)}_{\pi(\tau)}&=F\underline{s}_{\pi(\tau)}.
	\end{align}
	Consider the identity permutation with $\vec{\mu}=(0,...,0)$. The corresponding contribution is
	\begin{align}
		\langle\psi|\hat{A}(t)|\psi\rangle_{\text{RPA},I}^{(0,...,0)}&=\sum_{\{\underline{s}_\tau\}_{\tau=1}^{t}}\langle\underline{s}_{t}|\hat{A}|\underline{s}_{t}\rangle\left(\prod_{\tau=1}^{t-1}(\mathcal{M}_1)_{F\underline{s}_{\tau+1},\underline{s}_\tau}\right)(\mathcal{M}_1)_{F\underline{s}_{1},\psi}\notag\\
		&=\sum_{\{\underline{s}_\tau\}_{\tau=1}^{t}}\langle\underline{s}_{t}|\hat{A}|\underline{s}_{t}\rangle\left(\prod_{\tau=1}^{t-1}(F\mathcal{M}_1)_{\underline{s}_{\tau+1},\underline{s}_\tau}\right)(F\mathcal{M}_1)_{\underline{s}_{1},\psi}\notag\\
		&=\sum_{\underline{s}_{t}}\langle\underline{s}_{t}|\hat{A}|\underline{s}_{t}\rangle \left(\mathcal{M}^t\right)_{\underline{s}_{t},\psi},
		\label{At_2kick_I_sigma_0}
	\end{align}
	where $(\mathcal{M}_1)_{F\underline{s}_{\tau+1},\underline{s}_\tau}=|V_{F\underline{s}_{\tau+1},\underline{s}_\tau}|^2$. Since $V$ is a unitary matrix, $\mathcal{M}_1$ is a doubly stochastic matrix. The matrix elements of $\hat{F}$ in the computational basis are
	\begin{align}
		\langle\underline{s}|\hat{F}|\underline{s}'\rangle&=F_{\underline{s},\underline{s}'}=\delta_{\underline{s}',F\underline{s}}.
	\end{align}
	Therefore, the matrix $F$ has only one nonzero element in each row and column. Additionally, this nonzero element is one. Therefore, $F$ is also a doubly stochastic matrix. Consequently, $\mathcal{M}=F\mathcal{M}_1$ is also a doubly stochastic matrix.
	Since $\hat{V}$ commutes with $\hat{N}$, the matrix $V $ takes a block diagonal form on the subspace $\mathcal{H}_N\oplus\mathcal{H}_{L-N}$ as follows
	\begin{align}
		V&=\begin{pmatrix}
			v_N& 0\\
			0& v_{L-N}
		\end{pmatrix},
	\end{align}
	where $v_N$ is the block of the matrix $V$ acting on $\mathcal{H}_N$. Since $\mathcal{M}_1$ is obtained by taking squared modulus of each element of $V$, $\mathcal{M}_1$ can be expressed as
	\begin{align}
		\mathcal{M}_1&=\begin{pmatrix}
			m_N& 0\\
			0& m_{L-N}
		\end{pmatrix}.
	\end{align}
	\begin{figure*}[t!]
		\centering
		\begin{subfigure}[t]{0.45\linewidth}
			\centering
			\includegraphics[width=\linewidth]{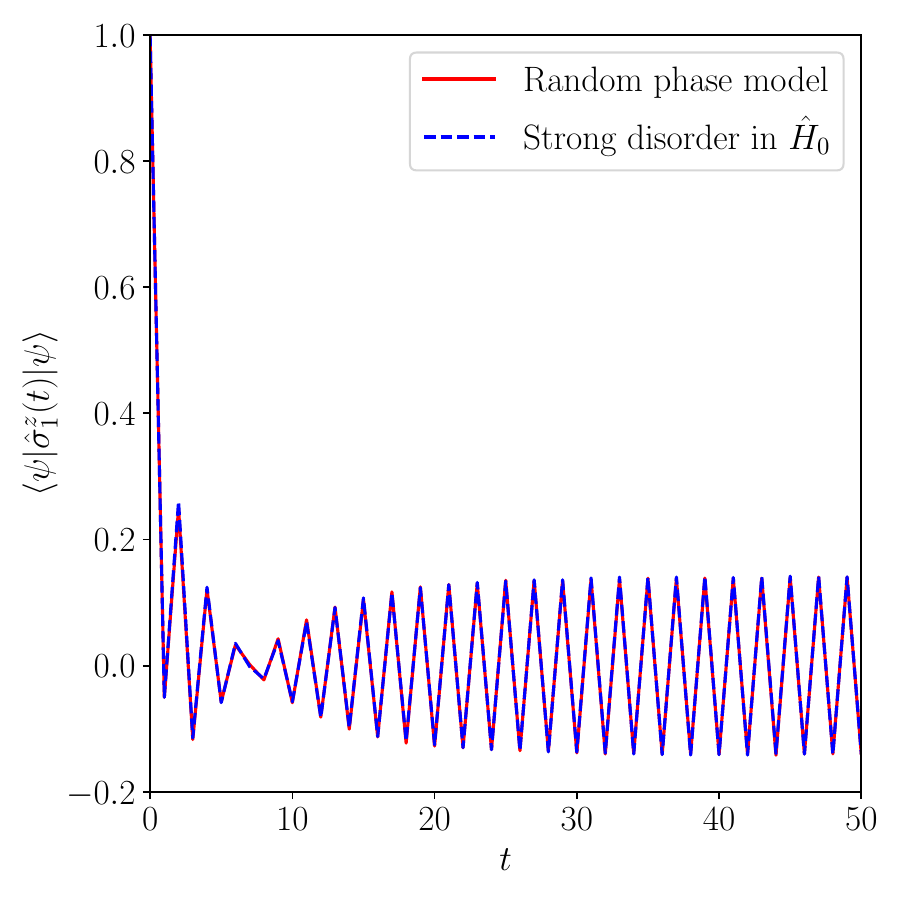}
			\caption{}
		\end{subfigure}
		\hfill
		\begin{subfigure}[t]{0.45\linewidth}
			\centering
			\includegraphics[width=\linewidth]{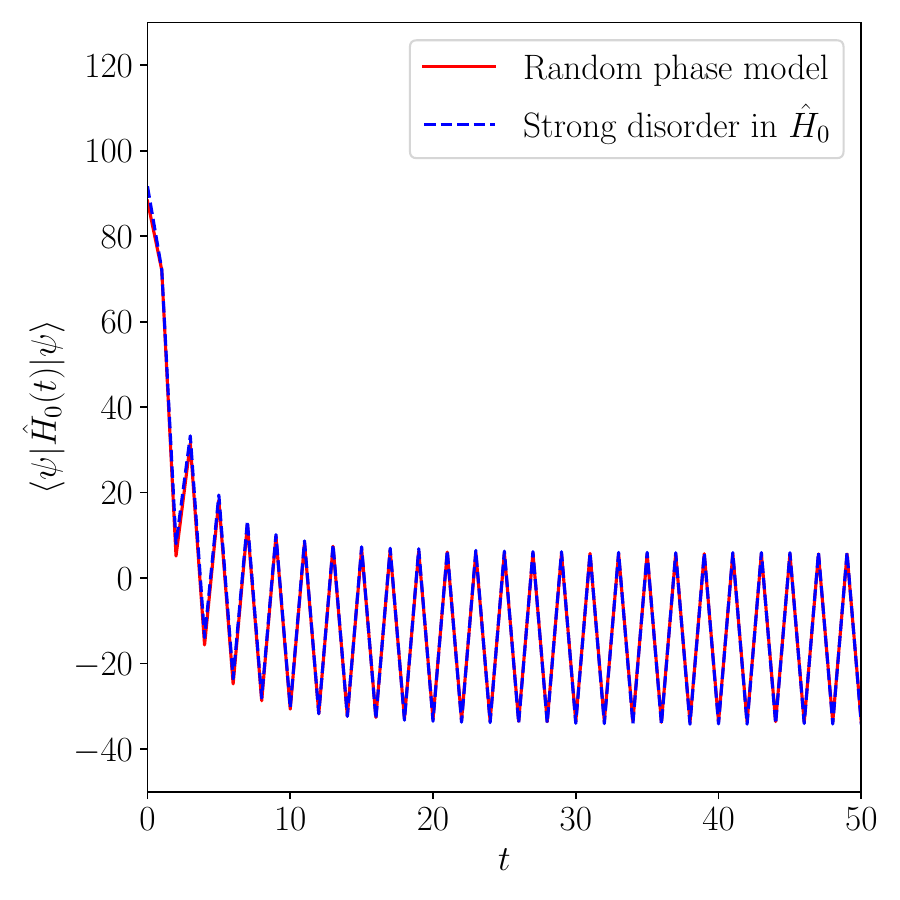}
			\caption{}
		\end{subfigure}
		\caption{\justifying{\small Time evolution of (a) local magnetization and (b) total energy from direct numerics for the Hamiltonian described by Eqs.~(\ref{H_2kick}), (\ref{H0_2kick}-\ref{H2_2kick}) and corresponding random phase model. Here $L=14,J=1,h=\pi/2,\epsilon=10,\Delta\epsilon=10,U_0=10,\Delta U_0=10,\alpha=1.5,|\psi\rangle=|1,1,1,1,1,1,-1,-1,-1,-1,-1,-1,-1,-1\rangle$. Averaging over 320 realizations of disorder is performed for direct numerics in each case. Simulation was done until $t=6000$.}}
		\label{RPM_vs_strong_dis_2kick}
	\end{figure*}
	\begin{figure*}[t!]
		\centering
		\begin{subfigure}[t]{0.45\linewidth}
			\centering
			\includegraphics[width=\linewidth]{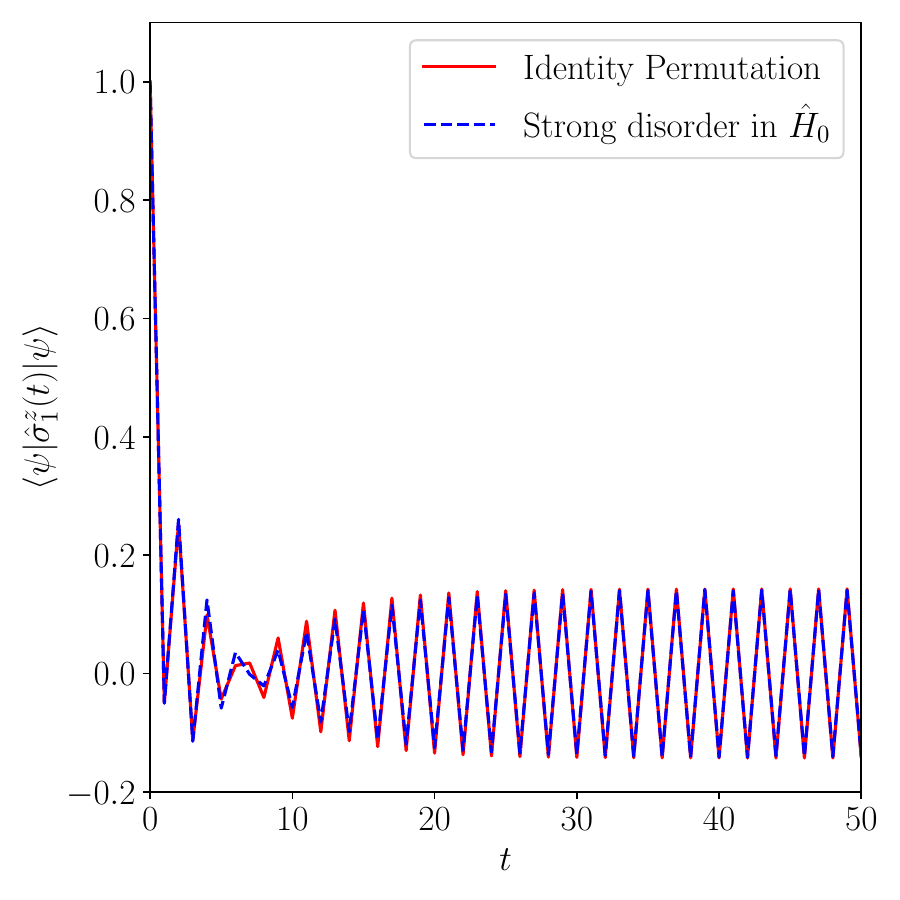}
			\caption{}
		\end{subfigure}
		\hfill
		\begin{subfigure}[t]{0.45\linewidth}
			\centering
			\includegraphics[width=\linewidth]{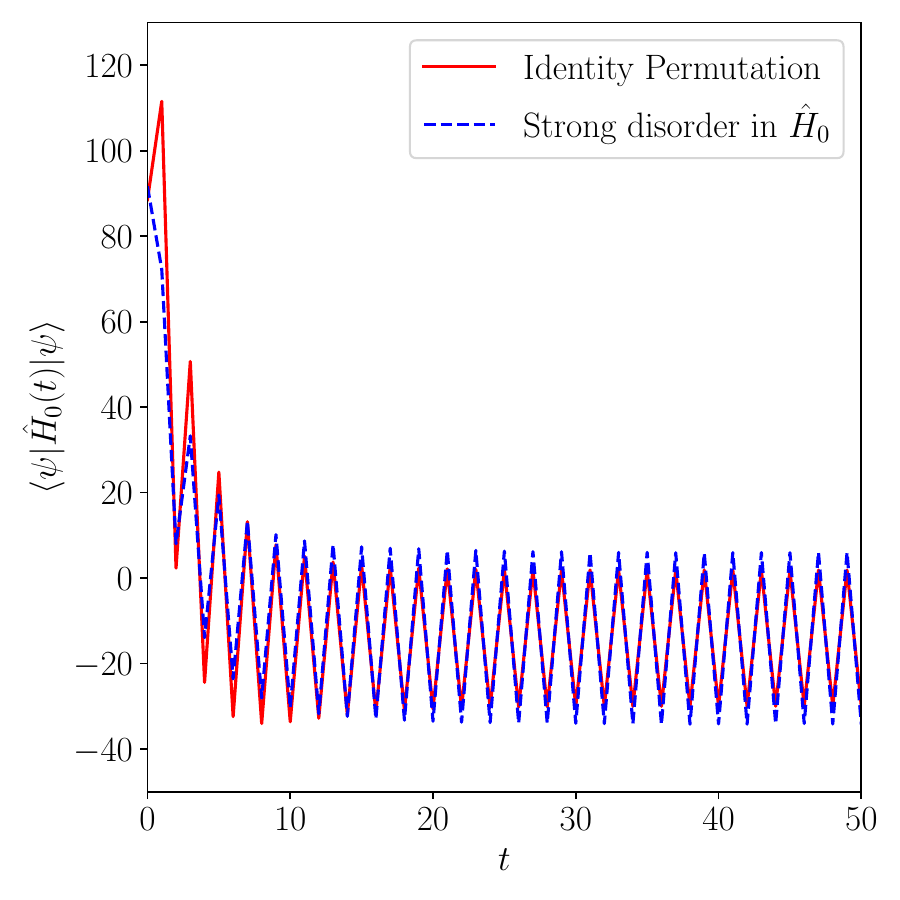}
			\caption{}
		\end{subfigure}
		\caption{\justifying{\small Time evolution of (a) local magnetization and (b) total energy from direct numerics and our analytics using the RPA for the Hamiltonian described by Eqs.~(\ref{H_2kick}) and (\ref{H0_2kick}-\ref{H2_2kick}). Here $L=14,J=1,h=\pi/2,\epsilon=10,\Delta\epsilon=10,U_0=10,\Delta U_0=10,\alpha=1.5,|\psi\rangle=|1,1,1,1,1,1,-1,-1,-1,-1,-1,-1,-1,-1\rangle$. Averaging over 320 realizations of disorder is performed for direct numerics in each case. Simulation was done until $t=6000$.}}
		\label{RPM_vs_strong_dis}
	\end{figure*}
	The operator $\hat{F}$ connects states from subspace $\mathcal{H}_N$ to states in subspace $\mathcal{H}_{L-N}$ and vise versa. Therefore, the matrix $F$ can be expressed as follows
	\begin{align}
		F&=\begin{pmatrix}
			0& I_{\mathcal{N}/2}\\
			I_{\mathcal{N}/2}& 0
		\end{pmatrix},
	\end{align}
	where $\mathcal{N}=\binom{L}{N}+\binom{L}{L-N}$. Therefore,
	\begin{align}
		\mathcal{M}=F\mathcal{M}_1=\begin{pmatrix}
			0& m_{L-N}\\
			m_N& 0
		\end{pmatrix}.
	\end{align}
	Since $\mathcal{M}$ is block-off-diagonal, it anticommutes with 
	\begin{align}
		\tau^z&=\begin{pmatrix}
			I_{\mathcal{N}/2}& 0\\
			0& -I_{\mathcal{N}/2}
		\end{pmatrix}.
	\end{align}
	Therefore, eigenvalues of $\mathcal{M}$ are symmetrically located about the origin in the complex plane.
	Thus, we have eigenvalues $1,-1,\lambda_1,-\lambda_1,...,\lambda_{\mathcal{N}/2-1},-\lambda_{\mathcal{N}/2-1}$. Therefore,
	\begin{align}
		\mathcal{M}^{t}&=\sum_{i=0}^{\mathcal{N}/2-1}\lambda_i^{t-1}|\lambda_i\rangle\langle\lambda_i|+(-\lambda_i)^{t-1}|-\lambda_i\rangle\langle-\lambda_i|,
	\end{align}
	where $\lambda_0=1$, $\langle \lambda_0|\equiv \sqrt{\frac{1}{\mathcal{N}}}(1,...,1,1,...,1)$, and $\langle -\lambda_0|\equiv \sqrt{\frac{1}{\mathcal{N}}}(1,...,1,-1,...,-1)$. Since $|\lambda_i|<1$, $\forall i\neq 0$, at long times we obtain
	\begin{align}
		\mathcal{M}^{t}&\simeq|\lambda_0\rangle\langle\lambda_0|+(-1)^{t}|-\lambda_0\rangle\langle-\lambda_0|\notag\\
		&=\begin{cases}
			\begin{pmatrix}
				\mathcal{M}_0& 0\\
				0& \mathcal{M}_0
			\end{pmatrix},\; \text{$t$ even},\\
			\begin{pmatrix}
				0& \mathcal{M}_0\\
				\mathcal{M}_0& 0
			\end{pmatrix},\; \text{$t$ odd}
		\end{cases},
		\label{Mt_2kick_long_time}
	\end{align}
	where $\mathcal{M}_0$ is a $\mathcal{N}/2\times \mathcal{N}/2$ matrix whose all elements are $2/\mathcal{N}$. Therefore, substituting Eq.~(\ref{Mt_2kick_long_time}) in Eq.~(\ref{At_2kick_I_sigma_0}) we obtain at long times
	\begin{align}
		\langle\psi|\hat{A}(t)|\psi\rangle_{\text{RPA},I}^{(0,...,0)}&=\frac{2}{\mathcal{N}}\begin{cases}
			\tr_{\mathcal{H}_N}\hat{A},\;\text{$t$ even},\\
			\tr_{\mathcal{H}_{L-N}}\hat{A},\; \text{$t$ odd}
		\end{cases},
	\end{align}
	where $\tr_{\mathcal{H}_N}\hat{A}$ and $\tr_{\mathcal{H}_{L-N}}\hat{A}$ are the traces of $\hat{A}$ over the subspaces $\mathcal{H}_N$ and $\mathcal{H}_{L-N}$, respectively.
	\begin{figure*}[t!]
		\centering
		\begin{subfigure}[t]{0.45\linewidth}
			\centering
			\includegraphics[width=\linewidth]{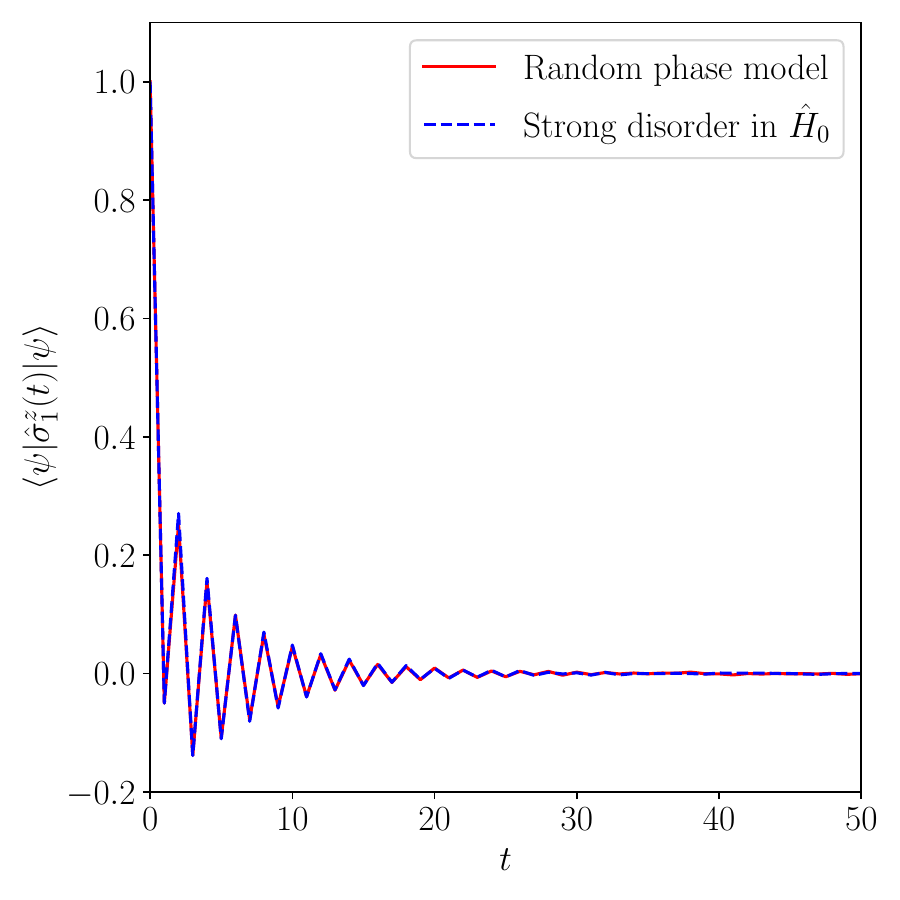}
			\caption{}
		\end{subfigure}
		\hfill
		\begin{subfigure}[t]{0.45\linewidth}
			\centering
			\includegraphics[width=\linewidth]{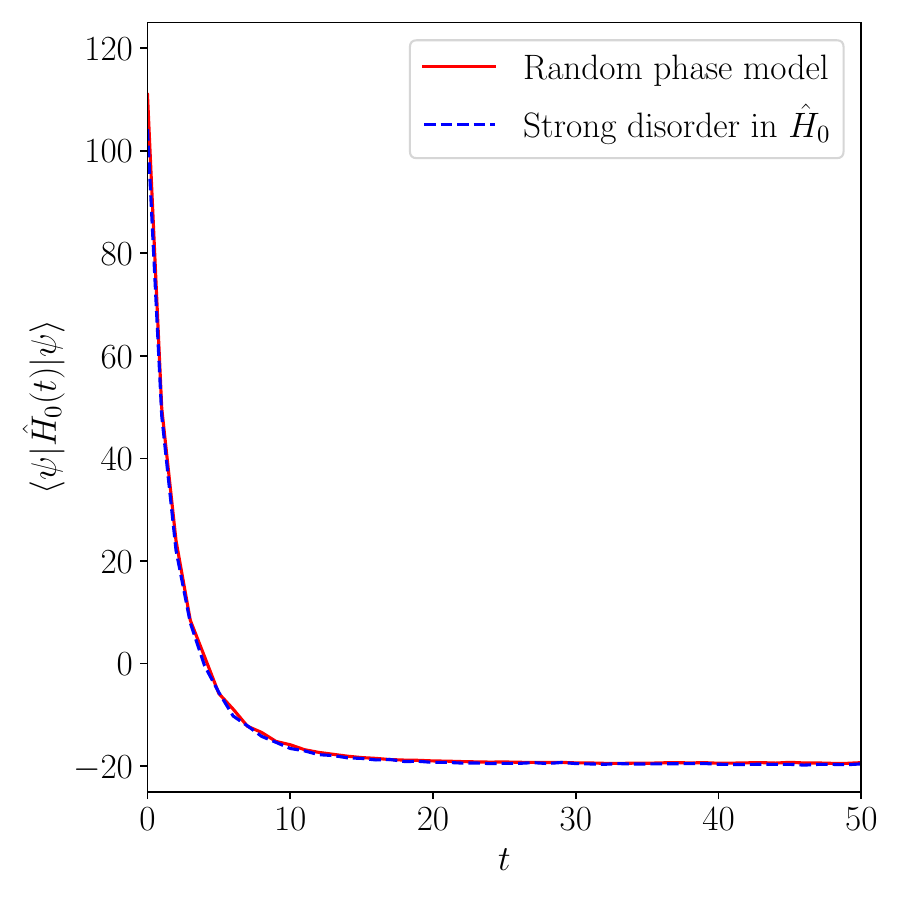}
			\caption{}
		\end{subfigure}
		\caption{\justifying{\small We show numerically calculated (a) local magnetization and (b) total energy as function of time for the Hamiltonian described by Eqs.~(\ref{H_2kick}), (\ref{H0_2kick}-\ref{H2_2kick}) in comparison with that calculated for the corresponding random phase model. Here $L=14,J=1,h=\pi/2,\epsilon=10,\Delta\epsilon=10,U_0=10,\Delta U_0=10,\alpha=1.5,|\psi\rangle=|1,1,1,1,1,1,1,-1,-1,-1,-1,-1,-1,-1\rangle$. Averaging over 320 realizations of disorder is performed for direct numerical simulation in each case. Simulation was done until $t=3000$.}}
		\label{RPM_vs_strong_dis_2kick_half_fill}
	\end{figure*}
	\begin{figure*}[t!]
		\centering
		\begin{subfigure}[t]{0.45\linewidth}
			\centering
			\includegraphics[width=\linewidth]{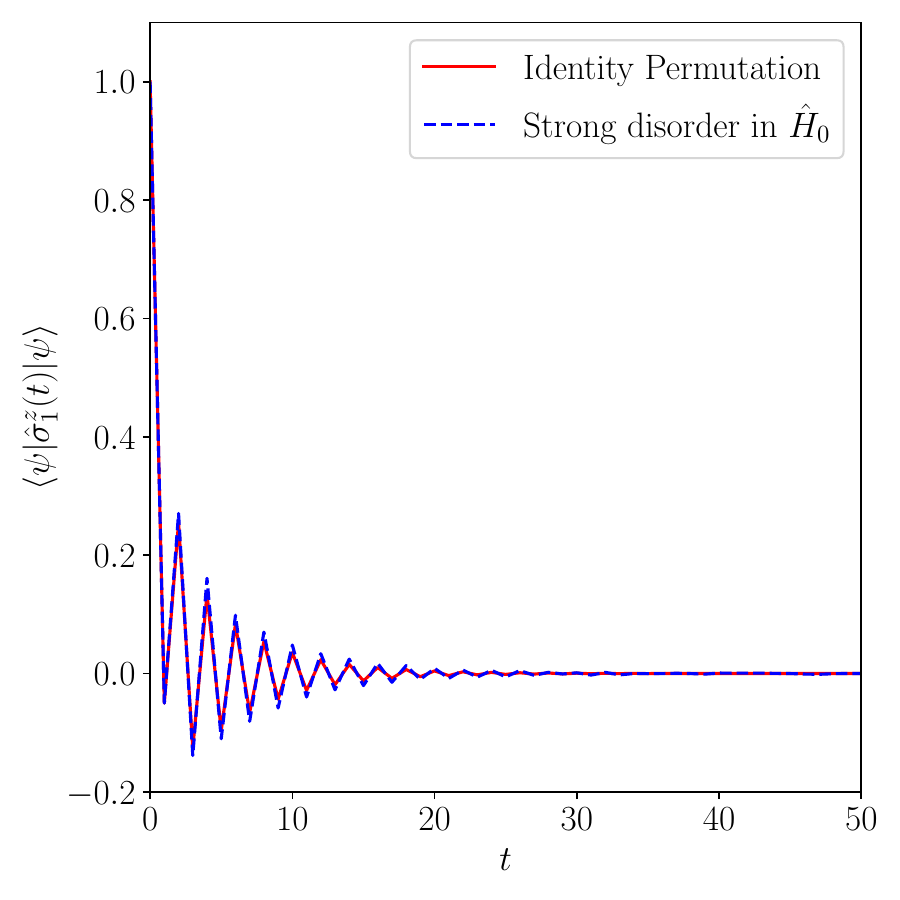}
			\caption{}
		\end{subfigure}
		\hfill
		\begin{subfigure}[t]{0.45\linewidth}
			\centering
			\includegraphics[width=\linewidth]{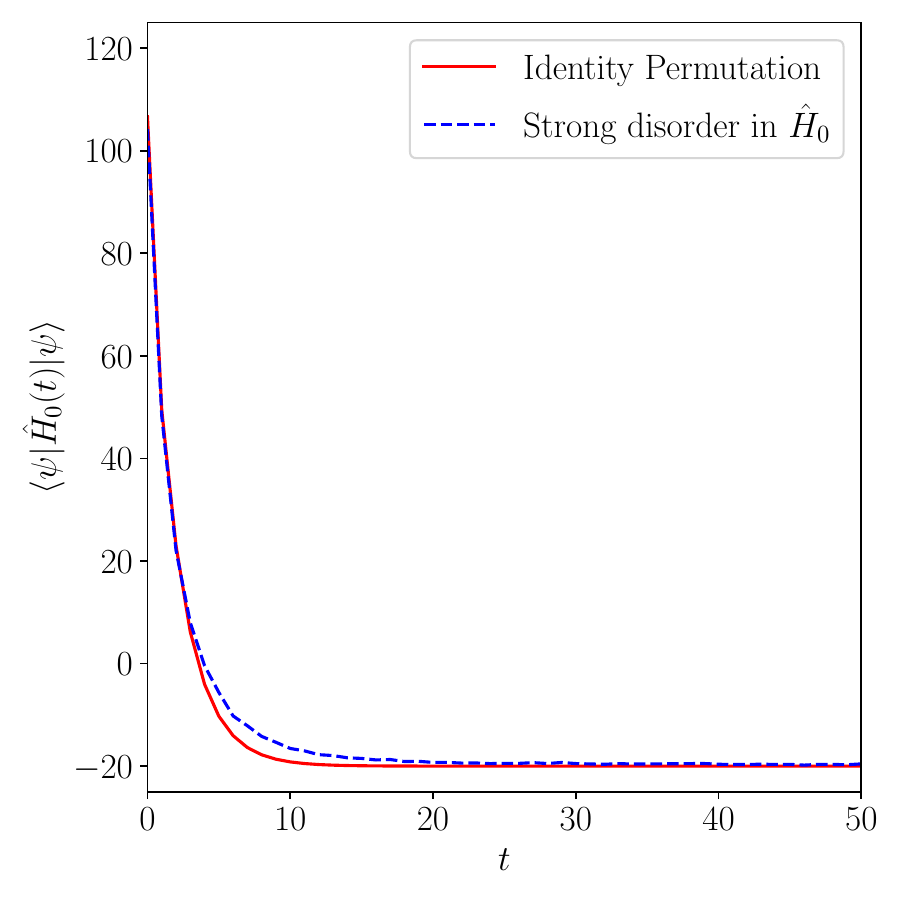}
			\caption{}
		\end{subfigure}
		\caption{\justifying{\small We show numerically calculated (a) local magnetization and (b) total energy as function of time for the Hamiltonian described by Eqs.~(\ref{H_2kick}), (\ref{H0_2kick}-\ref{H2_2kick}) in comparison with the analytical result given in Eq.~(\ref{At_2kick_I_sigma_0}). We take $L=14,J=1,g=\pi,\epsilon=10,\Delta\epsilon=10,U_0=10,\Delta U_0=10,\alpha=1.5,|\psi\rangle=|1,1,1,1,1,1,1,-1,-1,-1,-1,-1,-1,-1\rangle$. Averaging over 320 realizations of disorder is performed for direct numerical simulation in each case. Simulation was done until $t=3000$.}}
		\label{RPA_I_vs_strong_dis_2kick_half_fill}
	\end{figure*}
	\subsection{Spectral form factor in the discrete time crystal phase of the spin chain with two kicks per cycle}
	Now, we compute the spectral form factor (SFF) for the Hamiltonian described by Eqs.~(\ref{H_2kick}) and (\ref{H0_2kick}-\ref{H2_2kick}) in the discrete time crystal phase $h=\pi/2$. The SFF can be expressed in terms of the Floquet operator as
	\begin{align}
		K(t)&=\langle\tr \hat{U}^t\tr \hat{U}^{-t}\rangle.
		\label{SFF}
	\end{align}
	As discussed in the previous section, at this point in the parameter space, the Floquet operator $\hat{U}$ takes a block diagonal form in the computational basis. Therefore, we must compute SFF for individual blocks of $\hat{U}$.\\
	\textbf{Case 1}: Subspace $\mathcal{H}_N\oplus\mathcal{H}_{L-N}$.\\
	First, we expand $\tr \hat{U}^t$ in the computational basis in terms of matrix elements of $\hat{U}$ as follows
	\begin{align}
		\tr\hat{U}^t&=\sum_{\underline{s}_1}\langle\underline{s}_1|\hat{U}^t|\underline{s}_1\rangle.
		\label{tr_Ut_1}
	\end{align}
	Following Eq.~(\ref{Ut_2kick_elem_2}), we express Eq.~(\ref{tr_Ut_1}) as follows
	\begin{align}
		\tr \hat{U}^t&=i^{-Lt}\sum_{\underline{s}_1,...,\underline{s}_t}e^{-i\sum_{\tau=1}^t(\theta_{F\underline{s}_{\tau+1}}+\theta_{\underline{s}_\tau})}\prod_{\tau=1}^t V_{F\underline{s}_{\tau+1},\underline{s}_\tau},
	\end{align}
	\begin{figure*}[t!]
		\centering
		\begin{subfigure}[t]{0.45\linewidth}
			\centering
			\includegraphics[width=\linewidth]{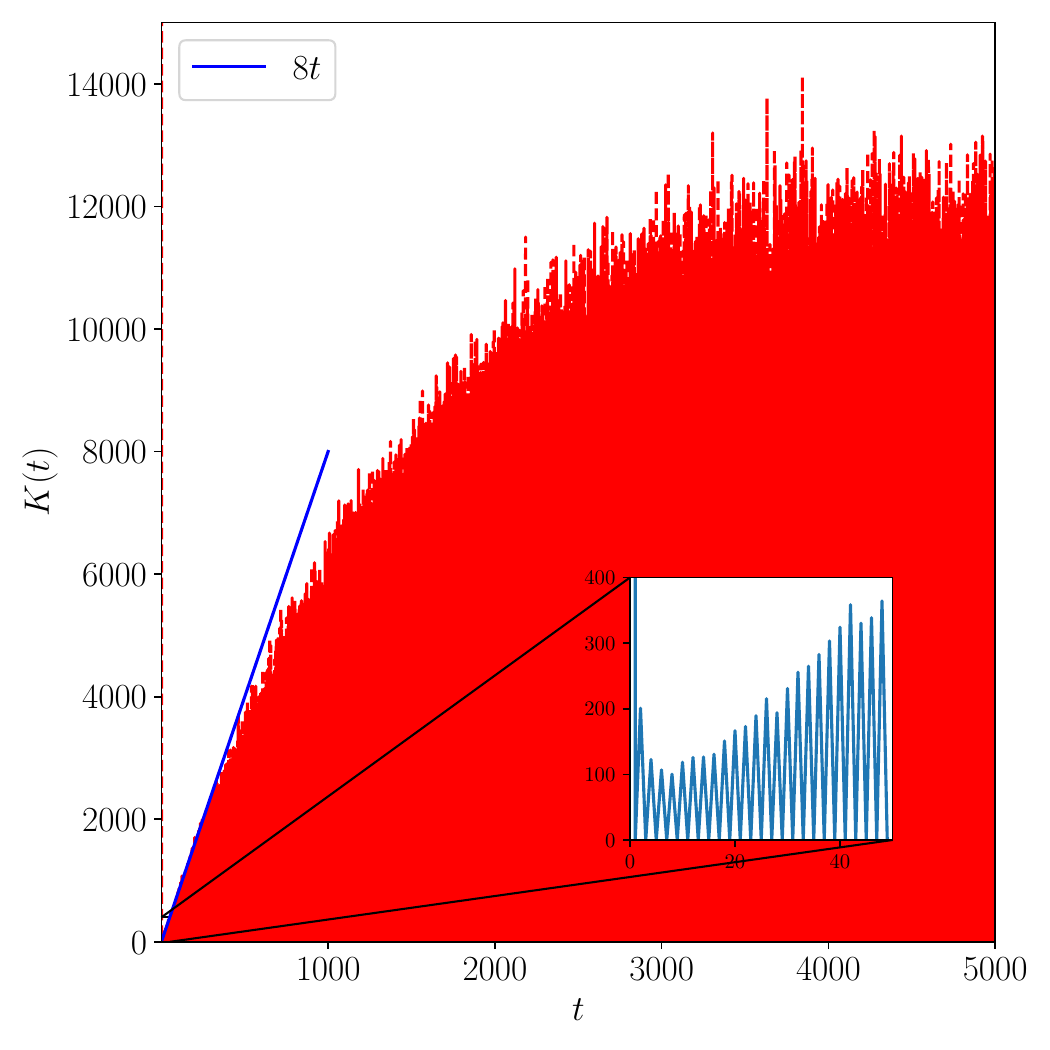}
			\caption{}
		\end{subfigure}
		\hfill
		\begin{subfigure}[t]{0.45\linewidth}
			\centering
			\includegraphics[width=\linewidth]{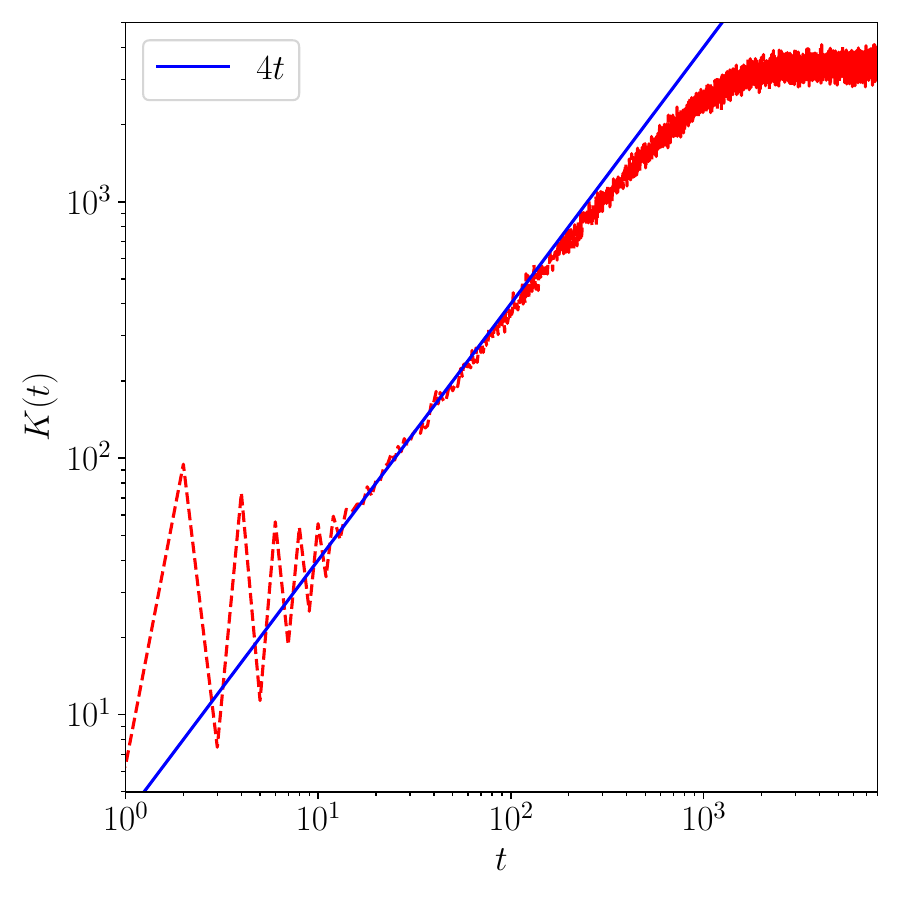}
			\caption{}
		\end{subfigure}
		\caption{\justifying{\small We show numerically calculated spectral form factor with (a) $N\neq L/2$ and (b) $N=L/2$ for the Hamiltonian described Eqs.~(\ref{H_2kick}) and (\ref{H0_2kick}-\ref{H2_2kick}). We find that the SFF in (a) and (b) behaves in accordance with Eq.~(\ref{K0t_N_not_L_by_2}) and Eq.~(\ref{K0t_N_equal_to_L_by_2}), respectively. We take $L=14,J=1,h=\pi,\epsilon=10,\Delta\epsilon=10,U_0=10,\Delta U_0=10,\alpha=1.5$. SFF in (a) and (b) is evaluated over the Hilbert space $\mathcal{H}_N\oplus\mathcal{H}_{L-N}$ and $\mathcal{H}_{L/2}$, respectively. Averaging over 320 realizations of disorder is taken in each case.}}
		\label{Direct_SFF_2kick}
	\end{figure*}
	where periodicity in time is assumed, $t+1\equiv 1$. Similarly,
	\begin{align}
		\tr \hat{U}^{-t}&=i^{Lt}\sum_{\underline{s}'_1,...,\underline{s}'_t}e^{i\sum_{\tau=1}^t(\theta_{F\underline{s}'_{\tau+1}}+\theta_{\underline{s}'_\tau})}\prod_{\tau=1}^t V_{F\underline{s}'_{\tau+1},\underline{s}'_\tau}
	\end{align}
	Therefore,
	\begin{align}
		K(t)&=\sum_{\{\underline{s}_\tau\}}\sum_{\{\underline{s}'_\tau\}}\langle e^{-i\sum_{\tau=1}^t(\theta_{\underline{s}_\tau}+\theta_{F\underline{s}_{\tau}}-\theta_{\underline{s}'_\tau}-\theta_{F\underline{s}'_\tau})}\rangle\prod_{\tau=1}^t V_{F\underline{s}_{\tau+1},\underline{s}_\tau}V^*_{F\underline{s}'_{\tau+1},\underline{s}'_\tau}.
	\end{align}
	Doing ensemble average using RPA, we obtain
	\begin{align}
		\underline{s}'_{\tau}=\underline{s}^{(\mu_\tau)}_{\pi(\tau)},\;\mu_\tau=0,1.
	\end{align}
	Therefore,
	\begin{align}
		K(t)&=\sum_{\{\underline{s}_\tau\}}\sum_\pi\sum_{\vec{\mu}}\prod_{\tau=1}^t V_{F\underline{s}_{\tau+1},\underline{s}_\tau}V^*_{F\underline{s}^{(\mu_{\tau+1})}_{\pi(\tau+1)},\underline{s}^{(\mu_\tau)}_{\pi(\tau)}}.
	\end{align}
	\begin{figure*}[t!]
		\centering
		\begin{subfigure}[t]{0.3\linewidth}
			\centering
			\includegraphics[width=\linewidth]{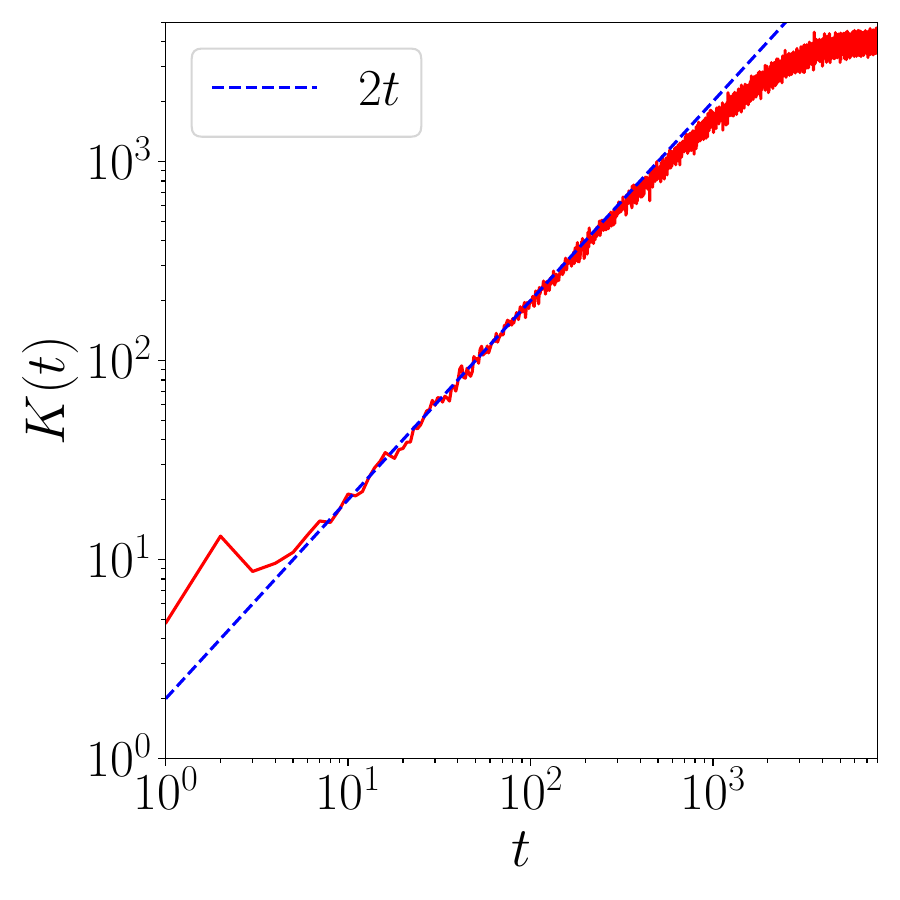}
			\caption{}
		\end{subfigure}
		\hfill
		\begin{subfigure}[t]{0.3\linewidth}
			\centering
			\includegraphics[width=\linewidth]{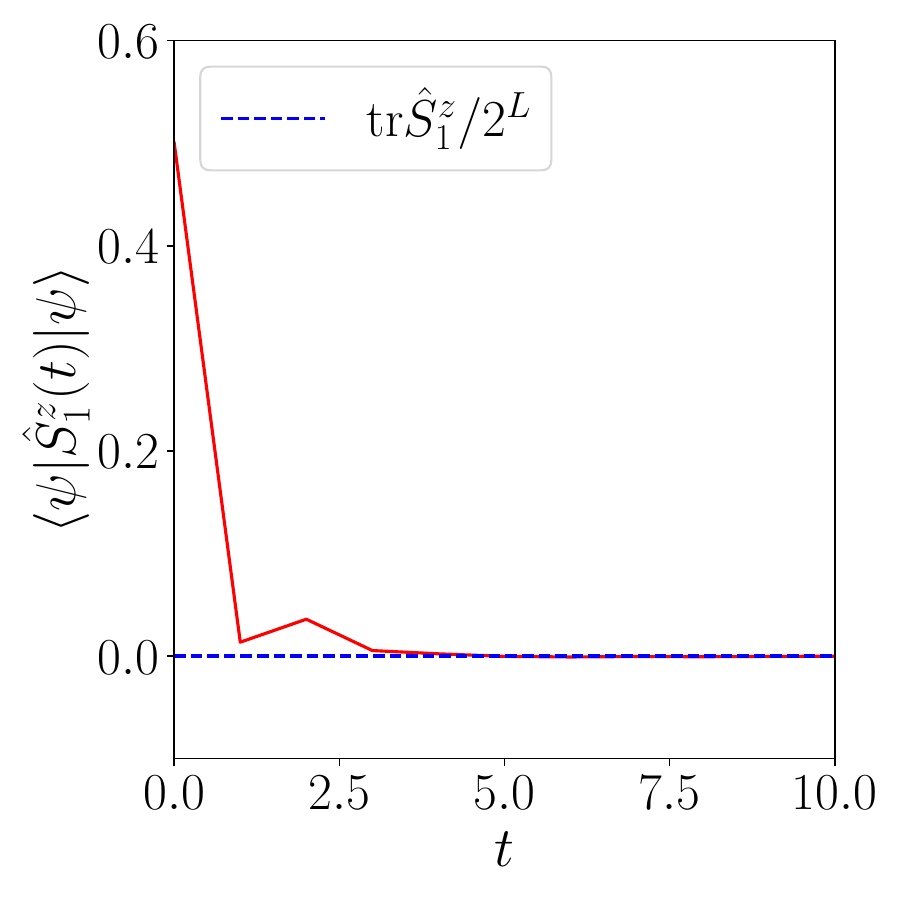}
			\caption{}
		\end{subfigure}
		\hfill
		\begin{subfigure}[t]{0.3\linewidth}
			\centering
			\includegraphics[width=\linewidth]{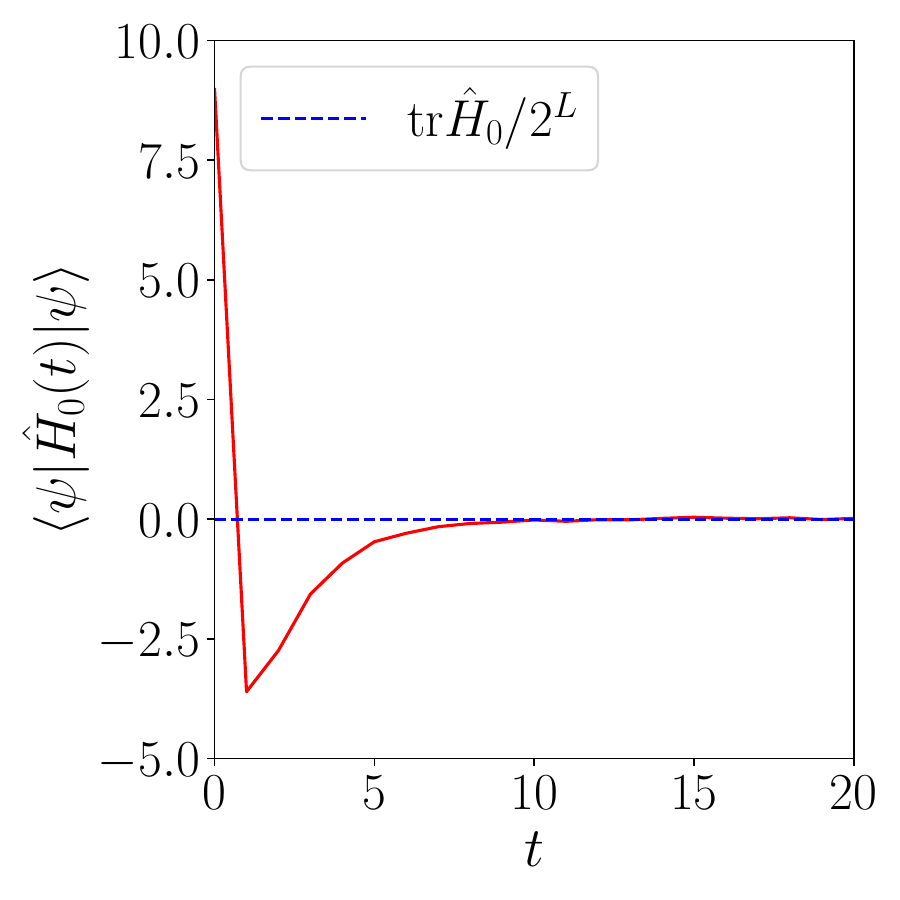}
			\caption{}
		\end{subfigure}
		\caption{\justifying{\small We show (a) spectral form factor, (b) local magnetization, and (c) total energy for the Hamiltonian described by Eqs.~(\ref{H_2kick}), (\ref{H0_2kick}), (\ref{H1_2kick}), and (\ref{H2_2kick}) for $h=0.5$ in the ergodic phase. Here $L=12, J=1, \epsilon=10, \Delta\epsilon=10, U_0=10, \Delta U_0=10, \alpha=1.5$, $|\psi\rangle=|1,1,1,1,1,-1,-1,-1,-1,-1,-1,-1\rangle$. Averaging over 320 realizations of disorder is performed in each case. Simulation was done until $t=8000$.}}
	\end{figure*}
	Due to periodicity in time, all cyclic variants of a permutation have identical contributions. Furthermore, due to the symmetric nature of the matrix $V$, all anticyclic variants of a permutation also have identical contributions that match those of cyclic variants. Therefore, considering identity permutation and its $t$ cyclic and $t$ anticyclic variants, we write leading-order SFF
	\begin{align}
		K^{(0)}(t)&=2t\sum_{\{\underline{s}_\tau\}}\sum_{\vec{\mu}}\prod_{\tau=1}^t V_{F\underline{s}_{\tau+1},\underline{s}_\tau}V^*_{F\underline{s}^{(\mu_{\tau+1})}_{\tau+1},\underline{s}^{(\mu_\tau)}_{\tau}}.
	\end{align}
	First, we consider $\vec{\mu}=(0,...,0)$
	\begin{samepage}
		\begin{align}
			\sum_{\{\underline{s}_\tau\}}\prod_{\tau=1}^t V_{F\underline{s}_{\tau+1},\underline{s}_\tau}V^*_{F\underline{s}_{\tau+1},\underline{s}_{\tau}}&=\sum_{\{\underline{s}_\tau\}}\prod_{\tau=1}^t(\mathcal{M}_1)_{F\underline{s}_{\tau+1},\underline{s}_\tau}\notag\\
			&=\sum_{\{\underline{s}_\tau\}}\prod_{\tau=1}^t(F\mathcal{M}_1)_{\underline{s}_{\tau+1},\underline{s}_\tau}\notag\\
			&=\tr (F\mathcal{M}_1)^t\notag\\
			&=\tr \mathcal{M}^t\notag\\
			&=\left(1+(-1)^t\right)+\left(\sum_{i=1}^{\mathcal{N}/2-1}\lambda_i^t+(-\lambda_i)^t\right).
		\end{align}
	\end{samepage}
	Before we study $\vec{\mu}=(1,...,1)$ case, we state that $\hat{V}$ commutes with $\hat{F}$. This statement will be proved later. However, we first study its consequences. Since
	\begin{align}
		V_{F\underline{s},F\underline{s}'}&=\langle F\underline{s}|\hat{V}|F\underline{s}'\rangle\notag\\
		&=\langle\underline{s}|\hat{F}\hat{V}\hat{F}|\underline{s}'\rangle\notag\\
		&=\langle\underline{s}|\hat{V}\hat{F}^2|\underline{s}'\rangle\notag\\
		&=\langle\underline{s}|\hat{V}|\underline{s}'\rangle,
	\end{align}
	where in the last line we used the property $\hat{F}^2=\hat{I}$. Therefore, $V_{F\underline{s},F\underline{s}'}=V_{\underline{s},\underline{s}'}$. Using this property, we obtain the contribution for the $\vec{\mu}=(1,...,1)$ case as follows
	\begin{align}
		\sum_{\{\underline{s}_\tau\}}\prod_{\tau=1}^t V_{F\underline{s}_{\tau+1},\underline{s}_\tau}V^*_{F^2\underline{s}_{\tau+1},F\underline{s}_{\tau}}=\sum_{\{\underline{s}_\tau\}}\prod_{\tau=1}^t |V_{F\underline{s}_{\tau+1},\underline{s}_\tau}|^2.
	\end{align}
	Thus,
	\begin{align}
		\sum_{\{\underline{s}_\tau\}}\prod_{\tau=1}^t V_{F\underline{s}_{\tau+1},\underline{s}_\tau}V^*_{F^2\underline{s}_{\tau+1},F\underline{s}_{\tau}}&=\left(1+(-1)^t\right)+\left(\sum_{i=1}^{\mathcal{N}/2-1}\lambda_i^t+(-\lambda_i)^t\right).
	\end{align}
	Therefore,
	\begin{align}
		K^{(0)}(t)&=4t\left(1+(-1)^t\right)+\left(\sum_{i=1}^{\mathcal{N}/2-1}\lambda_i^t+(-\lambda_i)^t\right).
	\end{align}
	Beyond the Thouless time
	\begin{align}
		K^{(0)}(t)&=\begin{cases}
			8t,\;\text{ $t$ even}\\
			0,\;\text{ $t$ odd}
		\end{cases}.
		\label{K0t_N_not_L_by_2}
	\end{align}
	Other $\vec{\mu}$ have contribution that decay exponentially with time. This is similar to the CSE case [\hyperlink{S1}{S1}].\\
	\textbf{Case 2}: Subspace $\mathcal{H}_{L/2}$. In this case, $m_N=m_{L-N}$, therefore, $V$ only has one block. Therefore, $\mathcal{M}$ does not have a block-off-diagonal structure. Consequently, the eigenvalues of $\mathcal{M}$ are just $1>|\lambda_1|\geq...\geq|\lambda_{\mathcal{N}-1}|$. Thus,
	\begin{align}
		K^{(0)}(t)&=4t(1+\sum_{i=1}^{\mathcal{N}-1}\lambda_i^t).
		\label{K0t_N_equal_to_L_by_2}
	\end{align}
	Beyond Thouless time, $K^{(0)}\simeq 4t$. This is different from the RMT form for the COE class by a factor of 2. This results from $\hat{F}$ commuting with $\hat{V}$. However, $\hat{F}$ does not commute with $\hat{U}$ as it does not commute with $\hat{H}_0$.
	\subsubsection{Commutation of $\hat{F}$ with $\hat{V}$}
	Since $\hat{V}=e^{-i\hat{H}_1}$, it is sufficient to show that $[\hat{F},\hat{H}_1]=0$. Consider a basis state $|\underline{s}\rangle=|s_1,...,s_L\rangle$. Let us say that domain walls appear immediately after sites $i_1,...,i_n$ in the state $|\underline{s}\rangle$. Since $\hat{H}_1$ just exchanges spins across a domain wall, we can write
	\begin{align}
		\hat{H}_1|\underline{s}\rangle&=J(|\underline{s};i_1\leftrightarrow i_1+1\rangle+...+|\underline{s};i_n\leftrightarrow i_n+1\rangle),
	\end{align}
	where $|\underline{s};i_{n'}\leftrightarrow i_{n'}+1\rangle$ is a state obtained by swapping spins $s_{i_{n'}}$ and $s_{i_{n'}+1}$ in the state $|\underline{s}\rangle$, $\forall n'\in\{1,...,n\}$. Since $\hat{F}$ does not affect domain walls
	\begin{align}
		\hat{F}\hat{H}_1|\underline{s}\rangle&=J(\hat{F}|\underline{s};i_1\leftrightarrow i_1+1\rangle+...+\hat{F}|\underline{s};i_n\leftrightarrow i_n+1\rangle)\notag\\
		&=J(|F\underline{s};i_1\leftrightarrow i_1+1\rangle+...+|F\underline{s};i_n\leftrightarrow i_n+1\rangle)\notag\\
		&=\hat{H}_1\hat{F}|\underline{s}\rangle.
	\end{align}
	Since this is true for any basis state, we conclude the following
	\begin{align}
		[\hat{F},\hat{H}_1]=0.
	\end{align}
	Therefore, $[\hat{F},\hat{V}]=0$.

	\section*{References}
	\noindent\hypertarget{S1}{[S1]} V. Kumar, T. Prosen, and D. Roy, Lead,
	Leading and beyond leading-order spectral form factor in chaotic quantum many-body systems across all Dyson symmetry classes,
	arXiv preprint arXiv:2502.04152 (2025).

\end{document}